\documentclass[fleqn,usenatbib]{mnras}

\usepackage{newtxtext,newtxmath}

\usepackage[T1]{fontenc}

\DeclareRobustCommand{\VAN}[3]{#2}
\let\VANthebibliography\thebibliography
\def\thebibliography{\DeclareRobustCommand{\VAN}[3]{##3}\VANthebibliography}

\usepackage{graphicx}	
\usepackage{amsmath}	
\usepackage{xcolor} 
\usepackage{caption} 
\usepackage{subcaption}
\usepackage{array}
\newcommand{\photodepthone}{729$^{+54}_{-53}$}

\newcommand{\photodepthtwo}{1243$^{+54}_{-51}$}

\newcommand{\photoradone}{2.24 $\pm$ 0.08}
\newcommand{\photoradoneshort}{2.24}

\newcommand{\photoradtwo}{2.93$^{+0.07}_{-0.06}$}
\newcommand{\photoradtwoshort}{2.93}

\newcommand{\photoaRone}{25.35$^{+0.22}_{-0.19}$}

\newcommand{\photoaRtwo}{39.92$^{+0.35}_{-0.30}$}

\newcommand{\photosemimajorone}{0.090 $\pm$ 0.001}
\newcommand{\photosemimajoroneshort}{0.090}

\newcommand{\photosemimajortwo}{0.141$^{+0.002}_{-0.001}$}
\newcommand{\photosemimajortwoshort}{0.141}

\newcommand{\photoincone}{87.98$^{+0.16}_{-0.12}$}

\newcommand{\photoinctwo}{88.75$^{+0.02}_{-0.03}$}

\newcommand{\photodurone}{1.80$^{+0.07}_{-0.08}$}

\newcommand{\photodurtwo}{2.19 $\pm$ 0.03}

\newcommand{\photoinsolone}{33.14$^{+2.60}_{-2.45}$}
\newcommand{\photoinsoloneshort}{33.1}

\newcommand{\photoinsoltwo}{13.37$^{+1.05}_{-0.99}$}
\newcommand{\photoinsoltwoshort}{13.4}

\newcommand{\photoeqmTone}{668 $\pm$ 13}
\newcommand{\photoeqmToneshort}{668}

\newcommand{\photoeqmTtwo}{532 $\pm$ 10}
\newcommand{\photoeqmTtwoshort}{532}

\newcommand{\photomassOBHvolatileone}{6.21$^{+1.56}_{-1.43}$}

\newcommand{\photomassOBHvolatiletwo}{9.47$^{+2.44}_{-2.21}$}

\newcommand{\photoKOBHvolatileone}{2.11$^{+0.53}_{-0.49}$}

\newcommand{\photoKOBHvolatiletwo}{2.54$^{+0.67}_{-0.60}$}

\newcommand{\photoperiodone}{10.924709 $\pm$ 0.000032}
\newcommand{\photoperiodoneshort}{10.92}

\newcommand{\photoepochone}{1416.3453$^{+0.0034}_{-0.0028}$}

\newcommand{\photopone}{0.027 $\pm$ 0.001}

\newcommand{\photobone}{0.86 $\pm$ 0.03}
\newcommand{\photoboneshort}{0.86}

\newcommand{\photoeccone}{0.11$^{+0.04}_{-0.03}$}
\newcommand{\photoecconeshort}{0.11}

\newcommand{\photoomegaone}{160.5$^{+76.9}_{-75.7}$}

\newcommand{\photoperiodtwo}{21.583298$^{+0.000052}_{-0.000055}$}
\newcommand{\photoperiodtwoshort}{21.58}

\newcommand{\photoepochtwo}{1430.8296$^{+0.0027}_{-0.0025}$}

\newcommand{\photoptwo}{0.035 $\pm$ 0.001}

\newcommand{\photobtwo}{0.90$\pm$ 0.01}
\newcommand{\photobtwoshort}{0.90}

\newcommand{\photoecctwo}{0.04 $\pm$ 0.01}
\newcommand{\photoecctwoshort}{0.04}

\newcommand{\photoomegatwo}{247.9$^{+38.8}_{-45.4}$}

\newcommand{\photorho}{2583.24$^{+68.01}_{-57.80}$}

\newcommand{\photoperiodratioshort}{1.976}

\newcommand{\photoqoneCHEOPS}{0.54 $\pm$ 0.05}

\newcommand{\photoqtwoCHEOPS}{0.50$^{+0.06}_{-0.05}$}

\newcommand{\photomfluxCHEOPSone}{5.2e-5 $\pm$ 1.5e-5}

\newcommand{\photomfluxCHEOPStwo}{2.2e-5 $\pm$ 1.2e-5}

\newcommand{\photomfluxCHEOPSthree}{-10.0e-5 $\pm$ 1.6e-5}

\newcommand{\photomfluxCHEOPSfour}{5.3e-5 $\pm$ 1.5e-5}

\newcommand{\photomfluxCHEOPSfive}{-6.7e-6 $^{+14.9\text{e-}6}_{-14.7\text{e-}6}$}

\newcommand{\photomfluxCHEOPSsix}{-10.0e-5$^{+1.5\text{e-}5}_{-1.4\text{e-}5}$}

\newcommand{\photosigmawCHEOPSone}{138.2$^{+22.0}_{-21.3}$}

\newcommand{\photosigmawCHEOPStwo}{2.7$^{+20.9}_{-2.3}$}

\newcommand{\photosigmawCHEOPSthree}{125.7$^{+22.9}_{-26.2}$}

\newcommand{\photosigmawCHEOPSfour}{128.1$^{+24.3}_{-26.7}$}

\newcommand{\photosigmawCHEOPSfive}{117.2$^{+20.8}_{-22.1}$}

\newcommand{\photosigmawCHEOPSsix}{106.6$^{+25.4}_{-28.0}$}

\newcommand{\photoqoneLCO}{0.39$^{+0.06}_{-0.05}$}

\newcommand{\photoqtwoLCO}{0.04 $\pm$ 0.04}

\newcommand{\photomfluxLCOone}{2.8e-5 $\pm$ 6.9e-5}

\newcommand{\photomfluxLCOtwo}{-2.2e-5$^{+8.4\text{e-}5}_{-8.8\text{e-}5}$}

\newcommand{\photosigmawLCOone}{901.5$^{+45.4}_{-42.1}$}

\newcommand{\photoqoneTESS}{0.38$^{+0.04}_{-0.05}$}

\newcommand{\photoqtwoTESS}{0.36$^{+0.09}_{-0.08}$}

\newcommand{\photomfluxTESSone}{0.00020$^{+0.00049}_{-0.00045}$}

\newcommand{\photosigmawTESSone}{193.8 $\pm$ 9.0}

\newcommand{\photomfluxTESStwo}{-4.1e-5$^{+6.7\text{e-}5}_{-7.3\text{e-}5}$}

\newcommand{\photosigmawTESStwo}{193.5$^{+8.4}_{-8.3}$}

\newcommand{\photoGPsigmaTESSone}{0.0012$^{+0.0004}_{-0.0002}$}

\newcommand{\photoGPrhoTESSone}{2.52$^{+0.76}_{-0.48}$}

\newcommand{\photoGPSzeroTESSone}{3.34e-10$^{+2.87\text{e-}10}_{-1.40\text{e-}10}$}

\newcommand{\photoGPomegazeroTESSone}{28.59$^{+0.07}_{-0.06}$}

\newcommand{\photoGPQTESSone}{0.93$^{+0.45}_{-0.32}$}

\newcommand{\photoGPsigmaTESStwo}{32.3e-5$^{+5.0\text{e-}5}_{-3.7\text{e-}5}$}

\newcommand{\photoGPrhoTESStwo}{0.59$^{+0.13}_{-0.09}$}

\newcommand{\photothetazeroCHEOPSone}{11.5e-5 $\pm$ 1.5e-5}

\newcommand{\photothetaoneCHEOPSone}{-4.4e-5$^{+1.5\text{e-}5}_{-1.4\text{e-}5}$}

\newcommand{\photothetatwoCHEOPSone}{-5.6e-5$^{+1.8\text{e-}5}_{-1.9\text{e-}5}$}

\newcommand{\photothetazeroCHEOPStwo}{3.0e-5$^{+1.2\text{e-}5}_{-1.3\text{e-}5}$}

\newcommand{\photothetaoneCHEOPStwo}{-4.5e-5$^{+1.2\text{e-}5}_{-1.3\text{e-}5}$}

\newcommand{\photothetazeroCHEOPSthree}{5.5e-5 $\pm$ 1.6e-5}

\newcommand{\photothetaoneCHEOPSthree}{4.9e-5$^{+1.6\text{e-}5}_{-1.5\text{e-}5}$}

\newcommand{\photothetatwoCHEOPSthree}{-4.5e-5$^{+1.4\text{e-}5}_{-1.5\text{e-}5}$}

\newcommand{\photothetathreeCHEOPSthree}{-3.1e-5 $\pm$ 1.5e-5}

\newcommand{\photothetafourCHEOPSthree}{3.9e-5 $\pm$ 2.2e-5}

\newcommand{\photothetazeroCHEOPSfour}{-4.4e-5 $\pm$ 1.4e-5}

\newcommand{\photothetaoneCHEOPSfour}{5.2e-5 $\pm$ 1.7e-5}

\newcommand{\photothetatwoCHEOPSfour}{6.4e-5$^{+2.3\text{e-}5}_{-2.2\text{e-}5}$}

\newcommand{\photothetathreeCHEOPSfour}{12.2e-5$^{+1.4\text{e-}5}_{-1.3\text{e-}5}$}

\newcommand{\photothetazeroCHEOPSfive}{6.6e-5 $\pm$ 1.3e-5}

\newcommand{\photothetaoneCHEOPSfive}{-5.5e-5 $\pm$ 1.2e-5}

\newcommand{\photothetatwoCHEOPSfive}{4.1e-5$^{+1.1\text{e-}5}_{-1.2\text{e-}5}$}

\newcommand{\photothetathreeCHEOPSfive}{-3.4e-5$^{+1.2\text{e-}5}_{-1.3\text{e-}5}$}

\newcommand{\photothetafourCHEOPSfive}{4.9e-5 $\pm$ 1.7e-5}

\newcommand{\photothetafiveCHEOPSfive}{-5.5e-5 $\pm$ 1.8e-5}

\newcommand{\photothetazeroCHEOPSsix}{-4.8e-5 $\pm$ 1.3e-5}

\newcommand{\photothetaoneCHEOPSsix}{3.7e-5$^{+1.3\text{e-}5}_{-1.4\text{e-}5}$}

\newcommand{\photothetatwoCHEOPSsix}{14.6e-5 $\pm$ 1.3e-5}

\newcommand{\photothetazeroLCOone}{39.5e-5$^{+6.4\text{e-}5}_{-6.6\text{e-}5}$}

\newcommand{\photothetaoneLCOone}{13.7e-5$^{+6.2\text{e-}5}_{-6.3\text{e-}5}$}

\newcommand{\photothetazeroLCOtwo}{0.00025$^{+0.00016}_{-0.00015}$}

\newcommand{\photothetaoneLCOtwo}{0.00035 $\pm$ 0.00015}

\title[Discovery of the HD\,15906 Multiplanet System]{\textit{TESS} and \textit{CHEOPS} Discover Two Warm Sub-Neptunes Transiting the Bright K-dwarf HD\,15906 \thanks{This article uses data from the \textit{CHEOPS} programme CH\_PR110048.}}

\author[A.~Tuson et al.]{
\parbox{\textwidth}{
A.~Tuson$^{1}$$^{\href{https://orcid.org/0000-0002-2830-9064}{\includegraphics[scale=0.5]{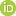}}}$\thanks{E-mail: \href{mailto:alt59@cam.ac.uk}{alt59@cam.ac.uk}},
D.~Queloz$^{1,2}$$^{\href{https://orcid.org/0000-0002-3012-0316}{\includegraphics[scale=0.5]{Figures/orcid.jpg}}}$,
H.~P.~Osborn$^{3,4}$$^{\href{https://orcid.org/0000-0002-4047-4724}{\includegraphics[scale=0.5]{Figures/orcid.jpg}}}$,
T.~G.~Wilson$^{5}$$^{\href{https://orcid.org/0000-0001-8749-1962}{\includegraphics[scale=0.5]{Figures/orcid.jpg}}}$,
M.~J.~Hooton$^{1,6}$$^{\href{https://orcid.org/0000-0003-0030-332X}{\includegraphics[scale=0.5]{Figures/orcid.jpg}}}$,
M.~Beck$^{7}$$^{\href{https://orcid.org/0000-0003-3926-0275}{\includegraphics[scale=0.5]{Figures/orcid.jpg}}}$,
M.~Lendl$^{7}$$^{\href{https://orcid.org/0000-0001-9699-1459}{\includegraphics[scale=0.5]{Figures/orcid.jpg}}}$,
G.~Olofsson$^{8}$$^{\href{https://orcid.org/0000-0003-3747-7120}{\includegraphics[scale=0.5]{Figures/orcid.jpg}}}$,
A.~Fortier$^{6,3}$$^{\href{https://orcid.org/0000-0001-8450-3374}{\includegraphics[scale=0.5]{Figures/orcid.jpg}}}$,
A.~Bonfanti$^{9}$$^{\href{https://orcid.org/0000-0002-1916-5935}{\includegraphics[scale=0.5]{Figures/orcid.jpg}}}$,
A.~Brandeker$^{8}$$^{\href{https://orcid.org/0000-0002-7201-7536}{\includegraphics[scale=0.5]{Figures/orcid.jpg}}}$,
L.~A.~Buchhave$^{10}$$^{\href{https://orcid.org/0000-0003-1605-5666}{\includegraphics[scale=0.5]{Figures/orcid.jpg}}}$,
A.~Collier~Cameron$^{5}$$^{\href{https://orcid.org/0000-0002-8863-7828}{\includegraphics[scale=0.5]{Figures/orcid.jpg}}}$,
D.~R.~Ciardi$^{11}$$^{\href{https://orcid.org/0000-0002-5741-3047}{\includegraphics[scale=0.5]{Figures/orcid.jpg}}}$,
K.~A.~Collins$^{12}$$^{\href{https://orcid.org/0000-0001-6588-9574}{\includegraphics[scale=0.5]{Figures/orcid.jpg}}}$,
D.~Gandolfi$^{13}$$^{\href{https://orcid.org/0000-0001-8627-9628}{\includegraphics[scale=0.5]{Figures/orcid.jpg}}}$,
Z.~Garai$^{14,15,16}$$^{\href{https://orcid.org/0000-0001-9483-2016}{\includegraphics[scale=0.5]{Figures/orcid.jpg}}}$,
S.~Giacalone$^{17}$$^{\href{https://orcid.org/0000-0002-8965-3969}{\includegraphics[scale=0.5]{Figures/orcid.jpg}}}$,
J.~Gomes~da~Silva$^{18}$,
S.~B.~Howell$^{19}$$^{\href{https://orcid.org/0000-0002-2532-2853}{\includegraphics[scale=0.5]{Figures/orcid.jpg}}}$,
J.~A.~Patel$^{8}$,
C.~M.~Persson$^{20}$$^{\href{https://orcid.org/0000-0003-1257-5146}{\includegraphics[scale=0.5]{Figures/orcid.jpg}}}$,
L.~M.~Serrano$^{13}$$^{\href{https://orcid.org/0000-0001-9211-3691}{\includegraphics[scale=0.5]{Figures/orcid.jpg}}}$,
S.~G.~Sousa$^{18}$$^{\href{https://orcid.org/0000-0001-9047-2965}{\includegraphics[scale=0.5]{Figures/orcid.jpg}}}$,
S.~Ulmer-Moll$^{7,6}$$^{\href{https://orcid.org/0000-0003-2417-7006}{\includegraphics[scale=0.5]{Figures/orcid.jpg}}}$,
A.~Vanderburg$^{4}$$^{\href{https://orcid.org/0000-0001-7246-5438}{\includegraphics[scale=0.5]{Figures/orcid.jpg}}}$,
C.~Ziegler$^{21}$,
Y.~Alibert$^{6}$$^{\href{https://orcid.org/0000-0002-4644-8818}{\includegraphics[scale=0.5]{Figures/orcid.jpg}}}$,
R.~Alonso$^{22,23}$$^{\href{https://orcid.org/0000-0001-8462-8126}{\includegraphics[scale=0.5]{Figures/orcid.jpg}}}$,
G.~Anglada$^{24,25}$$^{\href{https://orcid.org/0000-0002-3645-5977}{\includegraphics[scale=0.5]{Figures/orcid.jpg}}}$,
T.~Bárczy$^{26}$$^{\href{https://orcid.org/0000-0002-7822-4413}{\includegraphics[scale=0.5]{Figures/orcid.jpg}}}$,
D.~Barrado~Navascues$^{27}$$^{\href{https://orcid.org/0000-0002-5971-9242}{\includegraphics[scale=0.5]{Figures/orcid.jpg}}}$,
S.~C.~C.~Barros$^{18,28}$$^{\href{https://orcid.org/0000-0003-2434-3625}{\includegraphics[scale=0.5]{Figures/orcid.jpg}}}$,
W.~Baumjohann$^{9}$$^{\href{https://orcid.org/0000-0001-6271-0110}{\includegraphics[scale=0.5]{Figures/orcid.jpg}}}$,
T.~Beck$^{6}$,
W.~Benz$^{6,3}$$^{\href{https://orcid.org/0000-0001-7896-6479}{\includegraphics[scale=0.5]{Figures/orcid.jpg}}}$,
N.~Billot$^{7}$$^{\href{https://orcid.org/0000-0003-3429-3836}{\includegraphics[scale=0.5]{Figures/orcid.jpg}}}$,
X.~Bonfils$^{29}$$^{\href{https://orcid.org/0000-0001-9003-8894}{\includegraphics[scale=0.5]{Figures/orcid.jpg}}}$,
L.~Borsato$^{30}$$^{\href{https://orcid.org/0000-0003-0066-9268}{\includegraphics[scale=0.5]{Figures/orcid.jpg}}}$,
C.~Broeg$^{6,3}$$^{\href{https://orcid.org/0000-0001-5132-2614}{\includegraphics[scale=0.5]{Figures/orcid.jpg}}}$,
J.~Cabrera$^{31}$$^{\href{https://orcid.org/0000-0001-6653-5487}{\includegraphics[scale=0.5]{Figures/orcid.jpg}}}$,
S.~Charnoz$^{32}$$^{\href{https://orcid.org/0000-0002-7442-491X}{\includegraphics[scale=0.5]{Figures/orcid.jpg}}}$,
D.~M.~Conti$^{33}$$^{\href{https://orcid.org/0000-0003-2239-0567}{\includegraphics[scale=0.5]{Figures/orcid.jpg}}}$,
Sz.~Csizmadia$^{31}$$^{\href{https://orcid.org/0000-0001-6803-9698}{\includegraphics[scale=0.5]{Figures/orcid.jpg}}}$,
P.~E.~Cubillos$^{34,9}$,
M.~B.~Davies$^{35}$$^{\href{https://orcid.org/0000-0001-6080-1190}{\includegraphics[scale=0.5]{Figures/orcid.jpg}}}$,
M.~Deleuil$^{36}$$^{\href{https://orcid.org/0000-0001-6036-0225}{\includegraphics[scale=0.5]{Figures/orcid.jpg}}}$,
L.~Delrez$^{37,38}$$^{\href{https://orcid.org/0000-0001-6108-4808}{\includegraphics[scale=0.5]{Figures/orcid.jpg}}}$,
O.~D.~S.~Demangeon$^{18,28}$$^{\href{https://orcid.org/0000-0001-7918-0355}{\includegraphics[scale=0.5]{Figures/orcid.jpg}}}$,
B.-O.~Demory$^{3,6}$$^{\href{https://orcid.org/0000-0002-9355-5165}{\includegraphics[scale=0.5]{Figures/orcid.jpg}}}$,
D.~Dragomir$^{39}$$^{\href{https://orcid.org/0000-0003-2313-467X}{\includegraphics[scale=0.5]{Figures/orcid.jpg}}}$,
C.~D.~Dressing$^{17}$$^{\href{https://orcid.org/0000-0001-8189-0233}{\includegraphics[scale=0.5]{Figures/orcid.jpg}}}$,
D.~Ehrenreich$^{7,40}$$^{\href{https://orcid.org/0000-0001-9704-5405}{\includegraphics[scale=0.5]{Figures/orcid.jpg}}}$,
A.~Erikson$^{31}$,
Z.~Essack$^{41,4}$$^{\href{https://orcid.org/0000-0002-2482-0180}{\includegraphics[scale=0.5]{Figures/orcid.jpg}}}$,
J.~Farinato$^{30}$$^{\href{https://orcid.org/0000-0002-5840-8362}{\includegraphics[scale=0.5]{Figures/orcid.jpg}}}$,
L.~Fossati$^{9}$$^{\href{https://orcid.org/0000-0003-4426-9530}{\includegraphics[scale=0.5]{Figures/orcid.jpg}}}$,
M.~Fridlund$^{42,20}$$^{\href{https://orcid.org/0000-0002-0855-8426}{\includegraphics[scale=0.5]{Figures/orcid.jpg}}}$,
E.~Furlan$^{11}$$^{\href{https://orcid.org/0000-0001-9800-6248}{\includegraphics[scale=0.5]{Figures/orcid.jpg}}}$,
H.~Gill$^{17}$$^{\href{https://orcid.org/0000-0001-6171-7951}{\includegraphics[scale=0.5]{Figures/orcid.jpg}}}$,
M.~Gillon$^{37}$$^{\href{https://orcid.org/0000-0003-1462-7739}{\includegraphics[scale=0.5]{Figures/orcid.jpg}}}$,
C.~L.~Gnilka$^{19}$$^{\href{https://orcid.org/0000-0003-2519-6161}{\includegraphics[scale=0.5]{Figures/orcid.jpg}}}$,
E.~Gonzales$^{43}$,
M.~Güdel$^{44}$,
M.~N.~Günther$^{45}$$^{\href{https://orcid.org/0000-0002-3164-9086}{\includegraphics[scale=0.5]{Figures/orcid.jpg}}}$,
S.~Hoyer$^{36}$$^{\href{https://orcid.org/0000-0003-3477-2466}{\includegraphics[scale=0.5]{Figures/orcid.jpg}}}$,
K.~G.~Isaak$^{46}$$^{\href{https://orcid.org/0000-0001-8585-1717}{\includegraphics[scale=0.5]{Figures/orcid.jpg}}}$,
J.~M.~Jenkins$^{19}$$^{\href{https://orcid.org/0000-0002-4715-9460}{\includegraphics[scale=0.5]{Figures/orcid.jpg}}}$,
L.~L.~Kiss$^{47,48}$,
J.~Laskar$^{49}$$^{\href{https://orcid.org/0000-0003-2634-789X}{\includegraphics[scale=0.5]{Figures/orcid.jpg}}}$,
D.~W.~Latham$^{12}$$^{\href{https://orcid.org/0000-0001-9911-7388}{\includegraphics[scale=0.5]{Figures/orcid.jpg}}}$,
N.~Law$^{50}$,
A.~Lecavelier~des~Etangs$^{51}$$^{\href{https://orcid.org/0000-0002-5637-5253}{\includegraphics[scale=0.5]{Figures/orcid.jpg}}}$,
G.~Lo~Curto$^{52}$$^{\href{https://orcid.org/0000-0002-1158-9354}{\includegraphics[scale=0.5]{Figures/orcid.jpg}}}$,
C.~Lovis$^{7}$$^{\href{https://orcid.org/0000-0001-7120-5837}{\includegraphics[scale=0.5]{Figures/orcid.jpg}}}$,
R.~Luque$^{53}$$^{\href{https://orcid.org/0000-0002-4671-2957}{\includegraphics[scale=0.5]{Figures/orcid.jpg}}}$,
D.~Magrin$^{30}$$^{\href{https://orcid.org/0000-0003-0312-313X}{\includegraphics[scale=0.5]{Figures/orcid.jpg}}}$,
A.~W.~Mann$^{50}$$^{\href{https://orcid.org/0000-0003-3654-1602}{\includegraphics[scale=0.5]{Figures/orcid.jpg}}}$,
P.~F.~L.~Maxted$^{54}$$^{\href{https://orcid.org/0000-0003-3794-1317}{\includegraphics[scale=0.5]{Figures/orcid.jpg}}}$,
M.~Mayor$^{7}$,
S.~McDermott$^{55}$,
M.~Mecina$^{44}$,
C.~Mordasini$^{6,3}$,
A.~Mortier$^{56}$$^{\href{https://orcid.org/0000-0001-7254-4363}{\includegraphics[scale=0.5]{Figures/orcid.jpg}}}$,
V.~Nascimbeni$^{30}$$^{\href{https://orcid.org/0000-0001-9770-1214}{\includegraphics[scale=0.5]{Figures/orcid.jpg}}}$,
R.~Ottensamer$^{44}$,
I.~Pagano$^{57}$$^{\href{https://orcid.org/0000-0001-9573-4928}{\includegraphics[scale=0.5]{Figures/orcid.jpg}}}$,
E.~Pallé$^{22}$$^{\href{https://orcid.org/0000-0003-0987-1593}{\includegraphics[scale=0.5]{Figures/orcid.jpg}}}$,
G.~Peter$^{58}$$^{\href{https://orcid.org/0000-0001-6101-2513}{\includegraphics[scale=0.5]{Figures/orcid.jpg}}}$,
G.~Piotto$^{30,59}$$^{\href{https://orcid.org/0000-0002-9937-6387}{\includegraphics[scale=0.5]{Figures/orcid.jpg}}}$,
D.~Pollacco$^{60}$,
T.~Pritchard$^{61}$,
R.~Ragazzoni$^{30,59}$$^{\href{https://orcid.org/0000-0002-7697-5555}{\includegraphics[scale=0.5]{Figures/orcid.jpg}}}$,
N.~Rando$^{45}$,
F.~Ratti$^{45}$,
H.~Rauer$^{31,62,63}$$^{\href{https://orcid.org/0000-0002-6510-1828}{\includegraphics[scale=0.5]{Figures/orcid.jpg}}}$,
I.~Ribas$^{24,25}$$^{\href{https://orcid.org/0000-0002-6689-0312}{\includegraphics[scale=0.5]{Figures/orcid.jpg}}}$,
G.~R.~Ricker$^{4}$$^{\href{https://orcid.org/0000-0003-2058-6662}{\includegraphics[scale=0.5]{Figures/orcid.jpg}}}$,
M.~Rieder$^{6}$,
N.~C.~Santos$^{18,28}$$^{\href{https://orcid.org/0000-0003-4422-2919}{\includegraphics[scale=0.5]{Figures/orcid.jpg}}}$,
A.~B.~Savel$^{64}$$^{\href{https://orcid.org/0000-0002-2454-768X}{\includegraphics[scale=0.5]{Figures/orcid.jpg}}}$,
G.~Scandariato$^{57}$$^{\href{https://orcid.org/0000-0003-2029-0626}{\includegraphics[scale=0.5]{Figures/orcid.jpg}}}$,
R.~P.~Schwarz$^{12}$$^{\href{https://orcid.org/0000-0001-8227-1020}{\includegraphics[scale=0.5]{Figures/orcid.jpg}}}$,
S.~Seager$^{41,4,65}$$^{\href{https://orcid.org/0000-0002-6892-6948}{\includegraphics[scale=0.5]{Figures/orcid.jpg}}}$,
D.~Ségransan$^{7}$$^{\href{https://orcid.org/0000-0003-2355-8034}{\includegraphics[scale=0.5]{Figures/orcid.jpg}}}$,
A.~Shporer$^{4}$$^{\href{https://orcid.org/0000-0002-1836-3120}{\includegraphics[scale=0.5]{Figures/orcid.jpg}}}$,
A.~E.~Simon$^{6}$$^{\href{https://orcid.org/0000-0001-9773-2600}{\includegraphics[scale=0.5]{Figures/orcid.jpg}}}$,
A.~M.~S.~Smith$^{31}$$^{\href{https://orcid.org/0000-0002-2386-4341}{\includegraphics[scale=0.5]{Figures/orcid.jpg}}}$,
M.~Steller$^{9}$$^{\href{https://orcid.org/0000-0003-2459-6155}{\includegraphics[scale=0.5]{Figures/orcid.jpg}}}$,
C.~Stockdale$^{66}$$^{\href{https://orcid.org/0000-0003-2163-1437}{\includegraphics[scale=0.5]{Figures/orcid.jpg}}}$,
Gy.~M.~Szabó$^{15,16}$,
N.~Thomas$^{6}$,
G.~Torres$^{12}$$^{\href{https://orcid.org/0000-0002-5286-0251}{\includegraphics[scale=0.5]{Figures/orcid.jpg}}}$,
R.~Tronsgaard$^{10}$$^{\href{https://orcid.org/0000-0003-1001-0707}{\includegraphics[scale=0.5]{Figures/orcid.jpg}}}$,
S.~Udry$^{7}$$^{\href{https://orcid.org/0000-0001-7576-6236}{\includegraphics[scale=0.5]{Figures/orcid.jpg}}}$,
B.~Ulmer$^{58}$,
V.~Van~Grootel$^{38}$$^{\href{https://orcid.org/0000-0003-2144-4316}{\includegraphics[scale=0.5]{Figures/orcid.jpg}}}$,
R.~Vanderspek$^{4}$$^{\href{https://orcid.org/0000-0001-6763-6562}{\includegraphics[scale=0.5]{Figures/orcid.jpg}}}$,
J.~Venturini$^{7}$,
N.~A.~Walton$^{67}$$^{\href{https://orcid.org/0000-0003-3983-8778}{\includegraphics[scale=0.5]{Figures/orcid.jpg}}}$,
J.~N.~Winn$^{68}$$^{\href{https://orcid.org/0000-0002-4265-047X}{\includegraphics[scale=0.5]{Figures/orcid.jpg}}}$
and B.~Wohler$^{19,69}$$^{\href{https://orcid.org/0000-0002-5402-9613}{\includegraphics[scale=0.5]{Figures/orcid.jpg}}}$
}
\\
Affiliations can be found after the references.
}

\date{Accepted 2023 April 22. Received 2023 April 6; in original form 2023 January 27}

\pubyear{2023}

\begin{document}
\label{firstpage}
\pagerange{\pageref{firstpage}--\pageref{lastpage}}
\maketitle

\begin{abstract}
We report the discovery of two warm sub-Neptunes transiting the bright (G = 9.5 mag) K-dwarf HD\,15906 (TOI\,461, TIC\,4646810). This star was observed by the \textit{Transiting Exoplanet Survey Satellite (TESS)} in sectors 4 and 31, revealing two small transiting planets. The inner planet, HD\,15906\,b, was detected with an unambiguous period but the outer planet, HD\,15906\,c, showed only two transits separated by $\sim$ 734 days, leading to 36 possible values of its period. We performed follow-up observations with the \textit{CHaracterising ExOPlanet Satellite (CHEOPS)} to confirm the true period of HD\,15906\,c and improve the radius precision of the two planets. From \textit{TESS}, \textit{CHEOPS} and additional ground-based photometry, we find that HD\,15906\,b has a radius of \photoradone\ R$_\oplus$ and a period of \photoperiodone\ days, whilst HD\,15906\,c has a radius of \photoradtwo\ R$_\oplus$ and a period of \photoperiodtwo\ days. Assuming zero bond albedo and full day-night heat redistribution, the inner and outer planet have equilibrium temperatures of \photoeqmTone\ K and \photoeqmTtwo\ K, respectively. The HD\,15906 system has become one of only six multiplanet systems with two warm ($\lesssim$ 700 K) sub-Neptune sized planets transiting a bright star (G $\leq$ 10 mag). It is an excellent target for detailed characterisation studies to constrain the composition of sub-Neptune planets and test theories of planet formation and evolution.
\end{abstract}

\begin{keywords}
planets and satellites: detection -- techniques: photometric -- planets and satellites: fundamental parameters -- stars: fundamental parameters -- stars: individual: HD\,15906 (TOI\,461, TIC\,4646810)
\end{keywords}



\section{Introduction}\label{sec:intro}

Exoplanet population studies have shown that small planets between the size of Earth and Neptune (so-called super-Earths and sub-Neptunes) are the most ubiquitous in our galaxy \citep{Fressin2013,kunimoto_kepler}. However, there is a statistically significant drop in the occurrence rate of close-in planets (orbital period $\lesssim$ 100 days) with radii between 1.5 and 2.0 R$_\oplus$ \citep{fulton-gap,Fulton_petigura-18,vaneylen18}. One theory is that this radius gap represents a transition between predominantly rocky planets and planets with extended H/He envelopes. There are several possible explanations for how this could arise, including gas-poor formation \citep{lee2014,lee2016,lopez2018,lee2022}, core-powered mass loss \citep{ginzburg2018,gupta2019,gupta2020} and photoevaporation \citep{owen2013,owen2017,lopez2018}. More recently, \citet{Luque_density-gap} studied small planets transiting M-dwarfs and found that the radius gap might actually be a density gap separating rocky and water-rich planets. To test these theories we need small, well-characterised planets spanning a range of equilibrium temperatures, T$_{\text{eq}}$.

Warm (defined in this paper as T$_{\text{eq}} \lesssim$ 700 K) sub-Neptunes transiting bright stars are particularly interesting targets for detailed characterisation studies. These planets are amenable to observations to, for example, precisely measure their radii and masses and probe their atmospheres \citep[e.g.,][]{2014Kreidberg, 2019Benneke, K218_tsiaras, nu2lupid, 2021Scarsdale, Wilson2022, 2022Orell}. From measurements of a planet's mass and radius, the bulk density can be calculated and its internal composition inferred. This can help distinguish between different formation mechanisms for small planets \citep{bean2021}. Furthermore, since warm planets are less affected by radiation from their host star, they can retain their primordial atmospheres. Observations of these atmospheres and measurements of the carbon-to-oxygen ratio could therefore reveal their formation history \citep{2011Oberg,2014Madhu}. Multiplanet systems are especially powerful because they allow us to study planets that formed from a common protoplanetary disc, leading to additional constraints on formation and evolution models \citep[e.g.,][]{2011Lissauer, 2012Fang, 2018Weissa, VanEylen-2019, Weiss2022}. 

The \textit{Transiting Exoplanet Survey Satellite} \citep[\textit{TESS};][]{TESS-mission} is an all-sky transit survey searching for exoplanets around some of the brightest and closest stars. Since its launch in 2018, it has discovered a plethora of planets orbiting bright stars, including many super-Earth and sub-Neptune planets \citep[e.g.,][]{gandolfi2018,2019Vanderburg,2020Plavchan,2020Teske,2021Leleu,serrano2022}. However, due to the nature of its observing strategy, \textit{TESS} is limited in its ability to discover long-period exoplanets. During its two-year primary mission, \textit{TESS} observed the majority of the sky for $\sim$ 27 consecutive days. This means that planets with periods longer than $\sim$ 27\,days, and some planets with periods between $\sim$ 13 - 27\,days, would only have been observed to transit once, if at all. These single transit detections are known as ``monotransits'' and their orbital periods are unknown, although the shape of the transit allows the period to be constrained \citep[e.g.,][]{2015Wang,2016MNRAS.457.2273O}. In its extended mission, \textit{TESS} reobserved the sky approximately two years after the first observation and, as predicted by simulations \citep{cooke-2019,cooke-2020,cooke-2021}, a large fraction of primary mission monotransits were observed to transit a second time. The result was a sample of ``duotransits'' - planetary candidates with two observed transits separated by a large gap, typically two years. From the two non-consecutive transits, the period of the planet remains unknown, but there now exists a discrete set of allowed period aliases. These aliases can be calculated according to $P_{n} = T_{\text{diff}} / n$, where $T_{\text{diff}}$ is the time between the two transit events and $n \in$ \{1, 2, ... , $n_{\text{max}}$\}. The maximum value, $n_{\text{max}}$, is dictated by the non-detection of a third transit in the \textit{TESS} data. 

Both monotransits and duotransits are the observational signatures of long-period planets (P $\gtrsim$ 20\,days). However, follow-up photometric or spectroscopic observations are required to recover their true periods. The follow-up of monotransits requires a blind survey approach \citep[e.g.,][]{gill_2020c, Villanueva_mono, ulmer-moll_mono+duo}, whereas the period aliases of a duotransit allow more targeted follow-up observations \citep[e.g.,][]{ulmer-moll_mono+duo,grieves_duo}. So far, the majority of these follow-up efforts have focused on giant planets, partly because their deeper transits facilitate ground-based observations. It's vital that we also pursue follow-up of shallow duotransits to expand the sample of small, long-period planets, including warm sub-Neptunes.  

The \textit{CHaracterising ExOPlanet Satellite} \citep[\textit{CHEOPS};][]{CHEOPS-mission-Benz} is an ESA mission dedicated to the follow-up of known exoplanets. The effective aperture diameter of \textit{CHEOPS} ($\sim$ 30\,cm) is about three times larger than that of \textit{TESS} ($\sim$ 10\,cm), allowing it to achieve a higher per-transit signal-to-noise ratio \citep[SNR; e.g.,][]{bonfanti21}. Furthermore, \textit{CHEOPS} performs targeted photometric observations to observe multiple transits of a planet without the need for continuous monitoring. \textit{CHEOPS} is therefore very well-suited to the follow-up of small, long-period planets from \textit{TESS}. We have a dedicated \textit{CHEOPS} Guaranteed Time Observing (GTO) programme to recover the periods of \textit{TESS} duotransits, focusing on small planets that cannot be observed from the ground. We select most of our targets from the \textit{TESS} Objects of Interest (TOI) Catalog \citep{2021-Guerrero} and from our specialised duotransit pipeline \citep{2022pipeline}. Through our \textit{CHEOPS} programme, we have recovered the periods of two duotransits in the TOI\,2076 system \citep{TOI-2076}, one duotransit in the HIP\,9618 system \citep{Osborn2023}, one duotransit in the TOI\,5678 system \citep{UlmerMoll2023} and one duotransit in the HD\,22946 system \citep{Garai2023}. 

In this paper, we report the discovery of two warm sub-Neptunes transiting the bright (G = 9.5 mag) K-dwarf HD\,15906 (TOI\,461, TIC\,4646810). This paper is organised as follows. In Section \ref{sec:observations}, we provide details of the photometric and spectroscopic observations used in our analyses. In Section \ref{sec:stellar}, we describe our characterisation of the host star and in Section \ref{sec:all-analysis} we describe the analyses of the system. Section \ref{sec:all-results} presents the results of our analyses and in Section \ref{sec:vetting+validation} we validate the two planets. Finally, in Section \ref{sec:discussion}, we present a discussion of our findings and outlook for future observations. 

\section{Observations}\label{sec:observations}

\subsection{\textit{TESS} Photometry}\label{sec:tess-obs}
HD\,15906 was observed by \textit{TESS} (camera 1, CCD 1) at two-minute cadence in sector 4 (18 October to 15 November 2018) and sector 31 (21 October to 19 November 2020). During both sectors, the instrument suffered from operational anomalies causing interruptions in data collection. In sector 4, no data was collected between 1418.5 and 1421.2 (BJD - 2457000) due to an instrument shutdown and sector 31 ended $\sim$ 2 days earlier than scheduled due to a star tracker anomaly. No more \textit{TESS} observations are scheduled before the end of Cycle 6 (01 October 2024). 

\begin{figure*}
\begin{center}
\includegraphics[width=\textwidth]{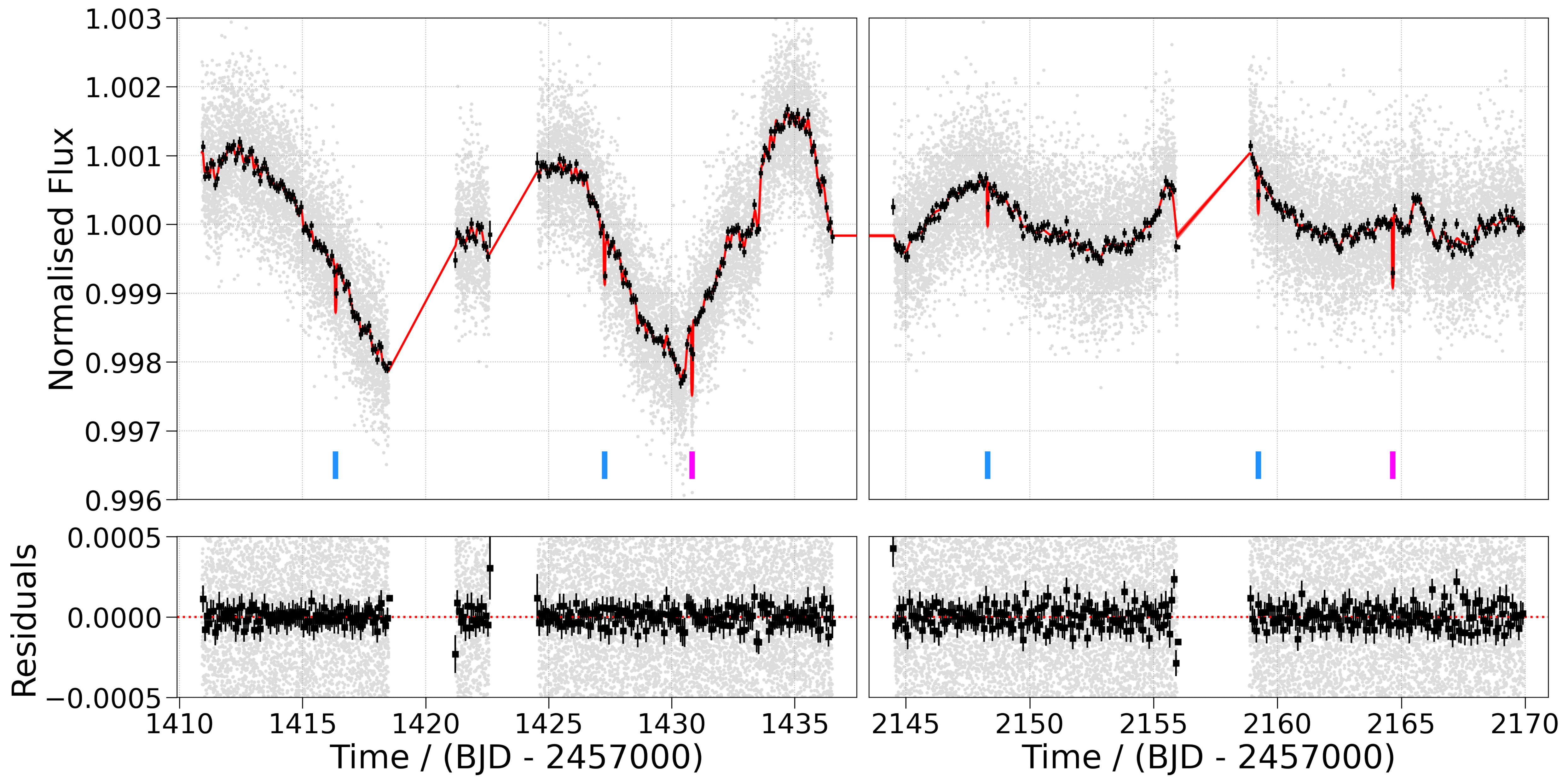}
\caption{\textit{TESS} PDCSAP lightcurve from sector 4 (left) and sector 31 (right). The 2 minute cadence data (grey) has been binned to 120 minutes (black squares) to guide the eye. The red line is the median model from the global photometric fit, described in Section \ref{sec:photo-analysis}, and the red shaded region (difficult to see on this scale) is the 1$\sigma$ uncertainty on the model. The blue and pink markers indicate the mid-transit times of the inner and outer planet, respectively. The lower panels show the residuals of the median model.}
\label{fig:tess_photo_fit}
\end{center}
\end{figure*}

The \textit{TESS} observations were reduced and analysed by the Science Processing Operations Center \citep[SPOC;][]{Jenkins2010,Jenkins2016} at the NASA Ames Research Center. We downloaded the lightcurve files, created by SPOC pipeline version 5.0.20-20201120, from the Mikulski Archive for Space Telescopes (MAST) portal\footnote{\url{https://mast.stsci.edu/portal/Mashup/Clients/Mast/Portal.html}}. These files include a Simple Aperture Photometry \citep[SAP;][]{twicken2010,morris2020} lightcurve and a Presearch Data Conditioning Simple Aperture Photometry \citep[PDCSAP;][]{Stumpe2012,Stumpe2014,Smith2012} lightcurve that has been corrected for instrumental systematics. For our analysis, we used the PDCSAP lightcurves. Following the advice in the \textit{TESS} Archive Manual\footnote{\label{footnote: tess-archive-manual}\url{https://outerspace.stsci.edu/display/TESS/TESS+Archive+Manual}}, we rejected all data points of lesser quality using the binary digits 1, 2, 3, 4, 5, 6, 8, 10, 13 and 15. We then rejected outliers from the lightcurve by calculating the mean absolute deviation (MAD) of the data from the median smoothed lightcurve and rejecting data greater than 5 x MAD away from the smoothed dataset. We repeated this process until no more outliers remained and the resulting \textit{TESS} lightcurve is shown in Figure \ref{fig:tess_photo_fit}. 

From the sector 4 data alone, the transiting planet search \citep[TPS;][]{Jenkins2002,Jenkins2010b,Jenkins2020} performed by the SPOC pipeline identified a single planet candidate. This planet candidate was announced as TOI\,461.01 in February 2019 with an epoch of 1416.3 (BJD - 2457000) and a period of 14.5 days. When the sector 31 data became available, we performed a by-eye search of the lightcurve and realised that TOI\,461.01 was actually a combination of two planetary signals. There was one multi-transiting planet candidate, with an epoch of 1416.3 (BJD - 2457000) and a period of 10.9 days, and one duotransit - a planet candidate with one transit in sector 4 and one transit in sector 31, separated by $T_{\text{diff}} \sim$ 733.8 days. When a multi-sector TPS was performed by SPOC in May 2021, it correctly identified the multi-transiting planet candidate and the ephemeris of TOI\,461.01 was updated accordingly. This planet candidate passed all of the SPOC vetting tests \citep{Twicken:DVdiagnostics2018,Li:DVmodelFit2019}, including the difference image centroid test, the odd-even depth test and the ghost diagnostic test, and the source of the transit signal was localised within 6.0 $\pm$ 4.2\arcsec\ of HD\,15906. The duotransit did not receive a TOI designation. 

The \textit{TESS} data contains four transits of the inner planet candidate (TOI\,461.01, hereafter called HD\,15906\,b) and two transits of the outer planet candidate (hereafter called HD\,15906\,c). From the \textit{TESS} data alone, the orbital period of the outer planet candidate was ambiguous. There existed a discrete set of 36 allowed period aliases, in the range 20.4 - 733.8 days (see Section \ref{sec:tess-only}), and follow-up observations were therefore required to recover the correct period.

\subsection{\textit{CHEOPS} Photometry}\label{sec:cheops-obs}
To recover the period of the outer planet candidate, we observed HD\,15906 through the \textit{CHEOPS} GTO programme CH\_PR110048 ("Duos - Recovering long period duo-transiting planets"). Our observing strategy was informed by our analysis of the \textit{TESS} data (see Section \ref{sec:tess-only}). We scheduled \textit{CHEOPS} observations of the 13 highest probability period aliases ($P$ $<$ 31 days), giving highest priority to the four most probable period aliases ($P$ $<$ 22.5 days). The first and second \textit{CHEOPS} visits did not reveal a transit and ruled out six period aliases in total. The third \textit{CHEOPS} visit revealed a transit and uniquely confirmed a period of $\sim$ 21.6 days for HD\,15906\,c. A fourth \textit{CHEOPS} visit, scheduled before the period had been confirmed, did not reveal a transit. We scheduled one additional observation of both HD\,15906\,b and c to improve radius precision and search for possible transit timing variations (TTVs). For all of our \textit{CHEOPS} observations, we used an exposure time of 60 seconds with no on-board image stacking, resulting in a final lightcurve cadence of 60 seconds. A summary of our six \textit{CHEOPS} observations is presented in Table \ref{tab:cheops-observations}.
\begin{table*}
\centering
  \caption{\textit{CHEOPS} observations of HD\,15906. See Section \ref{sec:cheops-detrend} for a description of the detrending terms.}
  \resizebox{\textwidth}{!}
  {\begin{tabular}{ccccccccc}
    \hline
    Visit & File Key & Start Time [UTC] & Dur. / hrs & Eff. / \% & Planet & Transit Observed? & Detrending Terms\\
    \hline
    1 & CH\_PR110048\_TG005901\_V0200 & 2021-09-21 12:41:29 & 8.10 & 71 & c & no & bg, t, cos($\phi$)\\
    2 & CH\_PR110048\_TG006201\_V0200 & 2021-09-29 20:02:09 & 8.10 & 74 & c & no & x, y\\
    3 & CH\_PR110048\_TG005301\_V0200 & 2021-09-30 19:07:09 & 8.10 & 73 & c & yes & bg, x, y, t, cos(3$\phi$)\\
    4 & CH\_PR110048\_TG005101\_V0200 & 2021-10-03 01:25:29 & 7.99 & 74 & c & no & bg, y, t, cos(3$\phi$)\\
    5 & CH\_PR110048\_TG009901\_V0200 & 2021-10-10 02:48:09 & 9.27 & 86 & b & yes & bg, x, y, t, cos(2$\phi$), sin(3$\phi$)\\
    6 & CH\_PR110048\_TG009801\_V0200 & 2021-11-12 22:11:30 & 8.39 & 74 & c & yes & bg, y, t\\
    \hline
  \end{tabular}}
\label{tab:cheops-observations}
\end{table*}

\begin{figure}
\begin{center}
\includegraphics[width=\columnwidth]{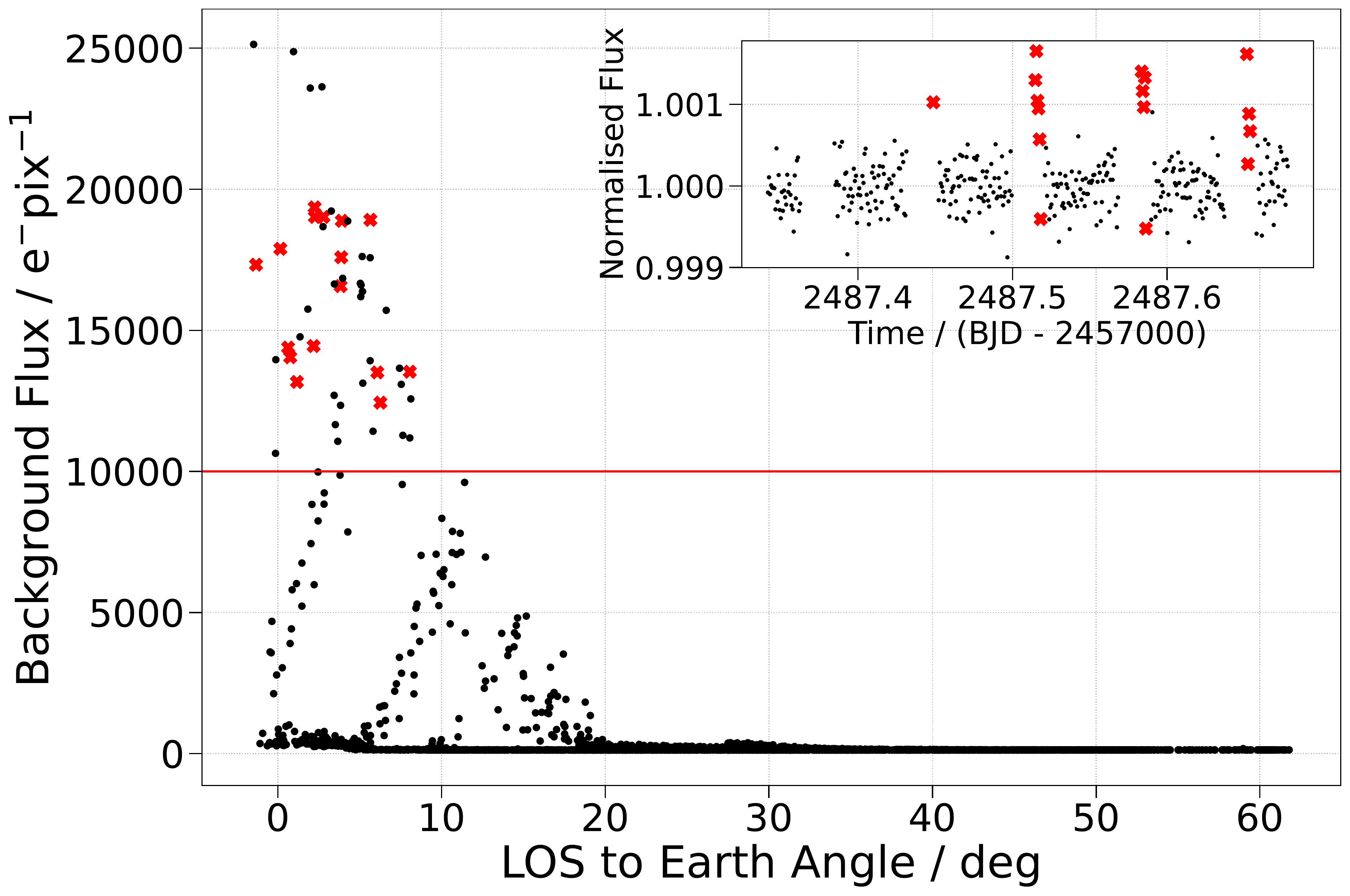}
\caption{Background flux versus angle between the instrument's line of sight (LOS) and the Earth limb for non-flagged data from all six \textit{CHEOPS} observations. The red horizontal line represents our background cut of 10\,000 e$^{-}$pix$^{-1}$ and the red crosses correspond to the data removed from \textit{CHEOPS} visit 2. Inset: \textit{CHEOPS} lightcurve from visit 2. Only non-flagged data is plotted and the points shown as red crosses were removed by the background cut.}
\label{fig:cheops-bg}
\end{center}
\end{figure}

Due to the fact \textit{CHEOPS} is in a low-Earth orbit, with an orbital period $\sim$ 98.7 minutes, our observations suffer from interruptions caused by high levels of stray light (from the illuminated Earth limb), occultations of the target by the Earth and passage of the satellite through the South Atlantic Anomaly \citep[SAA;][]{CHEOPS-mission-Benz}. These interruptions result in gaps in the \textit{CHEOPS} lightcurves, reducing the observing efficiency (time spent collecting data divided by the duration of the visit). The efficiencies of our six visits are included in Table \ref{tab:cheops-observations} and the inset of Figure \ref{fig:cheops-bg} shows examples of the lightcurve gaps. 

For each of our \textit{CHEOPS} visits, sub-array images and lightcurves were produced by the Data Reduction Pipeline \citep[DRP 13.1.0;][]{Hoyer-DRP}. The sub-array images are circular, with a diameter of 200 pixels ($\sim$ 200\arcsec), and are centred on the target star. They are calibrated and corrected for effects such as cosmic ray hits, smear trails caused by nearby stars and variations in background flux. From these images, the DRP uses aperture photometry to produce four lightcurves using circular apertures of different sizes. The DEFAULT, RINF and RSUP apertures are predefined with radii 25, 22.5 and 30 pixels respectively. The OPTIMAL aperture is selected per visit to minimise the effect of instrumental noise and contamination from nearby stars. We downloaded the \textit{CHEOPS} sub-array images and DRP lightcurves from the Data \& Analysis Center for Exoplanets \citep[DACE\footnote{\label{footnote:dace}\url{https://dace.unige.ch/dashboard}};][]{DACE}. Alongside the time, flux and flux error, the DRP lightcurves include a set of detrending vectors that can be used to model instrumental trends in the lightcurve. This includes the background flux, the smearing and contamination from nearby stars, the x and y centroid position of the target star and the roll angle of the satellite. \textit{CHEOPS} rolls around its pointing direction once per orbit, to maintain thermal stability, and every data point has an associated roll angle between 0 and 360 degrees.

We also extracted our own lightcurves from the \textit{CHEOPS} sub-array images using point-spread function (PSF) photometry. This technique is complementary to the aperture photometry performed by the DRP. We used the PSF Imagette Photometric Extraction (PIPE) package\footnote{\label{footnote:pipe}\url{https://github.com/alphapsa/PIPE}} \citep[see description in][]{deline-2022}, which was developed specifically for \textit{CHEOPS} data. PIPE photometry is less sensitive to contamination from nearby stars and the effects of smear trails are removed before extracting the flux \citep{LMS2022}. The PIPE lightcurves contain the time, flux and flux error, as well as the same detrending vectors as the DRP lightcurves, with the exception of smearing and contamination.

We performed preliminary transit fits of the DRP and PIPE lightcurves using \texttt{pycheops}\footnote{\label{footnote:pycheops}\url{https://github.com/pmaxted/pycheops}} \citep{pycheops-Maxted} and found that the planet parameters obtained in each case were fully compatible. We then compared the photometric precision of the DRP and PIPE lightcurves for each \textit{CHEOPS} visit. Firstly, we performed iterative outlier clipping as described in Section \ref{sec:tess-obs}. Then, we calculated the MAD of each clipped lightcurve, see Table \ref{tab:cheops-lcs-mad}. We found that for four of the six visits, including all three transit observations, the PIPE lightcurve had the lowest MAD. In the other two visits, the MAD of the PIPE lightcurve was comparable to the lowest value. We therefore chose to use the PIPE photometry for our analysis.  
\begin{table}
\centering
  \caption{Mean absolute deviation (MAD) of the clipped \textit{CHEOPS} lightcurves. The lightcurve with the lowest MAD for each visit is in bold.}
  {\begin{tabular}{cccccc}
    \hline
    Visit & \multicolumn{5}{c}{MAD / ppm}\\
    {} & DEFAULT & OPTIMAL & RINF & RSUP & PIPE\\
    \hline
    1 & 228.8 & 239.8 & \textbf{225.2} & 239.2 & 231.3\\
    2 & 210.2 & 275.5 & 220.4 & 221.8 & \textbf{208.2}\\
    3 & 291.9 & 346.0 & 348.4 & 326.3 & \textbf{217.4}\\
    4 & \textbf{236.5} & 289.9 & 258.3 & 260.3 & 247.4\\
    5 & 230.3 & 348.9 & 235.3 & 279.8 & \textbf{223.6}\\
    6 & 211.4 & 237.1 & 227.6 & 214.2 & \textbf{209.5}\\
    \hline
  \end{tabular}}
\label{tab:cheops-lcs-mad}
\end{table}

To prepare the PIPE lightcurves for our analysis, we performed a series of cuts to the data. Firstly, we rejected all flagged data. PIPE assigns flags to data of lesser quality, for example due to outliers in centroid position or a large number of bad pixels in the frame. Next, we performed a cut to remove data with high background flux. Some of the \textit{CHEOPS} lightcurves showed sharp spikes in the target's flux immediately before and/or after the data gaps (see an example in the inset of Figure \ref{fig:cheops-bg}). These spikes coincide with the target star approaching the illuminated Earth limb, causing high levels of scattered light and an increase in the background flux. This can be seen in Figure \ref{fig:cheops-bg}, where we have plotted the background flux against the angle between the instrument's line of sight (LOS) and the Earth limb for all six \textit{CHEOPS} visits. Notice that not all of the observations with a small angle have a high background flux; it is only when the star approaches the Earth's day side that there is a significant increase in scattered light. We removed all data with background flux $>$ 10\,000 e$^{-}$pix$^{-1}$ because this adequately reduced the spikes in the lightcurves whilst retaining as much data as possible. After the background cut, we removed remaining outliers from the lightcurves using the same iterative MAD clipping described in Section \ref{sec:tess-obs}. In total, these three cuts rejected 42/346 ($\sim$ 12\%), 33/358 ($\sim$ 9\%), 36/356 ($\sim$ 10\%), 47/353 ($\sim$ 13\%), 43/476 ($\sim$ 9\%) and 31/375 ($\sim$ 8\%) data points from each respective \textit{CHEOPS} visit. 

Following these steps, the PIPE photometry still contained trends correlated with instrumental parameters such as background flux, centroid position and roll angle. Rather than pre-detrending the data, we chose to fit a joint transit and detrending model, see Section \ref{sec:photo-analysis}. 

\subsection{LCOGT Photometry}\label{sec:ground-photometry}
We conducted ground-based photometric follow-up observations of HD\,15906 as part of the \textit{TESS} Follow-up Observing Program\footnote{\url{https://tess.mit.edu/followup/}} \citep[TFOP;][]{collins:2019} Sub Group 1. 

We used the \textit{TESS} Transit Finder, a customised version of the {\tt Tapir} software package \citep{Jensen:2013}, to schedule our transit observations. We observed full predicted transit windows of HD\,15906\,b in Pan-STARRS $z$-short band using the Las Cumbres Observatory Global Telescope \citep[LCOGT;][]{Brown:2013} 1.0\,m network nodes at Siding Spring Observatory and McDonald Observatory on 27 August 2021 and 1 November 2021, respectively. See Table \ref{tab:lco-observations} for a summary of these observations. The 1.0\,m telescopes are equipped with 4096 $\times$ 4096 SINISTRO cameras having an image scale of 0.389\arcsec pix$^{-1}$, resulting in a 26\arcmin\ $\times$ 26\arcmin\ field of view. We used an exposure time of 30 seconds and, with the full frame readout time of $\sim$ 30 seconds, the final image cadence was $\sim$ 60 seconds. The images were calibrated with the standard LCOGT {\tt BANZAI} pipeline \citep{McCully:2018}. The telescopes were intentionally defocused in an attempt to improve photometric precision, resulting in a typical HD\,15906 full width half maximum (FWHM) of 6.5\arcsec. Differential photometric data were extracted using {\tt AstroImageJ} \citep{Collins:2017}. We used a circular photometric aperture with radius 9.3\arcsec\ to exclude all flux from the nearest known \textit{Gaia} Data Release 3 stars \citep[\textit{Gaia} DR3;][]{Gaia-mission,gaiacollaboration_2022}. A transit-like event was detected in both LCOGT lightcurves and they were included in the analysis described in Section \ref{sec:photo-analysis}.
\begin{table}
\centering
  \caption{LCOGT observations of HD\,15906\,b. See Section \ref{sec:lco-detrend} for a description of the detrending terms.}
  \resizebox{\columnwidth}{!}
  {\begin{tabular}{ccccc}
    \hline
    Visit & Observatory & Start Time [UTC] & Dur. / hrs & Detrending Terms\\
    \hline 
    1 & Siding Spring & 2021-08-27 13:47:07 & 5.7 & airmass, FWHM\\
    2 & McDonald & 2021-11-01 03:36:18 & 3.8 & airmass, FWHM\\
    \hline
  \end{tabular}}
\label{tab:lco-observations}
\end{table} 

\subsection{WASP Photometry}\label{sec:wasp-obs}
\begin{figure}
\begin{center}
\includegraphics[width=\columnwidth]{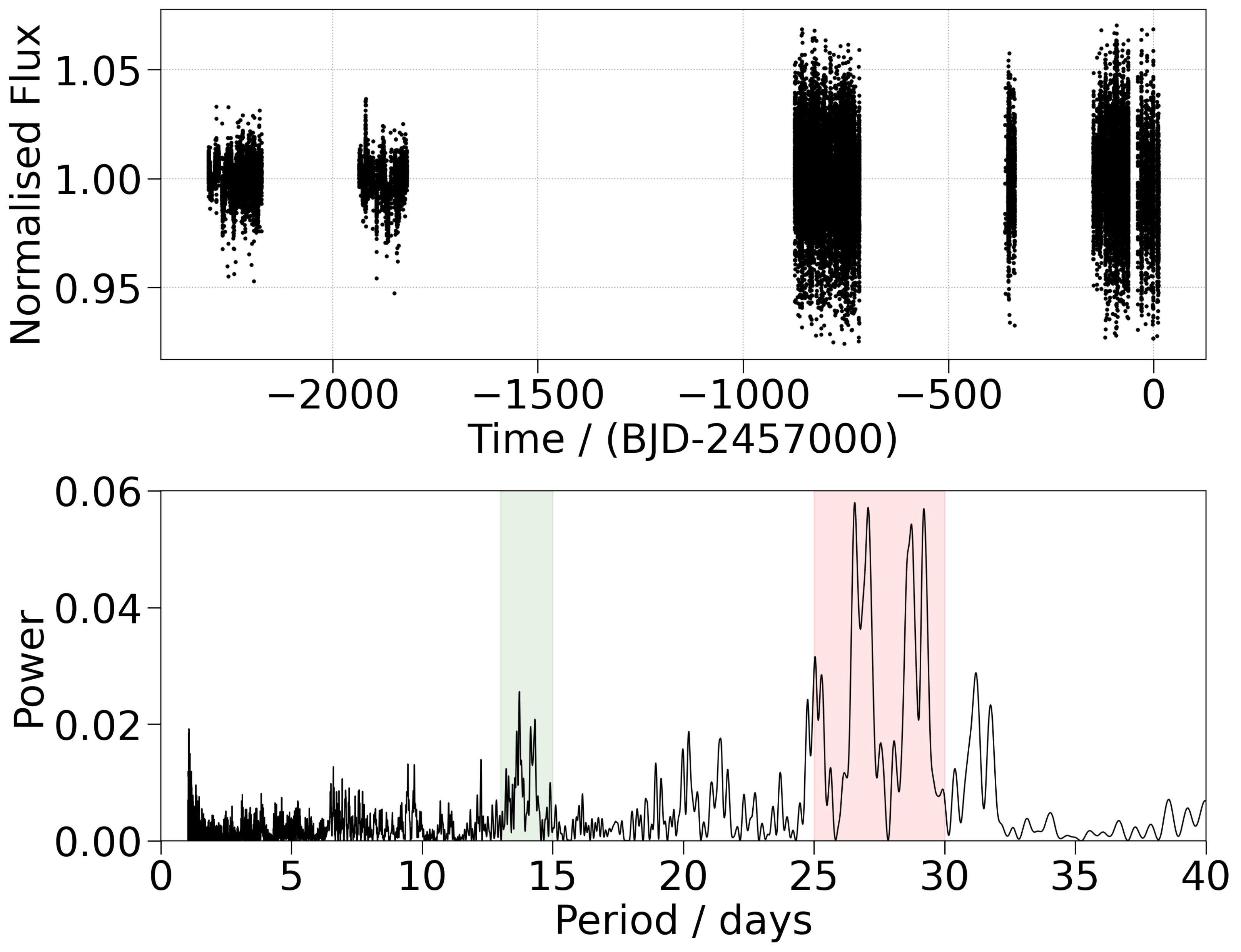}
\caption{Archival WASP photometry. Upper: Normalised WASP lightcurve spanning more than 6 years. Lower: GLS periodogram of the WASP lightcurve. The strongest peaks are in the range 25-30\,days (red highlight), followed by 13-15\,days (green highlight).}
\label{fig:wasp_photometry}
\end{center}
\end{figure}

HD\,15906 was observed 38\,740 times by the Wide Angle Search for Planets at the South
African Astronomical Observatory \citep[WASP-South;][]{Pollacco2006} between 19 August 2008 and 19 December 2014. The photometry was extracted and detrended for systematic effects following the methods described in \citet{2006cameron}. Based upon a visual inspection of the lightcurve, we removed data with a normalised flux greater than 1.07 or less than 0.93 and we removed data with a relative flux error greater than 0.03. These cuts removed 5\,231/38\,740 ($\sim$ 14\%) data points and the resulting lightcurve is shown in Figure \ref{fig:wasp_photometry}. With an average flux error of $\sim$ 9 ppt, we do not detect the transits of HD\,15906\,b or c in the WASP data. Furthermore, there were no additional transits detected in the lightcurve. Thanks to the long baseline, the WASP photometry is used to estimate the stellar rotation period (see Section \ref{sec:stellar_age}).

\subsection{HARPS Spectroscopy}\label{sec:harps-obs}
The High Accuracy Radial velocity Planet Searcher \citep[HARPS;][]{Mayor-2003-HARPS} is a high-resolution (R = 115\,000) fibre-fed spectrograph installed on the 3.6\,m telescope at the European Southern Observatory (ESO) in La Silla, Chile. It has been operational since 2003 and the optical fibres were upgraded in 2015, leading to an offset in the measured radial velocities (RVs) \citep{Lo-Curto-2015}.

HARPS observed HD\,15906 18 times between 3 November 2003 and 9 February 2018. There were 15 observations taken before the fibre upgrade and 3 observations taken after the upgrade. The exposure times of the observations ranged from 358 to 900 seconds and the average SNR at 550\,nm was 53.7. The data spans $\sim$ 5212 days, with an average separation of $\sim$ 307 days between each observation. The HARPS spectra are publicly available on the ESO Science Archive Facility.

For our analysis of the HD\,15906 system, we used the RVs presented in \citet{Trifonov-HARPS-RVs}. Specifically, we used the columns `RV\_mlc\_nzp' and `e\_RV\_mlc\_nzp' for the RV and RV error, respectively. These RVs were extracted by the SpEctrum Radial Velocity AnaLyser \citep[SERVAL;][]{SERVAL} pipeline, where the extraction was done independently before and after the fibre upgrade and a correction was made for the nightly zero-point. The data have a root mean square (RMS) of 10.70 ms$^{-1}$ and the average RV uncertainty is 1.51 ms$^{-1}$. We present these RVs in Table \ref{tab:rv-observations}. 
\begin{table}
\centering
  \caption{HARPS and FIES RVs of HD\,15906.}
  \resizebox{\columnwidth}{!}
  {\begin{tabular}{cccc}
    \hline
    Time / BJD & RV / ms$^{-1}$ & RV Error / ms$^{-1}$ & Instrument\\
    \hline 
    2452946.74714 & -2.799 & 2.005 & HARPS\\
    2453315.66562 & 10.217 & 1.243 & HARPS\\
    2453316.79132 & 3.188 & 2.672 & HARPS\\
    2453321.79052 & -7.562 & 1.354 & HARPS\\
    2454390.73395 & -3.929 & 1.621 & HARPS\\
    2454438.60542 & 10.151 & 1.369 & HARPS\\
    2454752.74485 & 11.626 & 1.375 & HARPS\\
    2455217.57723 & 4.607 & 1.662 & HARPS\\
    2455491.79108 & 15.727 & 1.631 & HARPS\\
    2455876.61897 & -9.996 & 1.475 & HARPS\\
    2456161.82258 & -0.028 & 1.137 & HARPS\\
    2456169.84113 & -11.997 & 1.422 & HARPS\\
    2456233.78781 & -10.516 & 1.116 & HARPS\\
    2456271.65665 & -1.351 & 1.135 & HARPS\\
    2456309.54995 & 0.427 & 1.614 & HARPS\\
    2457349.78566 & -29.043 & 2.260 & HARPS\\
    2457354.71407 & 11.282 & 1.051 & HARPS\\
    2458158.55270 & -5.683 & 1.121 & HARPS\\
    2458742.62217 & 2.65 & 4.90 & FIES\\
    2458745.71138 & 0.00 & 5.32 & FIES\\
    2458751.64001 & -8.37 & 14.61 & FIES\\
    2458753.70368 & -13.85 & 4.65 & FIES\\
    2458757.57125 & 11.98 & 4.64 & FIES\\
    2458765.57899 & 3.41 & 3.27 & FIES\\
    2458768.66127 & -3.25 & 5.16 & FIES\\
    \hline
  \end{tabular}}
\label{tab:rv-observations}
\end{table}

\subsection{FIES Spectroscopy}\label{sec:fies-obs}
As part of TFOP, we observed HD\,15906 seven times using the FIbre-fed \'{E}chelle Spectrograph \citep[FIES;][]{Telting14} at the Nordic Optical Telescope \citep[NOT;][]{djupvik10} between 15 September 2019 and 12 October 2019. For each observation, we used the high-resolution fibre (R $\sim$ 67\,000) and an exposure time of 1800 seconds. We extracted the spectra and derived multi-order RVs following \citet{Buchhave2010}. The SNR per resolution element at 550\,nm ranges between 20 and 105 with a median of 97. The RMS of the RV data is 7.88 ms$^{-1}$ and the average uncertainty is 6.08 ms$^{-1}$. These FIES RVs are included in Table \ref{tab:rv-observations}.

\section{Stellar Characterisation}\label{sec:stellar}

\subsection{Atmospheric Properties}\label{sec:stellar_spectroscopic}
As described in Section \ref{sec:harps-obs}, HD\,15906 was observed by HARPS 18 times between 2003 and 2018, with 15 observations made before the 2015 fibre upgrade. We retrieved the 15 pre-upgrade HARPS spectra from the ESO Science Archive Facility and co-added them to create a single master spectrum. This was used to perform the following spectroscopic analyses.

We performed an equivalent width (EW) analysis using ARES+MOOG to derive the stellar atmospheric parameters ($T_{\mathrm{eff}}$, $\log g$, microturbulence, [Fe/H]). We followed the same methodology described in \citet{Santos-13, Sousa-14, Sousa-21}. We used the latest version of ARES\footnote{\url{https://github.com/sousasag/ARES}} \citep{Sousa-07, Sousa-15} to measure the EWs of the iron lines in the master HARPS spectrum. We used a minimisation process to find ionisation and excitation equilibrium and converge to the best set of spectroscopic parameters. The iron abundances were computed using a grid of Kurucz model atmospheres \citep{Kurucz-93} and the radiative transfer code MOOG \citep{Sneden-73}. We also derived a more accurate trigonometric surface gravity using recent \textit{Gaia} data following the same procedure as described in \citet{Sousa-21}. The quoted errors for $T_{\mathrm{eff}}$, $\log g$ and [Fe/H] are ``accuracy'' errors, that is they have been corrected for systematics following the discussion presented in Section 3.1 of \citet{Sousa-2011}. The final spectroscopic parameters and their errors are included in Table \ref{tab:stellar-params} and we find that HD\,15906 is a K-dwarf.

We also performed an independent spectral synthesis with SME version 5.2.2\footnote{\url{http://www.stsci.edu/~valenti/sme.html}} \citep[Spectroscopy Made Easy;][]{vp96,pv2017}. A detailed description of the modelling can be found in \citet{2018A&A...618A..33P}. We used the ATLAS12 stellar atmosphere grid \citep{Kurucz2013} and atomic and molecular line data from VALD\footnote{\url{http://vald.astro.uu.se}} \citep[Vienna Atomic Line Database;][]{Ryabchikova2015}. The macro- and micro-turbulent velocities were held fixed to 1.5 kms$^{-1}$ and 0.5 kms$^{-1}$, respectively. The resulting $T_\mathrm{eff}$, $\log g$ and abundances were in excellent agreement with the ARES+MOOG analysis. We additionally derived the projected rotational velocity, $v \sin i_\star$ = 2.7 $\pm$ 0.7 kms$^{-1}$.

\begin{table}
\caption{Stellar properties of HD\,15906.}
\resizebox{\columnwidth}{!}
  {\begin{tabular}{lcc}
    \hline
    \multicolumn{3}{c}{HD\,15906}\\
    \hline
    \multicolumn{3}{l}{Alternative Identifiers}\\
    TOI & \multicolumn{2}{l}{461} \\
    TIC & \multicolumn{2}{l}{4646810} \\
    TYC & \multicolumn{2}{l}{5282-297-1} \\
    2MASS & \multicolumn{2}{l}{J02330530-1021062} \\
    \textit{Gaia} DR3 & \multicolumn{2}{l}{5175239363214344960} \\
    \hline
    Parameter & Value & Source \\
    \hline
    \multicolumn{3}{l}{Astrometric Properties}\\ 
    RA (J2016; hh:mm:ss.ss) & 02:33:05.09 & 1 \\
    Dec (J2016; dd:mm:ss.ss) & -10:21:07.89 & 1 \\
    $\mu_{\alpha}$ / mas yr$^{-1}$ & -172.92 $\pm$ 0.02 & 1 \\
    $\mu_{\delta}$ / mas yr$^{-1}$ & -92.22 $\pm$ 0.02 & 1 \\
    RV / kms$^{-1}$ & -3.64 $\pm$ 0.25 & 1 \\
    Parallax / mas & 21.834 $\pm$ 0.019 & 1* \\
    Distance / pc & 45.80 $\pm$ 0.04 & 6; inverse parallax \\
    U / kms$^{-1}$ & 37.87 $\pm$ 0.20 & 6 \\
    V / kms$^{-1}$ & 9.56 $\pm$ 0.01 & 6 \\
    W / kms$^{-1}$ & -17.25 $\pm$ 0.35 & 6 \\
    \hline
    \multicolumn{3}{l}{Photometric Properties} \\  
    G / mag & 9.484 $\pm$ 0.003 & 1 \\
    G$_{\rm BP}$ / mag & 9.999 $\pm$ 0.003 &  1 \\
    G$_{\rm RP}$ / mag & 8.817 $\pm$ 0.004 &  1 \\
    \textit{TESS} / mag &  8.872 $\pm$ 0.006 &  2 \\
    V / mag & 9.76 $\pm$ 0.03 & 3 \\
    B / mag & 10.79 $\pm$ 0.06 & 3 \\
    J / mag & 8.035 $\pm$ 0.018 & 4 \\
    H / mag & 7.557 $\pm$ 0.031 & 4 \\
    K / mag & 7.459 $\pm$ 0.023 & 4 \\
    W1 / mag & 7.345 $\pm$ 0.032 & 5 \\
    W2 / mag & 7.459 $\pm$ 0.020 & 5 \\
    \hline
    \multicolumn{3}{l}{Bulk Properties} \\
    $T_{\mathrm{eff}}$ / K & 4757 $\pm$ 89 & 6; ARES+MOOG \\
    $\log g$ / cms$^{-2}$ & 4.49 $\pm$ 0.05 & 6; ARES+MOOG \\\relax
    [Fe/H] / dex & 0.02 $\pm$ 0.04 & 6; ARES+MOOG \\
    $v \sin i_{\star}$ / kms$^{-1}$ & 2.7 $\pm$ 0.7 & 6; SME \\
    $\log R'_{\mathrm{HK}}$ & -4.694 $\pm$ 0.065 & 6; HARPS spectra \\
    E(B - V) & 0.023 $\pm$ 0.018 & 6; IRFM \\
    $R_{\star}$ / $R_{\odot}$ & 0.762 $\pm$ 0.005 & 6; IRFM \\
    $M_{\star}$ / $M_{\odot}$ & 0.790$_{-0.036}^{+0.020}$ & 6; isochrones \\
    $\rho_{\star}$ / $\rho_{\odot}$ & 1.79 $\pm$ 0.07 & 6; from $R_{\star}$ and $M_{\star}$ \\
    $\rho_{\star}$ / gcm$^{-3}$ & 2.52 $\pm$ 0.10 & 6; from $R_{\star}$ and $M_{\star}$ \\
    $L_{\star}$ / $L_{\odot}$ & 0.27 $\pm$ 0.02 & 6; from $R_{\star}$ and $T_{\mathrm{eff}}$ \\
    \hline
  \end{tabular}}
\footnotesize{\textbf{References}: 1 - \textit{Gaia} DR3 \citep{gaiacollaboration_2022}. 2 - \textit{TESS} Input Catalog Version 8 \citep[TICv8;][]{TICv8}. 3 - Tycho-2 \citep{Tycho2}. 4 - 2MASS \citep{skrutskie_2006}. 5 - \textit{WISE} \citep{wright_2010}. 6 - this work, see Section \ref{sec:stellar}. *\textit{Gaia} DR3 parallax corrected according to \citet{lindegren_2021}.}
\label{tab:stellar-params}
\end{table}

\subsection{Stellar Mass and Radius}\label{sec:stellar_params}

We determined the stellar radius, $R_{\star}$, of HD\,15906 from the stellar angular diameter and the offset corrected \textit{Gaia} DR3 parallax \citep{lindegren_2021} using a Markov-Chain Monte Carlo Infrared Flux Method \citep[MCMC IRFM;][]{blackwell_1977,schanche_2020}. We used the stellar spectral parameters as priors to construct model spectral energy distributions (SEDs) using atmospheric models from stellar catalogues. From this, we derived the stellar bolometric flux and angular diameter by comparing synthetic photometry, computed by convolving the model SEDs over broadband bandpasses of interest, to the observed data taken from the most recent data releases for the following bandpasses; \textit{Gaia} G, G$_{\rm BP}$, and G$_{\rm RP}$, Two Micron All-Sky Survey (2MASS) J, H, and K, and \textit{Wide-field Infrared Survey Explorer (WISE)} W1 and W2 \citep{skrutskie_2006,wright_2010,gaiacollaboration_2022}. To account for systematic model uncertainties in our stellar radius error, we used stellar atmospheric models taken from a range of ATLAS catalogues \citep{Kurucz-93,castelli_2003} and combined them in a Bayesian modelling averaging framework. Within the MCMC IRFM we attenuated the SED to correct for potential extinction and report the determined E(B-V) in Table \ref{tab:stellar-params}. We combined the retrieved angular diameter with the offset-corrected \textit{Gaia} DR3 parallax and found $R_{\star}$ = 0.762 $\pm$ 0.005 $R_{\odot}$.

We then determined the stellar mass, $M_{\star}$, by inputting $T_{\mathrm{eff}}$, [Fe/H], and $R_{\star}$ into two different stellar evolutionary models, PARSEC\footnote{\url{http://stev.oapd.inaf.it/cgi-bin/cmd}} v1.2S \citep[PAdova and TRieste Stellar Evolutionary Code;][]{marigo17} and CLES \citep[Code Liégeois d'Évolution Stellaire;][]{scuflaire08}. We employed the isochrone placement algorithm \citep{bonfanti15,bonfanti16} to interpolate the input parameters within pre-computed grids of PARSEC isochrones and tracks and we retrieved a first estimate of the stellar mass, $M_{\star\mathrm{,PD}}$ = 0.772 $\pm$ 0.037\,$M_{\odot}$. A second estimate was computed through the CLES code, which builds the best-fit evolutionary track of the star by applying the Levenberg-Marquadt minimisation scheme \citep[e.g.,][]{salmon21} and we found $M_{\star\mathrm{,LG}}$ = 0.797 $\pm$ 0.014\,$M_{\odot}$. To account for model-related uncertainties, we added in quadrature an uncertainty of 4\% to the mass estimates obtained from each set of models \citep[see][]{bonfanti21}. We note that the two outcomes are well within 1$\sigma$. We also checked their mutual consistency through the $\chi^2$-based criterion broadly presented in \citet{bonfanti21} and obtained a p-value\,=\,0.49, which is greater than the normally adopted significance level of 0.05, as expected. For each mass estimate, we built the corresponding Gaussian probability density function, as described in \citet{bonfanti21}, and we combined them to obtain a final mass value of $M_{\star}$ = 0.790$_{-0.036}^{+0.020} M_{\odot}$, as presented in Table \ref{tab:stellar-params}.

\subsection{Stellar Age}\label{sec:stellar_age}
The isochrone fitting described in Section \ref{sec:stellar_params} also provided an estimate of the stellar age. However, the stellar mass is sufficiently low that the slow evolutionary speed of the star along its tracks led to an uninformative age of 6.8$_{-6.3}^{+6.9}$\,Gyr. To try and constrain the stellar age more precisely, we used gyrochronology, empirical $\log R'_{\mathrm{HK}}$ relations and kinematics.  

For the gyrochronology, we first estimate the stellar rotation period, P$_{\rm rot}$. The \textit{TESS} photometry (Figure \ref{fig:tess_photo_fit}) shows flux modulation, likely caused by stellar activity, that can be used to do this. We conducted a generalised Lomb-Scargle \citep[GLS;][]{Lomb-LS,Scargle-LS,2009GLS} analysis on the \textit{TESS} SAP and PDCSAP photometry and found strong peaks at 11-12\,days and 25-27\,days. However, this analysis is adversely affected by the short $\sim$ 27\,day baseline of the \textit{TESS} lightcurves. The archival WASP photometry has a much longer baseline that can be used to derive an independent estimate of the stellar rotation period. We performed a GLS analysis on the WASP lightcurve, the results of which are shown in Figure \ref{fig:wasp_photometry}. The strongest peaks are in the range 25-30\,days, with the maximum power at 26.6\,days corresponding to a best-fit photometric amplitude of $\sim$ 4\,ppt. The next strongest peaks are in the range 13-15\,days, with a maximum power at 13.7\,days and an amplitude of $\sim$ 3\,ppt. This shorter rotation period is supported by our value of $v \sin i_{\star}$. Assuming $\sin i_{\star}$ = 1 and using the stellar radius in Table \ref{tab:stellar-params} leads to an upper limit of the rotation period, P$_{\rm rot}$ = 14.3 $\pm$ 3.7\,days. Finally, from a GLS analysis of the HARPS and FIES RVs (see Section \ref{sec:rv-analysis}), we found that the peak power was at 12.27\,days with a false alarm probability (FAP) of less than 1\%. It's possible that this corresponds to the stellar rotation period, however, due to the very sparse sampling of the RVs, this value is unreliable. The stellar rotation period remains somewhat ambiguous, but the evidence favours a value in the range 11-15\,days. Using the gyrochronological relations of \citet{Barnes2007} and (B-V) from Table \ref{tab:stellar-params}, these P$_{\rm rot}$ values yield a stellar age in the range 0.29-0.52\,Gyr. We note that the longer P$_{\rm rot}$ values (25-30\,days) would translate to an age of 1.39-1.97\,Gyr. However, more recent studies have shown that the relations of \citet{Barnes2007} might lead to an incorrect age estimate for low-mass stars because they do not account for the stalling period during spin-down \citep[e.g.,][]{Curtis2020}. Based upon a sample of benchmark stellar clusters, a rotation period of 11-15\,days for a star with a similar effective temperature as HD\,15906 is consistent with an age up to $\sim$ 1\,Gyr.

Next, we computed values of $\log R'_{\mathrm{HK}}$ from each of the 18 HARPS spectra using ACTIN\footnote{\url{https://github.com/gomesdasilva/ACTIN2}} \citep{actin} to extract the \ion{Ca}{ii} index and following the method described in \citet{GomesdaSilva2021} for the $\log R'_{\mathrm{HK}}$ calibration. We found an average value of -4.694 $\pm$ 0.065 and, using the empirical relations of \citet{Mamajek2008}, this converts into a stellar age of 1.9 $\pm$ 0.7\,Gyr.

Finally, we computed the kinematic age using the method developed in \citet{Almeida-Fernandes2018} and the Galactic $UVW$ velocities that we determined from the \textit{Gaia} DR3 proper motions, offset-corrected parallax \citep{lindegren_2021}, and stellar RV, using the method outlined in \citet{Johnson1987}. We found a stellar age of 1.9$^{+6.0}_{-0.7}$\,Gyr, favouring an older star.

\begin{figure}
\begin{center}
\includegraphics[width=\columnwidth]{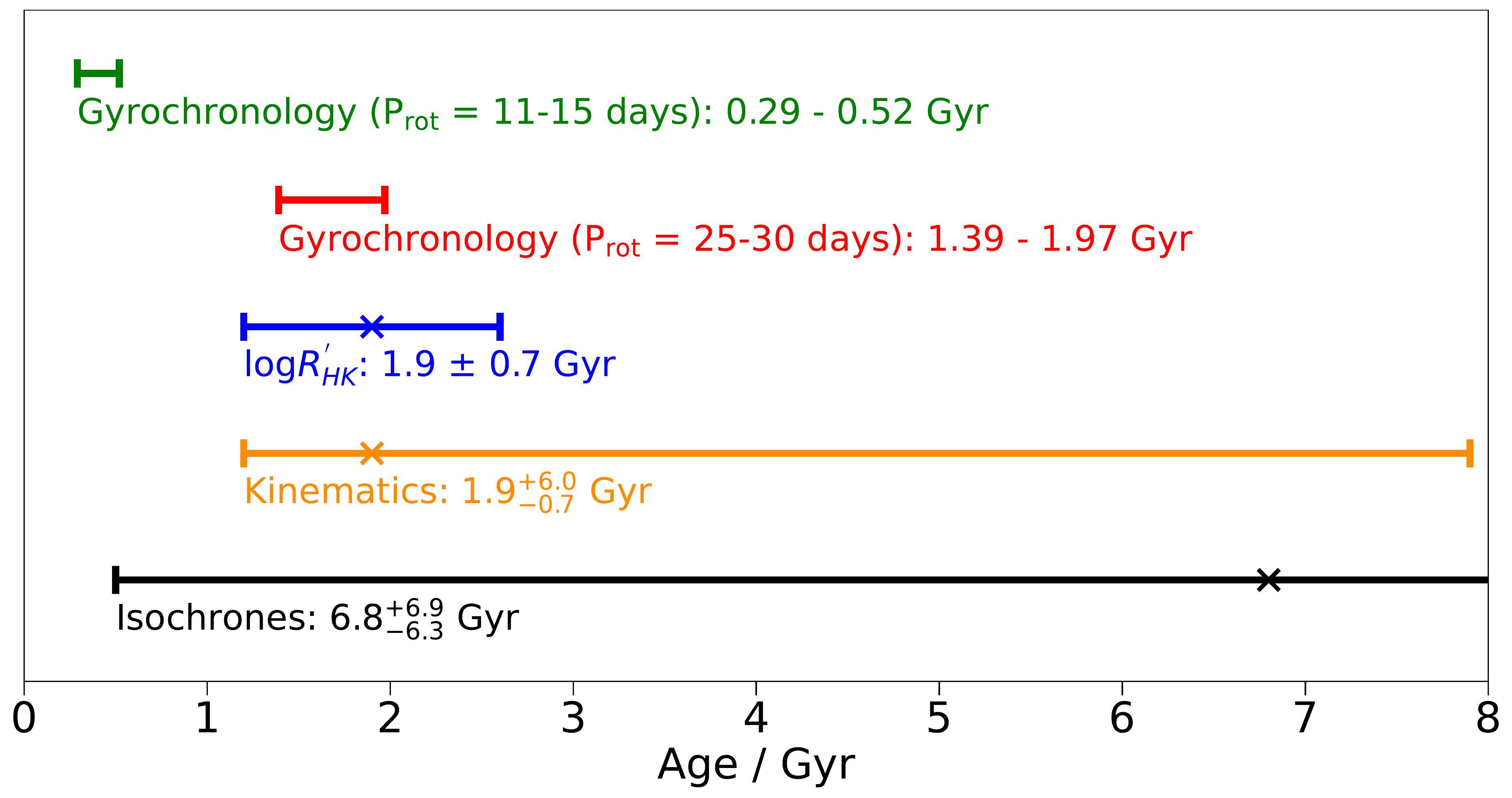}
\caption{A comparison of stellar age estimates obtained from isochrone fitting, $\log R'_{\mathrm{HK}}$ relations, kinematics and gyrochronology.}
\label{fig:stellar-ages}
\end{center}
\end{figure}

In Figure \ref{fig:stellar-ages}, we present a comparison of the age estimates derived by our various methods. The age estimates derived from $\log R'_{\mathrm{HK}}$ and kinematics are consistent and they are in agreement with the gyrochronological age implied by a rotation period of 25-30\,days. The favoured rotation period of 11-15\,days yields a much younger age, however we reiterate that gyrochronology is not necessarily accurate for low-mass stars. We conclude that the stellar age is ambiguous based on the current data.

\section{Analysis}\label{sec:all-analysis}

\subsection{\textit{TESS} Only Analysis}\label{sec:tess-only}
Before pursuing \textit{CHEOPS} follow-up observations of HD\,15906, we used \texttt{MonoTools}\footnote{\label{footnote:monotools}\url{https://github.com/hposborn/MonoTools}} \citep{Osborn2022code} to perform an analysis of the \textit{TESS} data. \texttt{MonoTools} is designed for the analysis of planets with unknown periods, including duotransits. It can be used to derive the allowed period aliases and their corresponding probabilities, crucial for scheduling follow-up observations.

We built a \texttt{MonoTools} model using the stellar parameters presented in Table \ref{tab:stellar-params}, one periodic planet and one duotransit. We defined initial guesses for transit depth, duration, and mid-transit time for the two planets using a visual inspection of the \textit{TESS} lightcurve. Since this is a multiplanet transiting system, we selected the eccentricity distribution from \citet{Van-Eylen-2015}. We also included a Gaussian Process \citep[GP;][]{2006GPs,2014Gibson} with a simple harmonic oscillator (SHO) kernel from \texttt{celerite} \citep{celerite} to model the correlated noise in the lightcurve. We sampled the posterior probability distribution using the No-U-Turn Sampler \citep[NUTS;][]{NUTS-MC}, a variant of Hamiltonian Monte Carlo, implemented via \texttt{pyMC3} \citep{pymc3}.

\begin{figure*}
\begin{center}
\includegraphics[width=\textwidth]{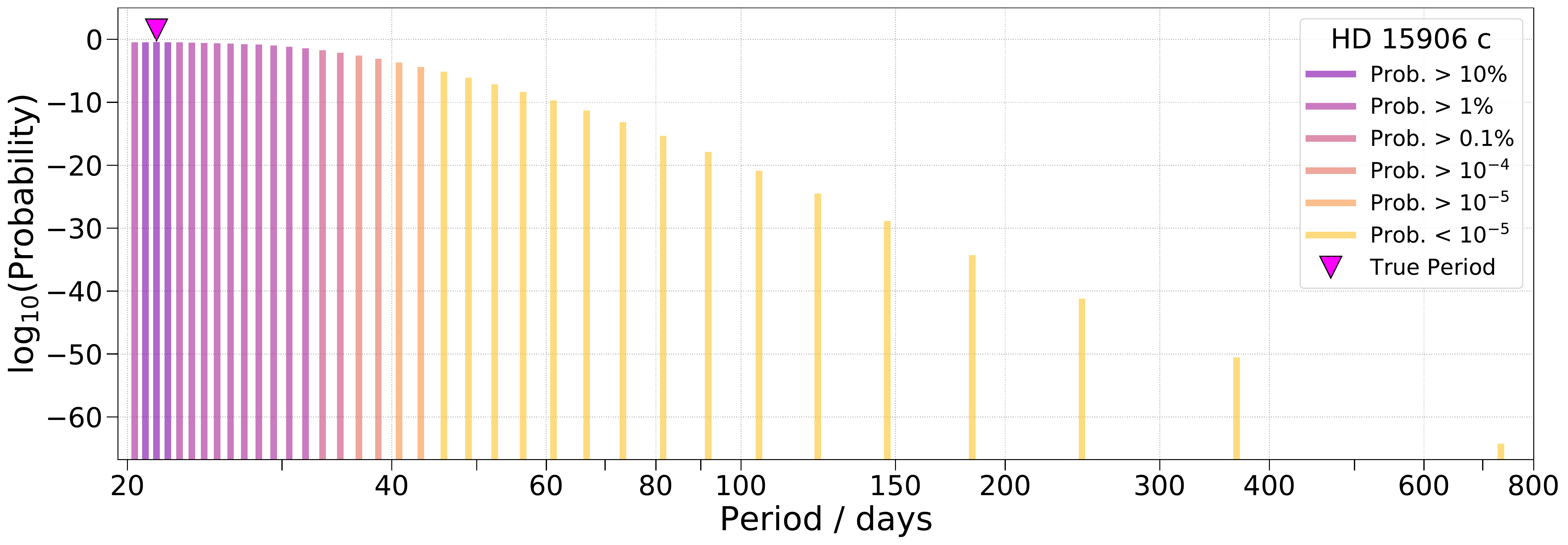}
\caption{From the \textit{TESS} data alone, HD\,15906\,c was a duotransit with 36 possible period aliases. This plot shows these aliases and their corresponding probabilities, derived using \texttt{MonoTools}. The true period, as determined by \textit{CHEOPS} follow-up observations, had the highest probability.}
\label{fig:period_aliases}
\end{center}
\end{figure*}

We found that the duotransit, HD\,15906\,c, had 36 possible period aliases, with a minimum value, P$_{\text{min}}$, of 20.384 days. The probability of each period alias is shown in Figure \ref{fig:period_aliases}. These results were used to schedule our \textit{CHEOPS} follow-up observations, from which we successfully determined the true period of planet c to be $\sim$ 21.6 days (see Section \ref{sec:cheops-obs}).

\subsection{Global Photometric Analysis}\label{sec:photo-analysis}
Once we had confirmed the true period of HD\,15906\,c with \textit{CHEOPS}, we performed a joint fit of the \textit{TESS}, \textit{CHEOPS} and LCOGT photometric data using \texttt{juliet}\footnote{\label{footnote:juliet}\url{https://github.com/nespinoza/juliet}}\citep{juliet}. This package combines transit models from \texttt{batman} \citep{batman} with the option to include linear models and GPs to model instrumental noise and stellar variability. We created a model consisting of two transiting planets, using the following parameterisation:
\begin{itemize}
\item Orbital period, $P$, and mid-transit time, $T_0$, for both planets. We set broad uniform priors on $P$ and $T_0$ from a visual inspection of the \textit{TESS} and \textit{CHEOPS} lightcurves.
\item Planet-to-star radius ratio, $p$ = R$_{\text{P}}$ / R$_{\star}$, and impact parameter, $b$, for both planets. We set uniform priors to allow exploration of all physically plausible solutions.
\item Eccentricity, $e$, and argument of periastron, $\omega$, for both planets. We used the eccentricity prior from \citet{VanEylen-2019} for systems with multiple transiting planets -- the positive half of a Gaussian with $\mu$ = 0 and $\sigma$ = 0.083. We used a uniform prior for $\omega$, covering the full range of possible values. We decided to fit for eccentricity, rather than assuming a circular orbit, to ensure that the uncertainties on the other fitted parameters were not underestimated. We note that we repeated our final global photometric fit assuming a circular orbit, with $e$ fixed to zero and $\omega$ fixed to 90 degrees, and all of the fitted planet parameters were consistent within 1.2$\sigma$.
\item Stellar density, $\rho_{\star}$. Using Kepler's third law, this can be combined with $P$ to derive a value of $a/R_{\star}$ for each planet. This is preferred to fitting for $a/R_{\star}$ directly; not only does it reduce the number of fitted parameters, but it also ensures a consistent value of $\rho_{\star}$. We defined a normal prior on $\rho_{\star}$ using the values of $R_{\star}$ and $M_{\star}$ presented in Table \ref{tab:stellar-params}.
\item Quadratic limb darkening parameters, $q1$ and $q2$, for each instrument. We used the \citet{Kipping-2013} parameterisation of the quadratic limb darkening law and defined normal priors on $q1$ and $q2$ for each instrument. The mean was computed by interpolating tables of quadratic limb darkening coefficients \citep{Claret-TESS,Claret-CHEOPS,Claret-LCO}, based on the stellar parameters presented in Table \ref{tab:stellar-params}, and a standard deviation of 0.1 was used in all cases.
\end{itemize}

In addition to the transit models, we used linear models to detrend \textit{CHEOPS} and LCOGT against instrumental systematics, see Sections \ref{sec:cheops-detrend} and \ref{sec:lco-detrend}. We treated each \textit{CHEOPS} and LCOGT observation independently for the sake of this detrending. We also included a GP to model the variability in the \textit{TESS} lightcurve, see Section {\ref{sec:tess-detrend}}. For each instrument, we included a jitter term to account for white noise and a relative flux offset term. We fixed the dilution factor to 1 due to the lack of any bright contaminating sources (see Section \ref{sec:gaia_validation}). We used the \texttt{dynesty} package to sample the posterior probability of this model with static nested sampling, using 300 live points and stopping when the difference between the evidence and the estimated remaining evidence was less than 0.01 \citep{dynesty}. For a full list of the parameters and priors used in our global fit see Appendix \ref{sec:photo-appendix} and for the results of our modelling see Section \ref{sec:photo-results}.

\subsubsection{\textit{CHEOPS} Detrending}\label{sec:cheops-detrend}
The \textit{CHEOPS} lightcurves contain trends that are correlated with instrumental parameters such as background flux (bg) and centroid position (x, y). There are also periodic noise features that repeat once per \textit{CHEOPS} orbit due to the satellite rolling around its pointing direction. Detrending the lightcurve against the sine or cosine of the roll angle ($\phi$) can remove these periodic instrumental effects.

The \textit{CHEOPS} lightcurves also include stellar variability. From the \textit{TESS} LC we know that HD\,15906 shows stellar variability (see Figure \ref{fig:tess_photo_fit}). On the shorter timescale of a \textit{CHEOPS} visit ($\sim$ 8.3 hours), this stellar variability can be modelled with a linear trend in time (t). 

We included linear models in our global fit to account for these instrumental trends and stellar variability. However, for each \textit{CHEOPS} observation, it was important to only select the relevant detrending parameters. To do this we used the \texttt{pycheops} package \citep{pycheops-Maxted} and the method described in \citet{Swayne-2021}. Briefly, we defined 10 detrending parameters: x, y, t, bg, cos($\phi$), sin($\phi$), cos(2$\phi$), sin(2$\phi$), cos(3$\phi$) and sin(3$\phi$). For each \textit{CHEOPS} visit, we took the clipped lightcurve (see Section \ref{sec:cheops-obs}) and did an initial fit of a transit model with no detrending. We defined broad uniform priors on the transit parameters based on a visual inspection of the \textit{TESS} and \textit{CHEOPS} data. We used the RMS of the residuals from this initial fit to define normal priors on the detrending parameters, with $\mu$ = 0 and $\sigma$ = RMS. We added the 10 detrending parameters to the fit one-by-one, selecting the parameter with the lowest Bayes factor at each step. When there were no remaining parameters with Bayes factor $<$ 1, we stopped adding detrending parameters. In order to remove strongly correlated parameters, if any of the selected detrending parameters had a Bayes factor $>$ 1, we removed the parameter with the largest Bayes factor until no more parameters with Bayes factor $>$ 1 remained. The selected detrending parameters for each \textit{CHEOPS} visit are included in Table \ref{tab:cheops-observations}.

\subsubsection{LCOGT Detrending}\label{sec:lco-detrend}
We used {\tt AstroImageJ} to select the relevant detrending vectors for each LCOGT observation by jointly fitting a transit model and linear combinations of zero, one, or two detrending parameters from the available detrending vectors: airmass, time, sky background, FWHM, x-centroid, y-centroid, total comparison star counts, humidity and exposure time. The best zero, one, or two detrending vectors were retained if they reduced the Bayesian information criterion (BIC) for a fit by at least two per detrending parameter. We found that the airmass plus FWHM detrending pair provided the best improvement to the lightcurve fit for both LCOGT observations. We therefore included linear models for airmass and FWHM for each LCOGT observation in our global fit.

\subsubsection{\textit{TESS} Detrending}\label{sec:tess-detrend}
\begin{figure}
\begin{center}
\includegraphics[width=\columnwidth]{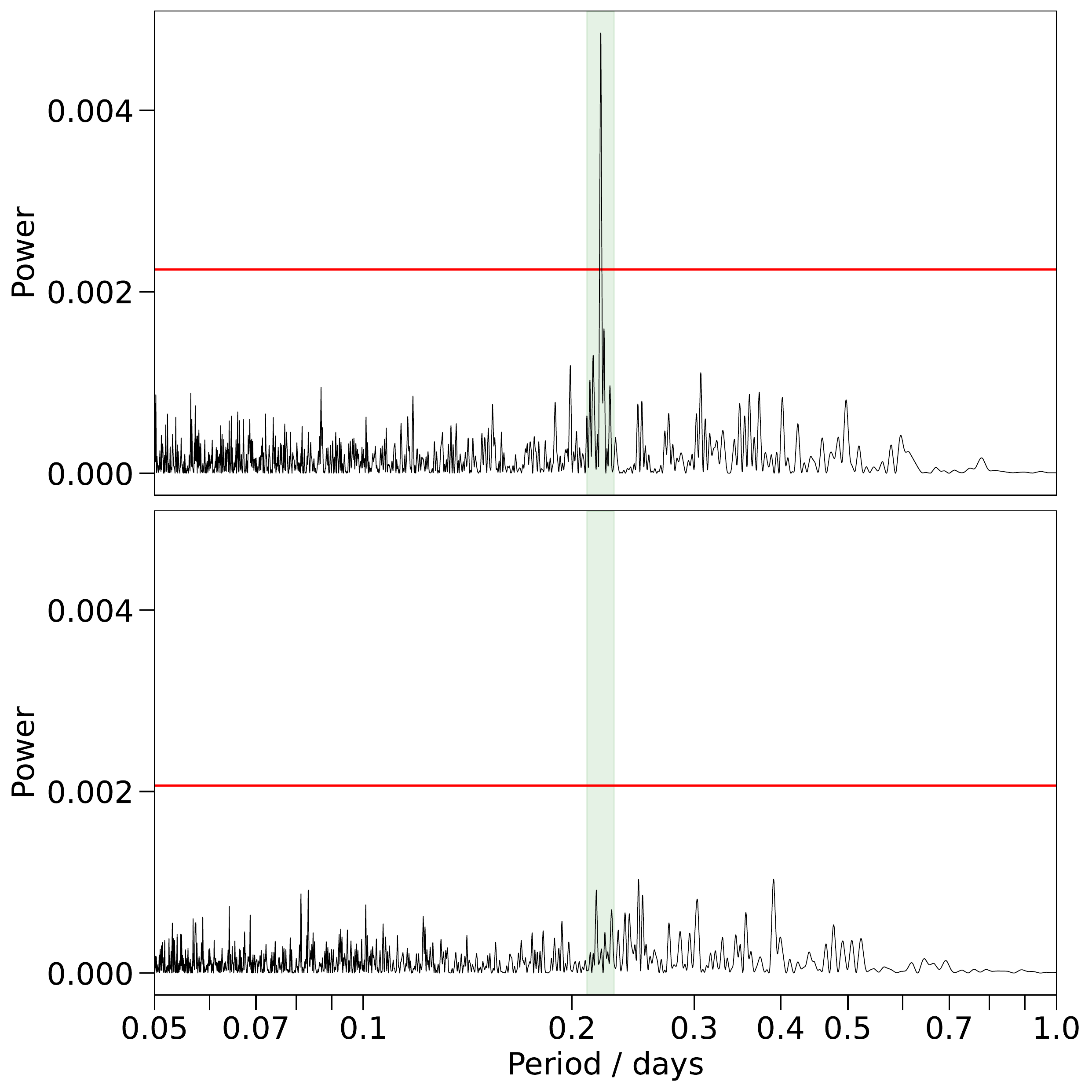}
\caption{GLS periodogram of the \textit{TESS} residuals from a global photometric fit using a GP with a Matérn-3/2 kernel to jointly model \textit{TESS} sector 4 and sector 31. Upper: GLS periodogram of sector 4 residuals. Lower: GLS periodogram of sector 31 residuals. The horizontal red line is the 1\% false alarm probability level in each case. The significant peak at 0.22004 days in the sector 4 residuals (highlighted in green) is not present in the sector 31 residuals.}
\label{fig:residual_periodogram}
\end{center}
\end{figure}

The \textit{TESS} lightcurves contain correlated noise, including stellar variability and residual instrumental systematics, that we model with a GP. We initially modelled sector 4 and sector 31 jointly, using a GP with an approximate Matérn-3/2 (M32) kernel implemented via \texttt{celerite} \citep{celerite}. Upon a visual inspection of the results from this fit, we noticed that the \textit{TESS} residuals contained a sinusoidal-like trend. We ran a GLS analysis on the \textit{TESS} residuals, treating the sector 4 and sector 31 data separately, and the resulting periodograms are presented in Figure \ref{fig:residual_periodogram}. We found a significant periodic signal in the \textit{TESS} sector 4 residuals, with a period of 0.22004 days and a FAP of 6.0 x 10$^{-11}$. This signal is persistent throughout the whole of sector 4 and the best-fit sinusoidal model has an amplitude of $\sim$ 57 ppm. There was no corresponding detection in the \textit{TESS} sector 31, \textit{CHEOPS} or LCOGT residuals. The periodic signal is present in the \textit{TESS} lightcurve itself, it was not introduced as a result of our detrending, and we discuss its origin in Section \ref{sec:discussion-tess-periodicity}.

A half-cycle of the periodic signal is a similar duration to the transits and it was therefore important to check if it was affecting the fitted planet parameters. We therefore performed two additional fits, changing only the \textit{TESS} detrending to account for this periodic signal. We made a custom GP kernel by adding together the M32 and SHO kernels from \texttt{celerite}. The M32 kernel was intended to capture the long-term variability and the SHO kernel was used to capture the short-term quasi-sinusoidal noise. We defined a normal prior on the natural frequency of the SHO kernel, $\omega_0$, using the peak and its width from the periodogram analysis. We performed one fit where we jointly modelled the sector 4 and sector 31 data with this kernel and we also performed a fit where we decoupled the sector 4 and sector 31 data. We used the M32 plus SHO kernel for sector 4 and the M32 kernel for sector 31, motivated by the fact we only detect the periodic trend in sector 4. After performing these two fits, we checked for periodicity in the \textit{TESS} sector 4 residuals. In both cases, the peak of the periodogram was still at 0.22004 days but with a FAP greater than 68\%. This confirms that the SHO kernel adequately removes the periodic trend from the sector 4 \textit{TESS} data. 

We checked the consistency of the fitted planet parameters between the three fits. The majority of the fitted planet parameters were consistent between all three of the fits within 1$\sigma$ and the remaining parameters were consistent within 2$\sigma$, except for the argument of periastron of the outer planet. There was a disagreement greater than 3$\sigma$ between the values from the joint M32 fit and the decoupled fit. Constraining eccentricity and the argument of periastron is challenging with photometry alone and we remind the reader that we only included them in our fit to ensure that the uncertainties on the other fitted parameters were not underestimated. We conclude that the fitted planet parameters are not significantly affected by the presence of the periodic signal in \textit{TESS} sector 4. 

We also compared the Bayes evidence (dlnZ) of the three fits (Table \ref{tab:bayes_evidences_photo-fits}). We found a decisive preference for both of the fits incorporating the SHO kernel over the original fit \citep{Bayesfactor}. The joint M32 plus SHO fit had the highest evidence, preferred over the original joint M32 fit with dlnZ = 106.8, and the decoupled fit of sector 4 and 31 was preferred over the original joint M32 fit with dlnZ = 89.2. 
\begin{table}
\centering
  \caption{Comparison of the Bayes evidence from three global photometric fits, where only the \textit{TESS} detrending was varied. The difference in Bayes evidence (dlnZ) between each fit and the original joint Matérn-3/2 (M32) fit is quoted, indicating a decisive preference for the fits incorporating a simple harmonic oscillator (SHO) kernel \citep{Bayesfactor}.}
  {\begin{tabular}{cc}
    \hline
    \textit{TESS} Detrending Model & dlnZ \\
    \hline
    Joint M32 + SHO GP & +106.8 \\
    Sector 4 M32 + SHO GP, Sector 31 M32 GP & +89.2 \\
    Joint M32 GP & 0.0 \\
    \hline
  \end{tabular}}
\label{tab:bayes_evidences_photo-fits}
\end{table}

Despite the fact the evidence favoured the model with the M32 plus SHO kernel jointly fit to sectors 4 and 31, the model where we decoupled sector 4 and sector 31 is more physically motivated. This is because we only detected the periodic signal in sector 4. We therefore chose the decoupled fit as our final global photometric fit and we present the results in Section \ref{sec:photo-results}.

We performed one last test to assess the dependence of our results on our chosen detrending model -- we repeated the decoupled fit, replacing the M32 kernels with SHO kernels. For sector 4, we used an SHO kernel for the short-term quasi-periodic signal summed with a second SHO kernel for the longer-term variability. For sector 31, we used a single SHO kernel. All of the fitted planet parameters were fully consistent with our final results (see Table \ref{tab:photo-planet-posteriors}) within 1$\sigma$, except for the eccentricity of the inner planet which was consistent within 2.3$\sigma$. We conclude that our results are not significantly influenced by the choice of GP kernel.

\subsection{Transit Timing Variation Analysis}\label{sec:ttv-analysis}
From the global photometric analysis, we found that HD\,15906\,b and c orbit close to a 2:1 period commensurability ($P_c$/$P_b$ = \photoperiodratioshort), an indication that the planets might be in mean motion resonance (MMR). Planets in or near a low-order period commensurability have amongst the largest amplitude TTVs \citep[e.g.,][]{2011VerasTTV,2018TTVbook}, so we therefore checked for TTVs in the HD\,15906 system.

\texttt{juliet} can incorporate TTVs into a photometric model, however, it expects that each instrument contains at least one transit of all the planets being fit. This is not true in our case -- none of the \textit{CHEOPS} or LCOGT observations contain a transit of both planets. We therefore had to perform a separate TTV fit for each planet. When fitting the inner planet, we included the \textit{TESS} data, \textit{CHEOPS} visit 5 and both LCOGT visits. For the outer planet, we included the \textit{TESS} data and \textit{CHEOPS} visits 3 and 6. In total, we had seven transits of the inner planet and four transits of the outer planet.

For the fit of each planet, we used a model consisting of one transiting planet and the same detrending as described in Section \ref{sec:photo-analysis}. The only difference in the transit model was that we fit for the individual transit times instead of $P$ and $T_0$. We set a uniform prior of width 0.1 days on each transit time based upon a visual inspection of the data. All other priors were unchanged from the global photometric analysis and we used \texttt{dynesty} to sample the posterior of the model with nested sampling. Our results are presented in Section \ref{sec:ttv-results}.

\subsection{Radial Velocity Analysis}\label{sec:rv-analysis}
\begin{figure}
\begin{center}
\includegraphics[width=\columnwidth]{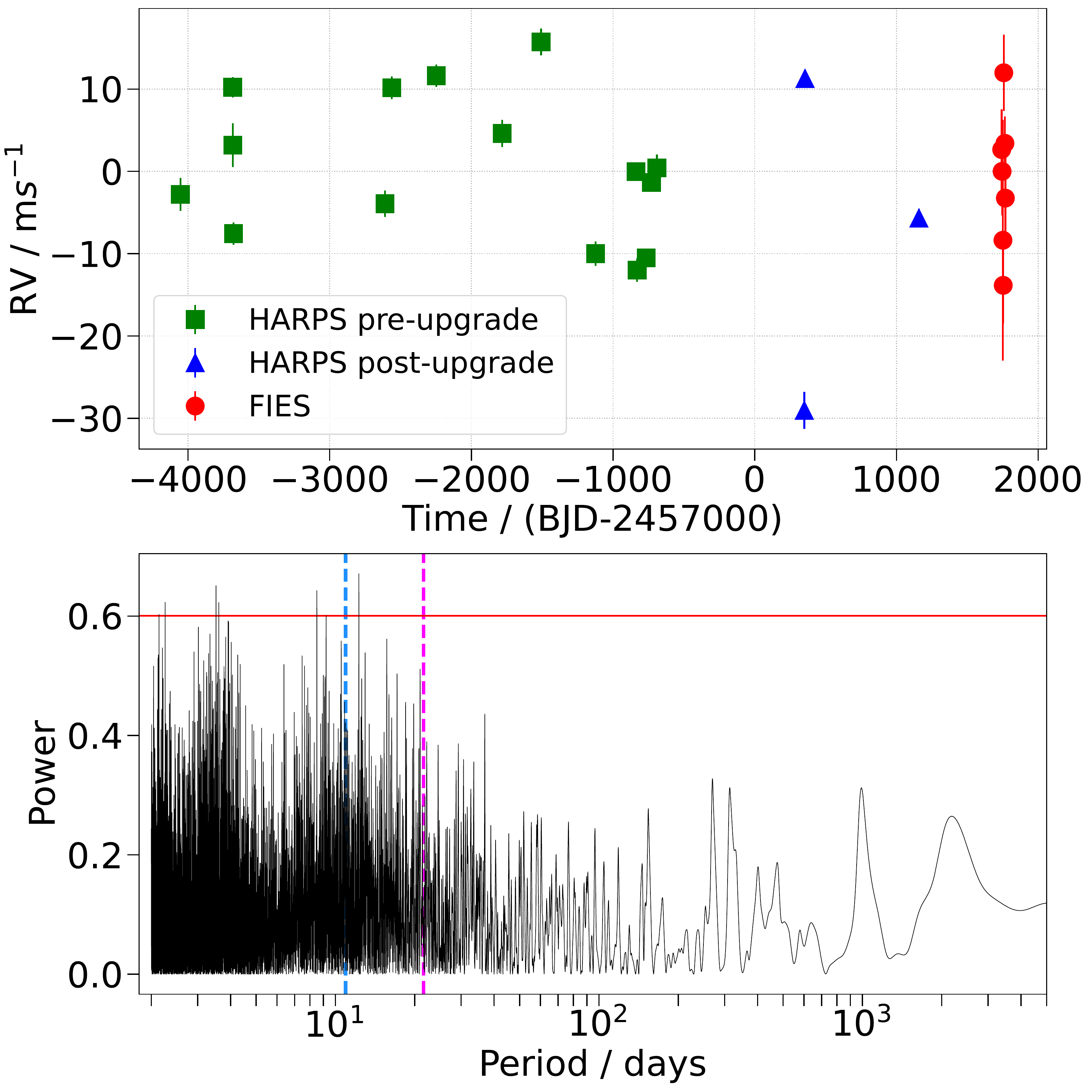}
\caption{Upper: HD\,15906 RV time-series, highlighting the sparsity of the data. HARPS data taken before/after the 2015 fibre upgrade is plotted (green squares/blue triangles) alongside the FIES data (red circles). Lower: GLS periodogram of the HARPS and FIES RV data. The photometrically derived orbital periods of the two planets, see Table \ref{tab:photo-planet-posteriors}, are indicated by the blue and pink vertical dashed lines and the red horizontal line represents the 1\% false alarm probability level. There are no significant peaks at the planet periods and the strongest peak is at 12.27 days.}
\label{fig:rv_periodogram}
\end{center}
\end{figure}

From HARPS and FIES, we have 25 sparsely sampled RV data points that show a relatively large scatter, see Figure \ref{fig:rv_periodogram}. We ran a GLS periodogram on the RV data and found no significant peaks at the planetary periods. The strongest peak was at 12.27 days and the best-fit sinusoid with this period had an amplitude of $\sim$ 10 ms$^{-1}$. It is possible that this signal is caused by stellar activity, but with such large gaps between each observation, the short-period peaks in the GLS periodogram are unreliable. We removed the best-fit sinusoid from the RV data and re-ran the GLS periodogram -- no additional peaks emerged.

To search for the planetary signals, we performed a series of fits to the HARPS and FIES RV data using \texttt{juliet}. For our first fit, we assumed that there were no planets in the system and we fit only for an offset and a white noise term for each instrument. We used a uniform prior for the offset, in the range -20 to 20 ms$^{-1}$, and a log-uniform prior for the white noise term, in the range 0.01 to 20 ms$^{-1}$. The HARPS data from before and after the fibre upgrade had to be treated as two independent instruments. However, we only had 3 data points from post-upgrade which was insufficient to constrain the instrumental parameters. We therefore excluded the 3 post-upgrade HARPS data points from our fits and we used \texttt{dynesty} to sample the posterior of the model.

We then added planets to our model. We performed one fit with only the inner planet, one fit with only the outer planet and finally a fit with both planets. We used a Keplerian for each planet, generated via {\tt RadVel} \citep{radvel}, with the following parameterisation:
\begin{itemize}
\item Orbital period, $P$, and mid-transit time, $T_0$. We fixed these to the solution from the global photometric fit (Table \ref{tab:photo-planet-posteriors}).  
\item Eccentricity, $e$, and argument of periastron, $\omega$. For simplicity, we fixed eccentricity to zero and $\omega$ to 90 degrees.
\item Semi-amplitude, $K$. We used a broad uniform prior to allow exploration of the range 0 to 20 ms$^{-1}$.
\end{itemize}

Finally, we took the model with both planets and added a GP with a quasi-periodic kernel \citep{celerite} to account for the stellar activity. This kernel is described by four hyperparameters: the amplitude, period, an additive factor impacting the amplitude and the scale of the exponential component. For the amplitude we used a uniform prior in the range 0 to 20 ms$^{-1}$ and for the period we defined a normal prior using the peak from the periodogram analysis ($\mu$ = 12.27 days, $\sigma$ = 0.1 days). The other two hyperparameters were allowed to vary uniformly over a broad range. With such a small number of sparsely sampled RVs, the GP was unlikely to yield a meaningful result but we chose to include it for completeness. The results of our RV modelling are presented in Section \ref{sec:rv-results}.

We note that we also tried a joint fit of the \textit{TESS}, \textit{CHEOPS} and LCOGT photometric data with the HARPS and FIES RV data using \texttt{juliet}. The photometric model was identical to that presented in Section \ref{sec:photo-analysis} and we used the RV model with two planets but no GP. However, due to the small number of sparse RVs, the fitted planet parameters were adversely affected compared to those from the global photometric model. Therefore, we decided to present independent analyses of the photometry and RVs in this paper. 

\section{Results}\label{sec:all-results}

\subsection{Global Photometric Results}\label{sec:photo-results}
\begin{figure*}
\begin{center}
\includegraphics[width=\textwidth]{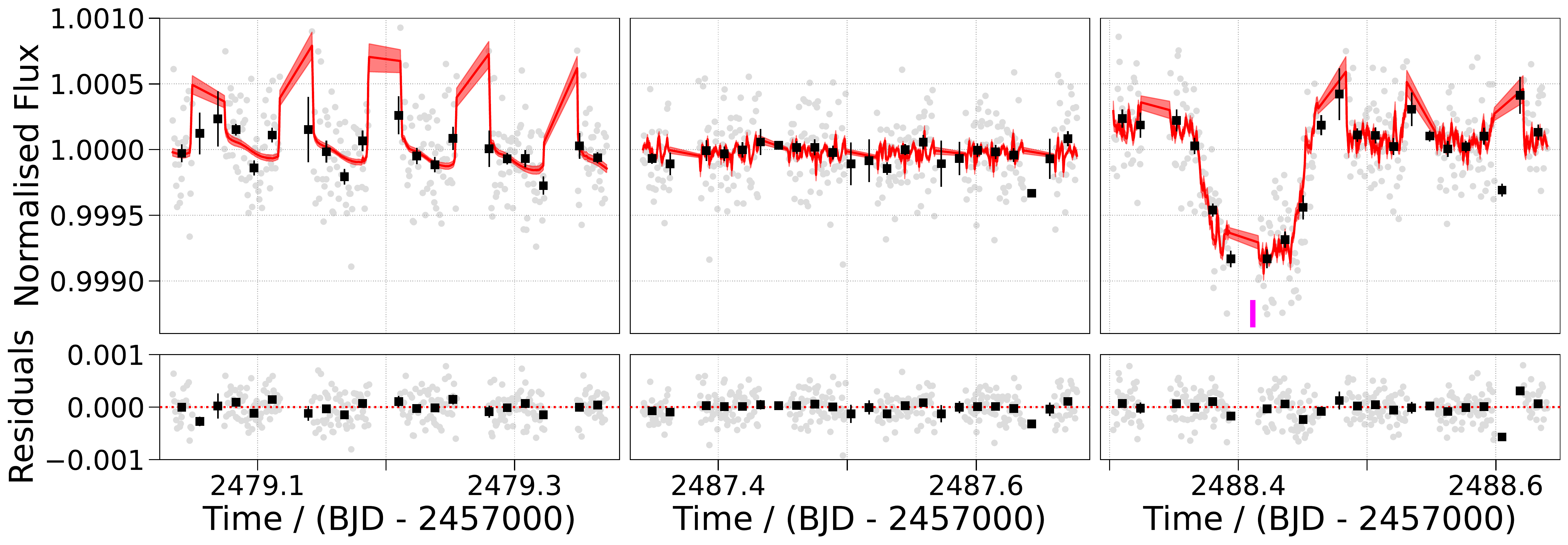}
\includegraphics[width=\textwidth]{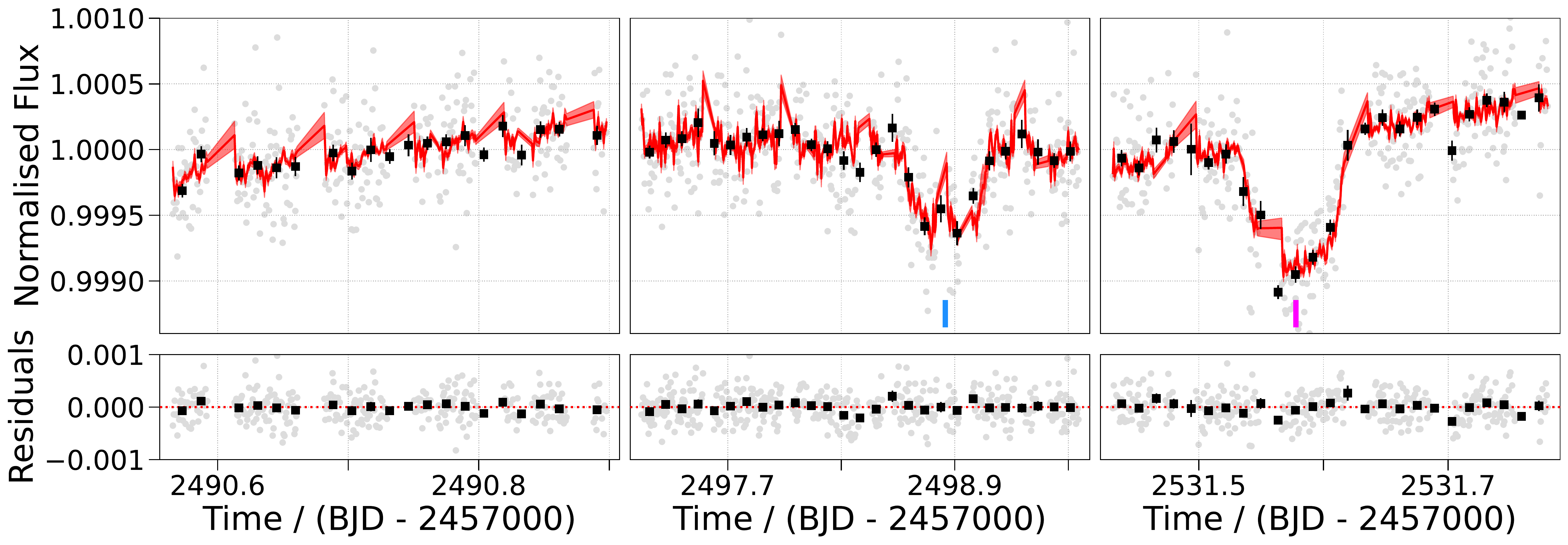}
\caption{Results of the global photometric fit. This plot shows the six \textit{CHEOPS} lightcurves, where the 60 second cadence data (grey) has been binned to 20 minutes (black squares) to guide the eye. The red line is the median model from the global photometric fit and the red shaded region is the 1$\sigma$ uncertainty on the model. The blue and pink markers indicate the mid-transit times of the inner and outer planets, respectively. The residuals of the model are included in the panel beneath each lightcurve.}
\label{fig:cheops_photo_fit}
\end{center}
\end{figure*}

\begin{figure*}
\begin{center}
\includegraphics[width=\textwidth]{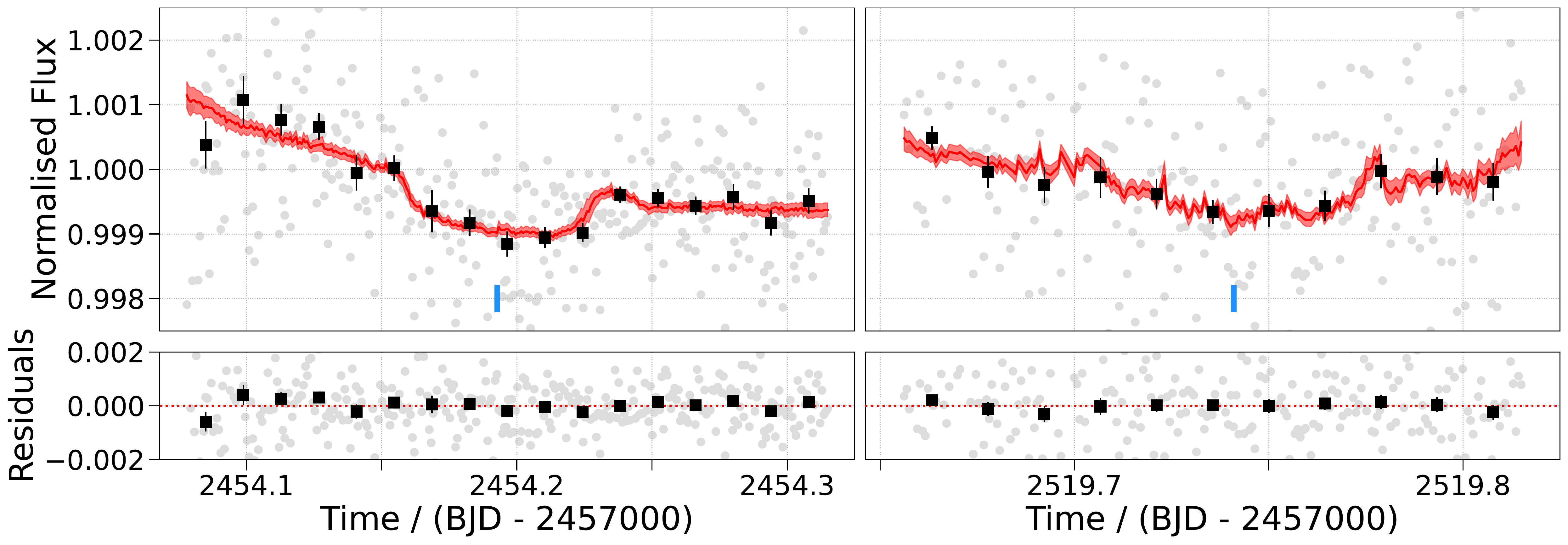}
\caption{Results of the global photometric fit. This plot shows the two LCOGT lightcurves, where the 60 second cadence data (grey) has been binned to 20 minutes (black squares) to guide the eye. The red line is the median model from the global photometric fit and the red shaded region is the 1$\sigma$ uncertainty on the model. The blue markers indicate the mid-transit times of the inner planet. The residuals of the model are included in the panel beneath each lightcurve.}
\label{fig:lco_photo_fit}
\end{center}
\end{figure*}

\begin{figure*}
\begin{center}
\includegraphics[width=\textwidth]{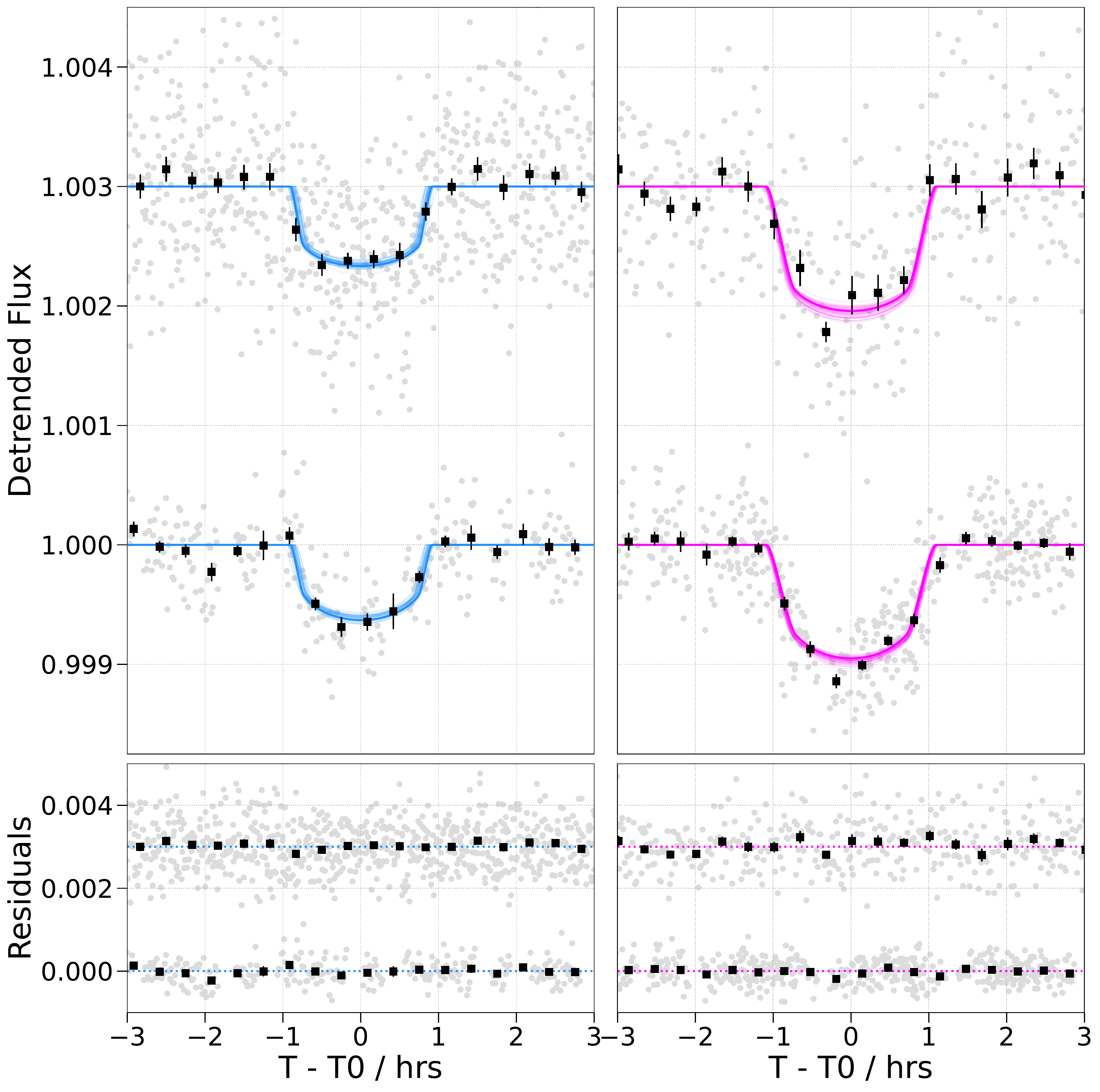}
\caption{Results of the global photometric fit. Upper: Phase-folded \textit{TESS} (top) and \textit{CHEOPS} (bottom) lightcurves for the inner (left) and outer (right) planet. The lightcurves have been detrended to remove the instrumental and stellar variability and the data (grey) has been binned to 20 minutes (black squares) to guide the eye. The median transit models for the inner (blue line) and outer (pink line) planet are included, along with 50 random samples drawn from the posterior distribution of the model. Lower: Residuals of the median transit models. Note that an arbitrary offset has been applied to the \textit{TESS} data and residuals for visibility purposes.}
\label{fig:cheops+tess_phase_photo_fit}
\end{center}
\end{figure*}

\begin{figure}
\begin{center}
\includegraphics[width=\columnwidth]{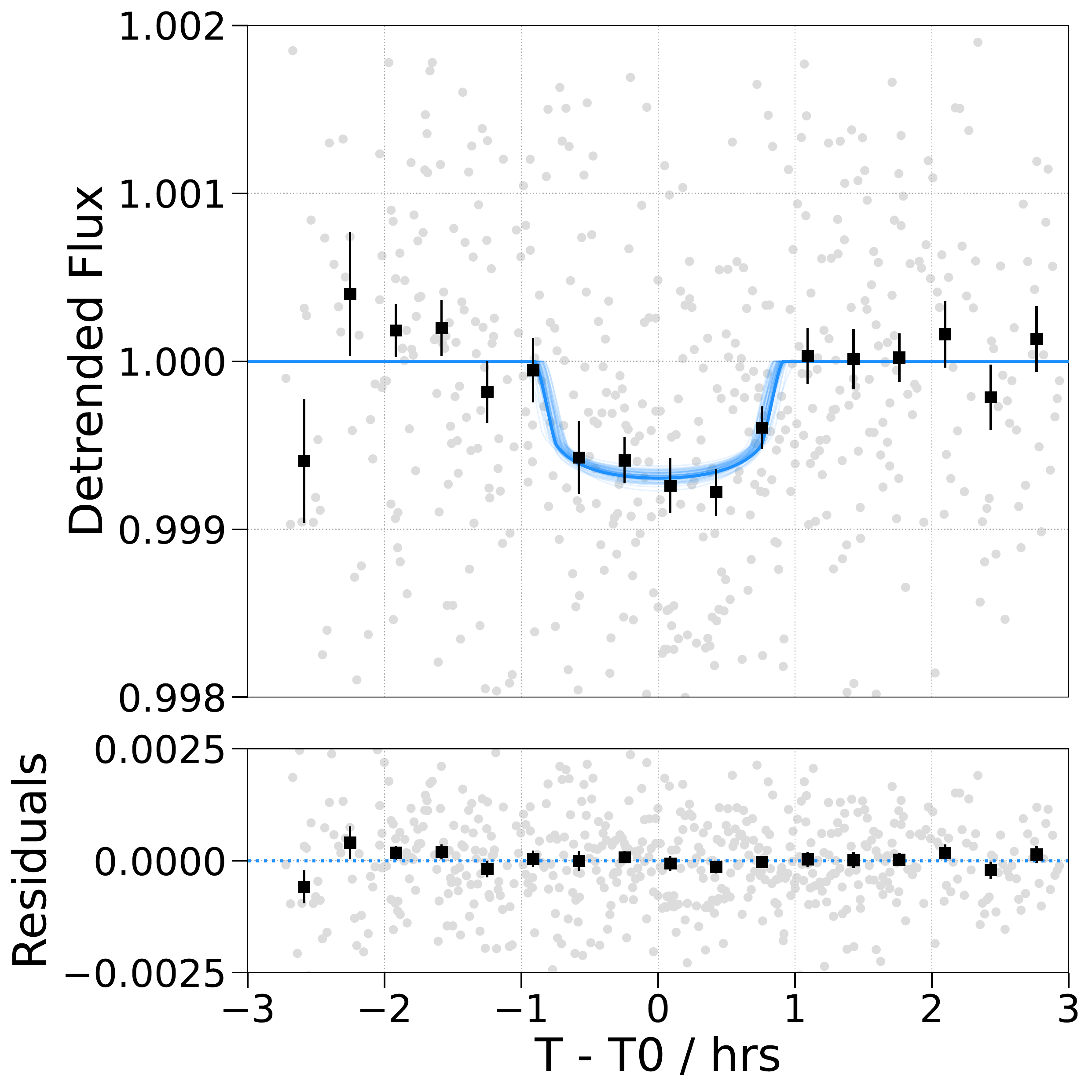}
\caption{Results of the global photometric fit. Upper: Phase-folded LCOGT lightcurve for the inner planet. The lightcurve has been detrended to remove instrumental effects and the data (grey) has been binned to 20 minutes (black squares) to guide the eye. The median transit model for the inner planet (blue line) is included, along with 50 random samples drawn from the posterior distribution of the model. Lower: Residuals of the median transit model.}
\label{fig:lco_phase_photo_fit}
\end{center}
\end{figure}

In Section \ref{sec:photo-analysis}, we described our joint fit of the \textit{TESS}, \textit{CHEOPS} and LCOGT photometry and we present the resulting fitted planetary parameters in Table \ref{tab:photo-planet-posteriors}. We also include the following derived planetary parameters: transit depth ($\delta$ = (R$_{\text{P}}$ / R$_{\star}$)$^2$), planet radius (R$_{\text{P}}$), semi-major axis (a), orbital inclination (i), total transit duration (T$_{\text{dur}}$), insolation flux (S$_{\text{P}}$) and equilibrium temperature assuming zero bond albedo and full day-night heat redistribution (T$_{\text{eq}}$).
\begin{table}
\centering
  \caption{Fitted and derived parameters for HD\,15906\,b and c from the global photometric fit presented in Section \ref{sec:photo-analysis}.}
  \resizebox{\columnwidth}{!}
  {\begin{tabular}{ccc}
    \hline
    Parameter & HD\,15906\,b & HD\,15906\,c\\
    \hline
    \multicolumn{3}{c}{Fitted Parameters}\\
    \hline
    $P$ / days & \photoperiodone & \photoperiodtwo \vspace{1mm}\\
    $T_0$ / (BJD - 2457000) & \photoepochone & \photoepochtwo \vspace{1mm}\\
    R$_{\text{P}}$ / R$_{\star}$ & \photopone & \photoptwo \vspace{1mm}\\
    $b$ & \photobone & \photobtwo \vspace{1mm}\\
    $e$ & \photoeccone & \photoecctwo \vspace{1mm}\\
    $\omega$ / deg & \photoomegaone & \photoomegatwo \vspace{1mm}\\
    $\rho_{\star}$ / kgm$^{-3}$ & \multicolumn{2}{c}{\photorho} \vspace{1mm}\\
    \hline
    \multicolumn{3}{c}{Derived Parameters}\\
    \hline
    $\delta$ / ppm & \photodepthone & \photodepthtwo \vspace{1mm}\\
    R$_{\text{P}}$ / R$_\oplus$ & \photoradone & \photoradtwo \vspace{1mm}\\
    a / R$_{\star}$ & \photoaRone & \photoaRtwo \vspace{1mm}\\
    a / AU & \photosemimajorone & \photosemimajortwo \vspace{1mm}\\
    i / deg & \photoincone & \photoinctwo \vspace{1mm}\\
    T$_{\text{dur}}$ / hrs & \photodurone & \photodurtwo \vspace{1mm}\\
    S$_{\text{P}}$ / S$_\oplus$ & \photoinsolone & \photoinsoltwo \vspace{1mm}\\
    T$_{\text{eq}}$ / K & \photoeqmTone & \photoeqmTtwo \\
    \hline
  \end{tabular}}
\label{tab:photo-planet-posteriors}
\end{table}
Figures \ref{fig:tess_photo_fit}, \ref{fig:cheops_photo_fit} and \ref{fig:lco_photo_fit} show the \textit{TESS}, \textit{CHEOPS} and LCOGT data alongside the global photometric model. Figure \ref{fig:cheops+tess_phase_photo_fit} shows the detrended \textit{TESS} and \textit{CHEOPS} data, phase-folded on each planet with the best-fitting transit model, and Figure \ref{fig:lco_phase_photo_fit} shows the same for the LCOGT data. For a full list of posterior values and the corner plots presenting the posterior distributions of the fitted planetary parameters, see Appendix \ref{sec:photo-appendix}. 

In Figure \ref{fig:cheops+tess_phase_photo_fit}, there is a small dip during the transit of the outer planet which occurs just before the mid-transit position in both the \textit{TESS} and \textit{CHEOPS} phase-folded lightcurves. Rather than being a significant feature, it is most likely a coincidence. In the \textit{CHEOPS} data, there is very poor coverage of this part of the transit and the dip is exaggerated by binning. In the \textit{TESS} data, the mid-transit dip is only present in the first of the two transits.

Our analysis has shown that HD\,15906\,b is a \photoradoneshort\ R$_\oplus$ planet orbiting its host star at a separation of \photosemimajoroneshort\ AU with a period of \photoperiodoneshort\ days. HD\,15906\,c is bigger (\photoradtwoshort\ R$_\oplus$) and orbits the host star at a larger separation (\photosemimajortwoshort\ AU) with a longer period (\photoperiodtwoshort\ days). The fit favoured slightly eccentric orbits (e$_{\text{b}}$ = \photoecconeshort, e$_{\text{c}}$ = \photoecctwoshort) with a high impact parameter (b$_{\text{b}}$ = \photoboneshort, b$_{\text{c}}$ = \photobtwoshort), but the transits of both planets are non-grazing. The inner and outer planet receive \photoinsoloneshort\ and \photoinsoltwoshort\ times the amount of flux that the Earth receives from the Sun and, assuming zero bond albedo and full day-night heat redistribution, they have equilibrium temperatures of \photoeqmToneshort\ and \photoeqmTtwoshort\ K. We remind the reader that we repeated our global photometric fit with zero eccentricity and all fitted planet parameters were consistent within 1.2$\sigma$. In this case, we derived planetary radii of 2.24 $\pm$ 0.07 R$_\oplus$ and 2.84 $\pm$ 0.05 R$_\oplus$ for HD\,15906\,b and c, respectively. 

\subsection{Transit Timing Variation Results}\label{sec:ttv-results}
In Section \ref{sec:ttv-analysis}, we described our TTV analysis of the HD\,15906 system. The fitted observed transit times for each planet are presented in Table \ref{tab:ttv-transit-posteriors}. From these values, \texttt{juliet} derived the best-fitting period and mid-transit time for each planet, assuming a linear ephemeris. These values, and all other fitted planet parameters, were fully consistent with the results of the global photometric model (see Section \ref{sec:photo-results}) within 2$\sigma$.
\begin{table}
\centering
  \caption{Observed mid-transit times for HD\,15906\,b and c from the TTV analysis presented in Section \ref{sec:ttv-analysis}}
  {\begin{tabular}{cc}
    \hline
    Mid-transit time / (BJD - 2457000) & Instrument\\
    \hline
    \multicolumn{2}{c}{HD\,15906\,b}\\
    \hline
     1416.3499$^{+0.0033}_{-0.0039}$ & \textit{TESS} \vspace{1mm} \\
     1427.2780$^{+0.0037}_{-0.0053}$ & \textit{TESS} \vspace{1mm} \\
     2148.2970$^{+0.0017}_{-0.0016}$ & \textit{TESS} \vspace{1mm} \\
     2159.2181$^{+0.0050}_{-0.0048}$ & \textit{TESS} \vspace{1mm} \\
     2454.1965$^{+0.0027}_{-0.0034}$ & LCOGT \vspace{1mm} \\
     2497.8933 $\pm$ 0.0009 & \textit{CHEOPS} \vspace{1mm} \\
     2519.7323$^{+0.0056}_{-0.0050}$ & LCOGT \vspace{1mm} \\
    \hline
    \multicolumn{2}{c}{HD\,15906\,c}\\
    \hline
     1430.8323$^{+0.0031}_{-0.0033}$ & \textit{TESS} \\
     2164.6570 $\pm$ 0.0022 & \textit{TESS} \\
     2488.4142 $\pm$ 0.0007 & \textit{CHEOPS} \\
     2531.5753 $\pm$ 0.0008 & \textit{CHEOPS} \\
    \hline
  \end{tabular}}
\label{tab:ttv-transit-posteriors}
\end{table}

\begin{figure}
\begin{center}
\includegraphics[width=\columnwidth]{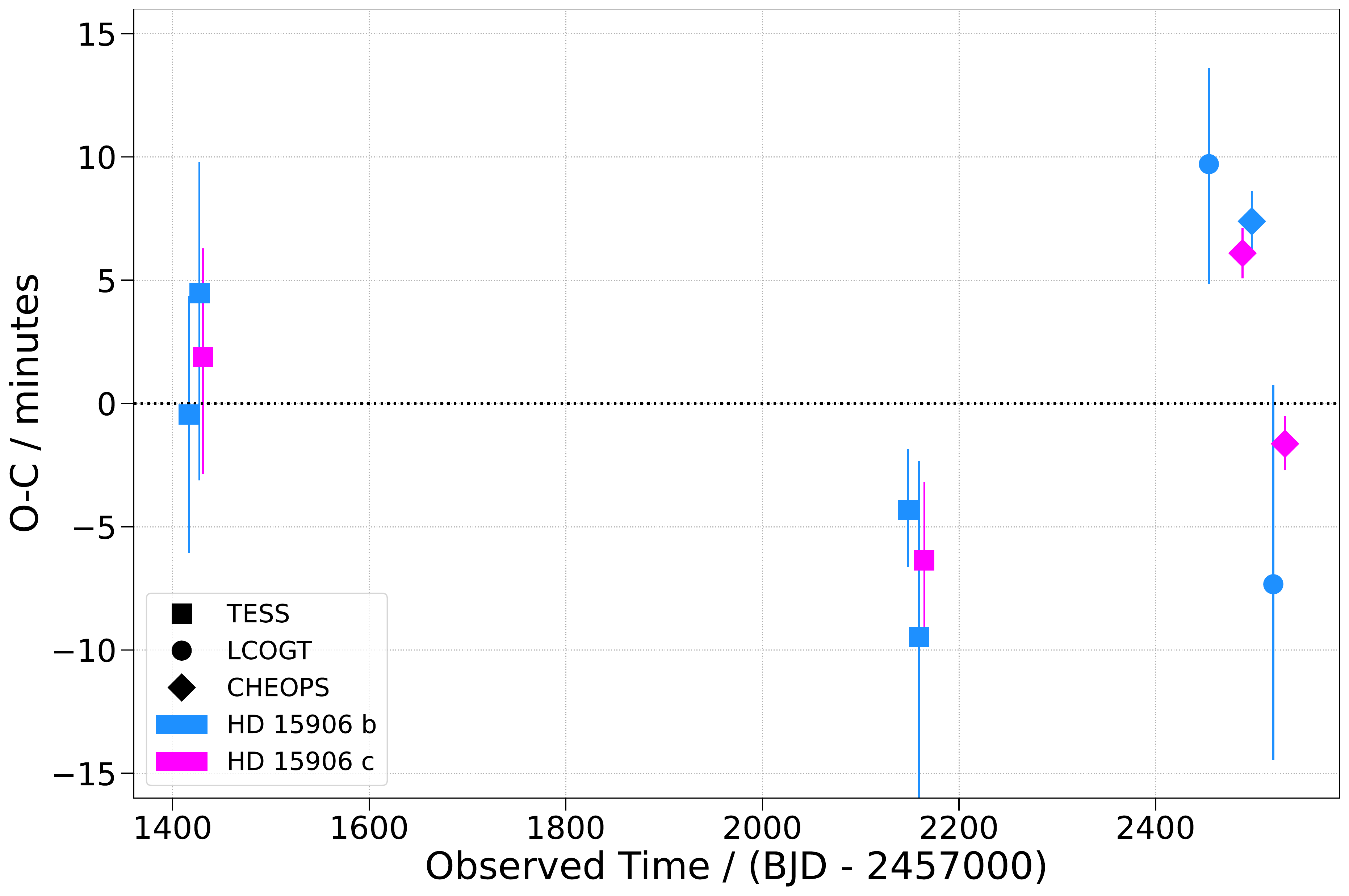}
\caption{Results of the TTV analysis. This plot shows the difference between the observed (O) transit time and the computed (C) transit time, assuming a linear ephemeris, for transits of the inner (blue) and outer (pink) planet from \textit{TESS} (square), \textit{CHEOPS} (diamond) and LCOGT (circle).}
\label{fig:ttv_results}
\end{center}
\end{figure}

Using the best-fitting period and mid-transit time, we computed the expected transit times for each planet. We then plotted an observed - computed (O-C) diagram, see Figure \ref{fig:ttv_results}, to show the TTVs. We found marginal evidence for TTVs -- the maximum TTV is $\sim$ 10 minutes, but nine of the eleven transits are consistent with no TTVs within 3$\sigma$. With only eleven transits of two planets and a gap of $\sim$ 2 years in the data, we did not attempt to model these TTVs. In Section \ref{sec:ttv_predictions}, we simulate the expected TTV signals for the two planets and compare these predictions with the observations.

\subsection{Radial Velocity Results}\label{sec:rv-results}
In an attempt to detect the two planetary signals in the HARPS and FIES data, we fit five models to the RVs (see Section \ref{sec:rv-analysis}). We tried a model with no planets, only the inner and outer planet, both planets and both planets plus a GP to model the stellar activity. In the fit with the GP, the posterior distributions of the GP hyperparameters were the same as the priors, which tells us the data was unable to constrain the GP model, as expected. In Table \ref{tab:bayes_evidences_rv-fits}, we present the Bayes evidence of each fit compared to the fit with no planets. The model with no planets had the highest evidence, with a substantial or strong preference over the other models \citep{Bayesfactor}, and we therefore conclude that the two transiting planets are not detected in the current HARPS and FIES RV data. However, we can still utilise this data for validation purposes, see Section \ref{sec:spec_validation}.
\begin{table}
\centering
  \caption{Comparison of the Bayes evidence from our HARPS and FIES RV fits. The difference in Bayes evidence (dlnZ) between each fit and the fit with no planets is quoted. The model with no planets was preferred over the more complex models.}
  {\begin{tabular}{cc}
    \hline
    Model & dlnZ \\
    \hline
    No Planets & 0.0 \\
    Inner Planet Only & -2.1 \\
    Outer Planet Only & -1.6 \\
    Two Planets & -3.9 \\
    Two Planets and GP & -3.7 \\
    \hline
  \end{tabular}}
\label{tab:bayes_evidences_rv-fits}
\end{table}

\section{Vetting and Validation}\label{sec:vetting+validation}
It is important to confirm that the transits we observed with \textit{TESS}, \textit{CHEOPS} and LCOGT were caused by planets orbiting HD\,15906. We therefore need to rule out false positive scenarios, including:
\begin{enumerate}
\item The target star is an eclipsing binary (EB).
\item The target star has a gravitationally associated companion star that is either an EB or has transiting planets. 
\item There is an aligned foreground or background star, not gravitationally associated with the target star, that is either an EB or has transiting planets. 
\item There is a nearby star, with a small angular separation from the target star but not gravitationally associated with it, that is either an EB or has transiting planets. 
\end{enumerate}
Furthermore, it is important to check for nearby unresolved stars because, if not accounted for, the blended flux can lead to underestimated planetary radii and improper characterisation of the host star \citep{ciardi2015,FH1,FH2}.

As mentioned in Section \ref{sec:tess-obs}, HD\,15906\,b passed all of the SPOC vetting tests. In addition, it has been shown that multiplanet systems are significantly less likely to be false positives than single planet systems, especially when the planets are smaller than 6 R$_\oplus$ \citep{Liss-2012,2021-Guerrero}. In this section, we use additional observational and statistical techniques to validate the HD\,15906 planetary system.

\subsection{High-Resolution Spectroscopy}\label{sec:spec_validation}
Using the HARPS and FIES data, we did not detect the RV signals induced by the two transiting objects (see Section \ref{sec:rv-results}). In this section, we use the HARPS data to rule out stellar masses for the transiting objects and place limits on the presence of a bound stellar companion.

\begin{figure}
\begin{center}
\includegraphics[width=\columnwidth]{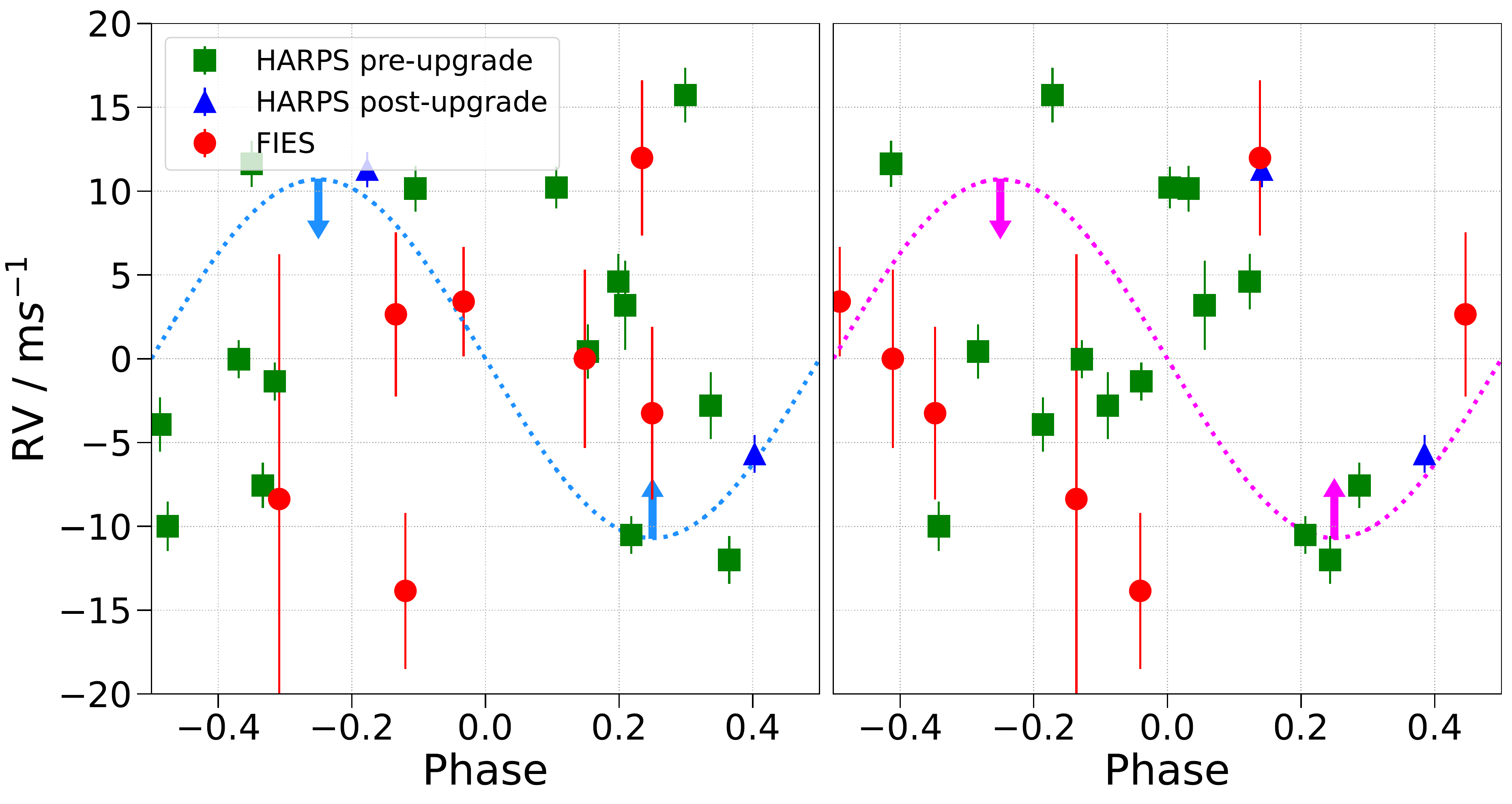}
\caption{HARPS (green squares and blue triangles) and FIES (red circles) RV data folded on the inner (left) and outer (right) planet. The transits occur at phase zero. A Keplerian model (dotted line) has been plotted on each axis to guide the eye and the arrows illustrate that this is an upper limit. The model represents a planet on a circular orbit with a semi-amplitude equivalent to the RMS of the HARPS data (10.70 ms$^{-1}$), a proxy for its maximum value.}
\label{fig:rv_data_phase}
\end{center}
\end{figure}

In Figure \ref{fig:rv_data_phase}, we show the HARPS and FIES RVs folded on HD\,15906\,b and c using the ephemerides obtained in the global photometric analysis (Table \ref{tab:photo-planet-posteriors}). 
The RMS of the HARPS data (10.70 ms$^{-1}$) can be used as a proxy for the maximum possible semi-amplitudes of the two transiting objects. Using the stellar mass presented in Table \ref{tab:stellar-params}, the orbital parameters presented in Table \ref{tab:photo-planet-posteriors} and a semi-amplitude of 10.70 ms$^{-1}$, HD\,15906\,b has an upper mass limit of $\sim$ 32 M$_\oplus$ and HD\,15906\,c has an upper mass limit of $\sim$ 39 M$_\oplus$. This confirms that the two transiting objects must be of planetary mass. 

Furthermore, under the assumption of a circular orbit and an orbital inclination of 90 degrees, the RMS of the HARPS data rules out a bound brown dwarf or star, with a mass greater than 13 M$_{\text{Jupiter}}$, out to $\sim$ 1500 AU. At the distance of HD\,15906, this corresponds to an angular separation of $\sim$ 32\arcsec. Even down to an orbital inclination of 10 degrees, we can rule out a brown dwarf or stellar companion out to $\sim$ 45 AU, corresponding to an angular separation of $\sim$ 1\arcsec. 

Finally, we checked for a linear drift in the RV data because this could be indicative of a long-period bound stellar companion. We chose the pre-upgrade HARPS data for this purpose because it has the longest baseline ($>$ 9 years). We used \texttt{juliet} to perform a fit of this data, using a model consisting of no planets, an offset, white noise and a linear trend. The best-fit gradient was consistent with zero within 1$\sigma$ and this supports the conclusion that HD\,15906 does not have a bound stellar companion.

\subsection{Archival Imaging}\label{sec:archival-imaging}
HD\,15906 is a high proper motion star \citep[$\mu$ = 195.97 mas yr$^{-1}$;][]{gaiacollaboration_2022}. We therefore made use of archival imaging to check for foreground or background objects at the star's present day position. 

\begin{figure*}
\begin{center}
\includegraphics[width=\textwidth]{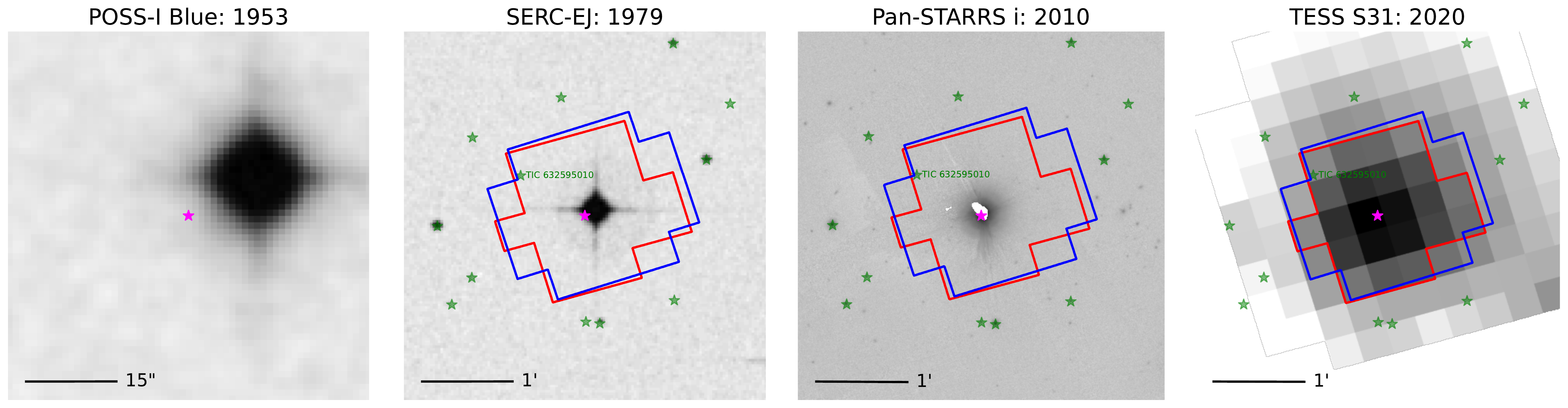}
\caption{Images of HD\,15906 spanning 67 years, from 1953 to 2020. Left to right: POSS-I, SERC-EJ, Pan-STARRS and \textit{TESS} sector 31. All images are shown on a scale of 4\arcmin\ $\times$ 4\arcmin, except for the POSS-I image which is zoomed in to 1\arcmin\ $\times$ 1\arcmin, and centred on the 2020 position of HD\,15906 (pink star). We overlaid the \textit{TESS} apertures from sector 4 (blue) and sector 31 (red) on the images, as well as the 2020 positions of all known stars from \textit{Gaia} DR3 \citep[green stars;][]{gaiacollaboration_2022}. Only one of these stars (TIC\,632595010; \textit{TESS} magnitude = 20.3) is within the \textit{TESS} apertures.}
\label{fig:archival-imaging}
\end{center}
\end{figure*}

HD\,15906 was observed on 11 November 1953 by the Oschin Schmidt Telescope, using a blue photographic emulsion \citep[$\lambda$ = 330 - 500 nm;][]{Monet-2003}, during the first Palomar Observatory Sky Survey (POSS-I). It was observed again on 21 September 1979 by the UK Schmidt Telescope, using a blue photographic emulsion \citep[$\lambda$ = 395 - 540 nm;][]{Monet-2003}, during the SERC-EJ survey. We downloaded these images from the Digitized Sky Survey (DSS)\footnote{\label{footnote:DSS}\url{https://archive.stsci.edu/cgi-bin/dss_form}} and plotted them in the first two panels of Figure \ref{fig:archival-imaging}. HD\,15906 was also observed in 2010 by the Panoramic Survey Telescope and Rapid Response System \citep[Pan-STARRS;][]{PS1-2016}. We downloaded the i-filter Pan-STARRS image from the MAST and plotted it in the third panel of Figure \ref{fig:archival-imaging}. Finally, HD\,15906 was observed during \textit{TESS} sector 31 in 2020. We downloaded the target pixel file (TPF) from the MAST and plotted the first good quality cadence in the final panel of Figure \ref{fig:archival-imaging}.

HD\,15906 moved $\sim$ 13\arcsec\ between the POSS-I observation in 1953 and \textit{TESS} sector 31 in 2020. Using the POSS-I image, we rule out a foreground or background star at the \textit{TESS} sector 31 position of HD\,15906 down to a \textit{TESS} magnitude of $\sim$ 18. A star this faint would be incapable of producing the transit signals we observe, even in the case of a full EB, and it would not significantly impact the derived planet parameters due to flux blending \citep{ciardi2015}. We therefore conclude that our results are not affected by an unresolved foreground or background star. 

\subsection{High-Resolution Imaging}\label{sec:high-res-obs}
\begin{figure*}
\begin{center}
\includegraphics[width=\textwidth]{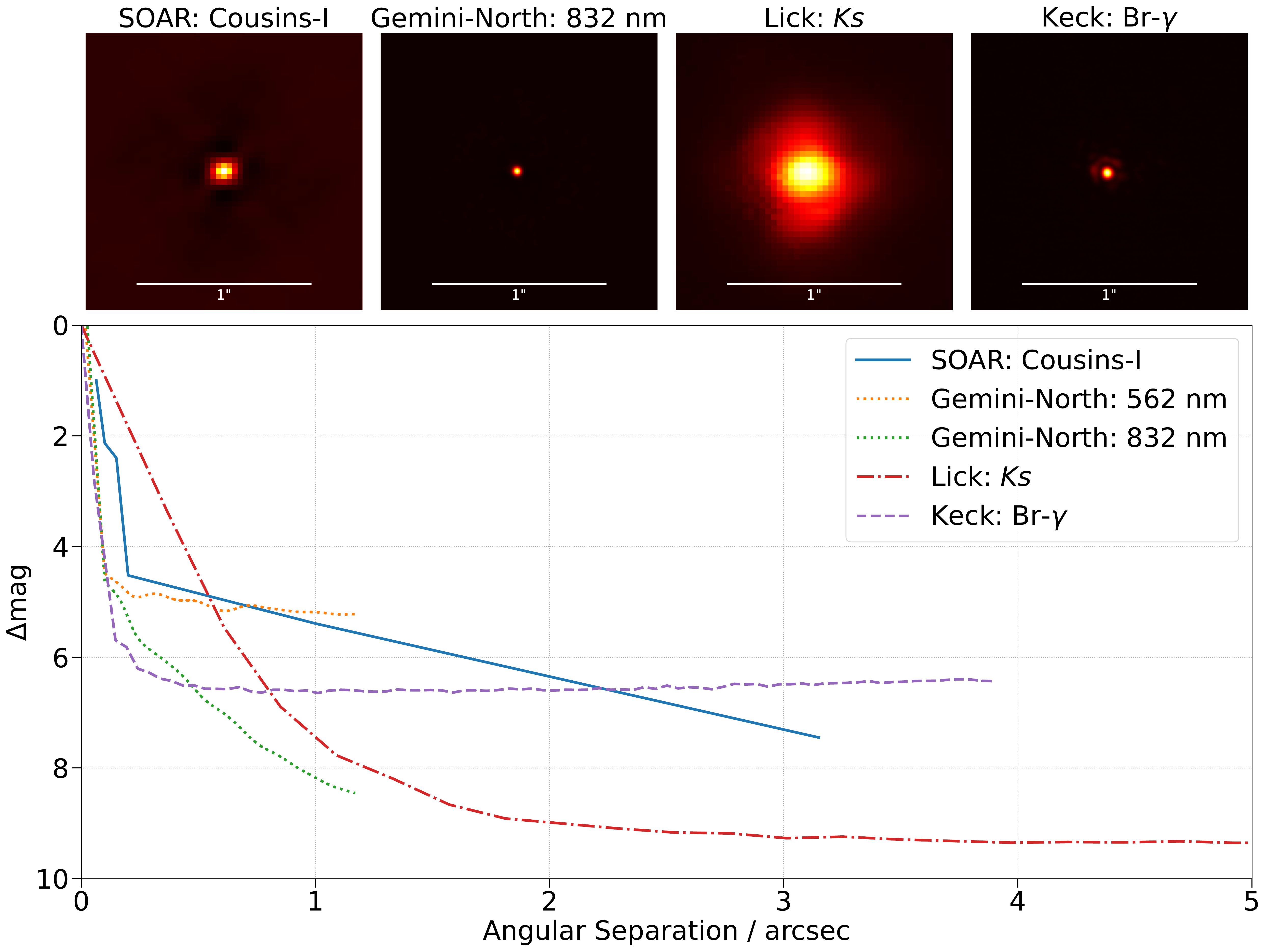}
\caption{High-resolution imaging of HD\,15906. Upper: From left to right are the speckle images from SOAR and Gemini-North and the NIR AO images from Lick and Keck. Each image is zoomed into a region of 1.6\arcsec\ $\times$ 1.6\arcsec centred on HD\,15906. Lower: Contrast curves from each observation.}
\label{fig:high-res-imaging}
\end{center}
\end{figure*}

High-resolution imaging was used to search for nearby stars, bound or unbound, that could be contaminating the photometry. We observed HD\,15906 with a combination of high-resolution resources, including near-infrared adaptive optics (NIR AO) imaging at the Keck and Lick Observatories and optical speckle imaging at Gemini-North and SOAR. While the optical observations tend to provide higher resolution, the NIR AO tend to provide better sensitivity, especially to lower-mass stars. The combination of the observations in multiple filters enables better characterisation of any companions that might be detected. The observations are described in detail in the following subsections and a summary is provided in Table \ref{tab:high-res-imaging}. Figure \ref{fig:high-res-imaging} shows the resulting images and contrast curves. No stellar companions were detected within the contrast and angular limits of each facility, essentially ruling out stars at least $\sim$ 7 magnitudes fainter than HD\,15906 between 0.5\arcsec\ and 10\arcsec. At small angular separations, where high-resolution imaging does not achieve a high contrast, we used high-resolution spectroscopy to rule out bound companions within $\sim$ 1\arcsec\ (see Section \ref{sec:spec_validation}). 
\begin{table}
\centering
  \caption{A summary of the high-resolution imaging observations of HD\,15906.}
  {\begin{tabular}{cccc}
    \hline
    Facility & Instrument & Filter & Date [UTC] \\
    \hline
    SOAR & HRCam & Cousins-I & 2019-07-14 \\
    Lick & ShARCS & $Ks$ & 2019-07-21 \\
    Gemini-North & 'Alopeke & 562 nm & 2019-10-15 \\
    Gemini-North & 'Alopeke & 832 nm & 2019-10-15 \\
    Keck & NIRC2 & Br-$\gamma$ & 2020-09-09 \\
    \hline
  \end{tabular}}
\label{tab:high-res-imaging}
\end{table}

\subsubsection{SOAR}\label{sec:soar-obs}
We searched for stellar companions to HD\,15906 with speckle imaging on the 4.1\,m Southern Astrophysical Research (SOAR) telescope \citep{Tokovinin-2018} on 14 July 2019. We observed in Cousins I-band, a similar visible bandpass to \textit{TESS}. This observation was sensitive to a star 5.3 magnitudes fainter than HD\,15906 at an angular distance of 1\arcsec\ from the target. More details of the observations within the SOAR \textit{TESS} survey are available in \citet{ziegler2020}. The 5$\sigma$ detection sensitivity and speckle auto-correlation functions from the observations are shown in Figure \ref{fig:high-res-imaging}. No nearby stars were detected within 3\arcsec\ ($\sim$ 137 AU, if bound) of HD\,15906 in the SOAR observations.

\subsubsection{Lick}\label{sec:lick-obs}
We observed HD\,15906 on 21 July 2019 using the Shane Adaptive optics infraRed Camera-Spectrograph (ShARCS) camera on the Shane 3\,m telescope at Lick Observatory \citep{2012SPIE.8447E..3GK,2014SPIE.9148E..05G,2014SPIE.9148E..3AM}. Observations were taken with the Shane AO system in natural guide star mode in order to search for nearby, unresolved stellar companions. We collected a single sequence of observations using a $Ks$ filter ($\lambda_0$ = 2.150 $\mu$m, $\Delta\lambda$ = 0.320 $\mu$m). We reduced the data using the publicly available \texttt{SImMER} pipeline\footnote{\url{https://github.com/arjunsavel/SImMER}}\citep{2020AJ....160..287S}. Our reduced image and corresponding contrast curve is shown in Figure \ref{fig:high-res-imaging}. The observations rule out stellar companions $\sim$ 4 magnitudes fainter than HD\,15906 at 0.5\arcsec\ ($\sim$ 23 AU, if bound) and $\sim$ 9 magnitudes fainter between 2\arcsec\ and 10\arcsec\ ($\sim$ 92 - 458 AU, if bound). 

\subsubsection{Gemini-North}\label{sec:gemN-obs}
HD\,15906 was observed on 15 October 2019 using the 'Alopeke speckle instrument on the Gemini-North 8\,m telescope \citep{SH,HF}. 'Alopeke provides simultaneous speckle imaging in two bands (562 nm and 832 nm) with output data products including a reconstructed image with robust contrast limits on companion detections. Three sets of 1000 $\times$ 0.06 second images were obtained and processed in our standard reduction pipeline \citep[see][]{Howell2011}. Figure \ref{fig:high-res-imaging} includes our final 5$\sigma$ contrast curves and the 832 nm reconstructed speckle image. We find that HD\,15906 has no companion stars brighter than 5-8 magnitudes below that of the target star within the angular and image contrast levels achieved. The angular region covered ranges from the 8\,m telescope diffraction limit (20 mas) out to 1.2\arcsec\ ($\sim$ 0.9 to 55 AU, if bound).

\subsubsection{Keck}\label{sec:keck-obs}
HD\,15906 was observed with NIR AO high-resolution imaging at the Keck Observatory on 9 September 2020. The observations were made with the NIRC2 instrument, which was positioned behind the natural guide star AO system \citep{wizinowich2000}, on the Keck-II telescope. We used the standard 3-point dither pattern to avoid the lower left quadrant of the detector which is typically noisier than the other three quadrants. The dither pattern step size was 3\arcsec\ and was repeated twice, with each dither offset from the previous dither by 0.5\arcsec. The camera was in the narrow-angle mode with a full field of view of $\sim$ 10\arcsec\ and a pixel scale of approximately 0.0099442\arcsec pix$^{-1}$. The observations were made in the narrow-band Br-$\gamma$ filter ($\lambda_0$ = 2.1686 $\mu$m, $\Delta\lambda$ = 0.0326 $\mu$m) with an integration time of 0.5 second with one co-add per frame for a total of 4.5 seconds on target. The AO data were processed and analysed with a custom set of IDL tools \citep[see description in][]{2021NIRC2} and the resolution of the final combined image, determined from the FWHM of the PSF, was 0.048\arcsec. The sensitivity of the combined AO image was determined according to \citet{furlan2017} and the resulting sensitivity curve for the Keck data is shown in Figure \ref{fig:high-res-imaging}. The image reaches a contrast of $\sim$ 7 magnitudes fainter than the host star between 0.5\arcsec\ and 4\arcsec\ ($\sim$ 23 to 183 AU, if bound) and no stellar companions were detected.

\subsection{\textit{Gaia} Assessment}\label{sec:gaia_validation}
We used \textit{Gaia} DR3 \citep{gaiacollaboration_2022} to show that there are no nearby, resolved stars bright enough to cause the transits we observe. The images presented in Figure \ref{fig:archival-imaging} show that there is only one \textit{Gaia} DR3 star within the \textit{TESS} optimal apertures. This is TIC\,632595010 with a \textit{TESS} magnitude of 20.3 ($>$ 10 mag fainter than HD\,15906) and a separation of $\sim$ 50\arcsec\ from HD\,15906. This star is not bright enough to be the source of the transit signals we see, even in the case of a full EB. Furthermore, as explained in Section \ref{sec:ground-photometry}, the LCOGT observations confirmed that the transit signals do not originate from any of the known \textit{Gaia} DR3 stars.

We also searched for wide stellar companions that may be bound members of the system. Based upon similar parallaxes and proper motions \citep{mugrauer2020,mugrauer2021}, there are no additional widely separated companions identified by \textit{Gaia}.

Finally, the \textit{Gaia} DR3 astrometry provides additional information on the possibility of inner companions that may have gone undetected by either \textit{Gaia} or the high-resolution imaging/spectroscopy. The \textit{Gaia} Renormalised Unit Weight Error (RUWE) is a metric, similar to a reduced chi-square, where values that are $\lesssim$ 1.4 indicate that the \textit{Gaia} astrometric solution is consistent with a single star whereas RUWE values $\gtrsim$ 1.4 may indicate an astrometric excess noise, possibly caused by the presence of an unseen companion \citep[e.g.,][]{ziegler2020}. HD\,15906 has a \textit{Gaia} DR3 RUWE value of 1.15, indicating that the astrometric fits are consistent with a single star model. 

\subsection{Statistical Validation}\label{sec:statistical-validation}
We finally used \texttt{TRICERATOPS} \citep[Tool for Rating Interesting Candidate Exoplanets and Reliability Analysis of Transits Originating from Proximate Stars;][]{triceratops} to statistically validate the two transiting planets in the HD\,15906 system. This Bayesian tool uses the stellar and planet parameters, the transit lightcurve and the high-resolution imaging to test the false positive scenarios listed at the start of Section \ref{sec:vetting+validation} and calculate the false positive probability (FPP) and the nearby false positive probability (NFPP) of \textit{TESS} planet candidates. The FPP is the probability that the observed transit is not caused by a planet on the target star and the NFPP is the probability that the observed transit originates from a resolved nearby star. To consider a planet candidate validated, it must have FPP $<$ 0.015 and NFPP $<$ 0.001. 

We ran \texttt{TRICERATOPS} on both HD\,15906\,b and c. We used the stellar parameters presented in Table \ref{tab:stellar-params}, the planet parameters presented in Table \ref{tab:photo-planet-posteriors}, the combined \textit{TESS}, \textit{CHEOPS} and LCOGT lightcurve and the high-resolution imaging contrast curves from Section \ref{sec:high-res-obs}. \texttt{TRICERATOPS} only accepts one contrast curve as input, so we ran the analysis with each of the five contrast curves and compared the results. In agreement with our analysis in Section \ref{sec:gaia_validation}, \texttt{TRICERATOPS} did not identify any nearby resolved stars that were bright enough to be the source of the transits. The results confirmed that the highest probability scenario was that of two planets transiting HD\,15906. The most probable form of false positive scenario for the inner planet was an unresolved background EB and for the outer planet was an unresolved bound companion that is an EB. With our archival imaging (Section \ref{sec:archival-imaging}) and high-resolution spectroscopy (Section \ref{sec:spec_validation}), that \texttt{TRICERATOPS} does not consider, these scenarios become less likely. For the Gemini-North and Keck contrast curves, both planets were validated with a negligible value of NFPP and FPP $<$ 0.015. With the SOAR and Lick contrast curves, both planets had a negligible value of NFPP, the outer planet had FPP $<$ 0.015 and the inner planet had a FPP just greater than 0.015 (0.0159 for Lick and 0.0166 for SOAR). According to the \texttt{TRICERATOPS} criteria, this means that the inner signal is likely a planet. However, \texttt{TRICERATOPS} does not account for the fact that multiplanet systems are more likely to be real \citep{Liss-2012,2021-Guerrero}, so the fact that the outer planet was validated means the inner planet may also be considered validated. We therefore conclude that both HD\,15906\,b and c are validated planets according to the \texttt{TRICERATOPS} criteria. 

\section{Discussion}\label{sec:discussion}
We have presented the discovery of the HD\,15906 multiplanet system. In this section, we discuss our results, compare the system to other confirmed exoplanets and assess the feasibility of future follow-up observations. 

\subsection{\textit{TESS} Periodicity}\label{sec:discussion-tess-periodicity}
In Section \ref{sec:tess-detrend}, we reported the detection of a sinusoidal-like signal in the \textit{TESS} sector 4 lightcurve of HD\,15906. This signal has a period of 0.22004 days ($\sim$ 5 hours) and the best-fit sinusoidal model has an amplitude of $\sim$ 57 ppm, equivalent to the transit depth expected for a planet with a radius of $\sim$ 0.63 R$_\oplus$. In this section, we provide a discussion of this signal and its origin.

The 0.22 day periodic signal is present in the \textit{TESS} sector 4 lightcurve, but not the sector 31 lightcurve. The signal is present in the sector 4 SAP and PDCSAP flux, but not in the background flux or centroid position. We checked for a periodic signal in the nearest star of comparable magnitude (TIC\,4646803; \textit{TESS} magnitude = 9.51, separation = 167\arcsec). This star was observed at 30 minute cadence in sector 4, so we searched the \textit{TESS}-SPOC lightcurve \citep{TESS-SPOC} and found no periodicity at 0.22 days. 

We also extracted our own HD\,15906 lightcurves from the \textit{TESS} TPFs for both sectors. This was done using a default quality bitmask and optimising the aperture mask to reduce the combined differential photometric precision (CDPP) noise in the resulting data. The extracted target fluxes were sky-corrected using a custom background mask. Detrending was done in two steps: scattered light was corrected for using a principal component analysis and any flux modulation caused by spacecraft jitter was removed by a linear model detrending using co-trending basis vectors and the mean and average of the engineering quaternions as the basis vectors. This second method has shown promise in cleaning up \textit{TESS} photometry previously \citep{nu2lupid}. Our lightcurves were consistent with the \textit{TESS} SAP and PDCSAP flux; our sector 4 lightcurve contained a 0.22 day periodicity and our sector 31 lightcurve did not. We can therefore confirm that the periodic signal is not dependent on lightcurve extraction technique. 

Furthermore, we performed experiments extracting lightcurves from apertures of different sizes and found that using an aperture of radius 1 pixel centred on HD\,15906 resulted in a significantly larger amplitude variability (roughly by a factor of two) than when we used an aperture of radius 4 pixels. This is not what we would expect for a signal originating from within a pixel of HD\,15906 (where we would expect the amplitude to stay roughly constant given the lack of nearby bright stars to dilute the flux) or from a blended star from larger distances (which should show larger amplitude in larger apertures).

We considered the possibility that the periodic signal is a form of stellar activity originating from HD\,15906. However, a variety of arguments suggested that this was unlikely. Firstly, the signal is strongly present in \textit{TESS} sector 4, but is undetectable in any other observations. The very short period of the signal strongly disfavours it being related to the rotation period of HD\,15906, given the star's narrow spectral lines and amenability to precise RV measurements. The period ($\sim$ 5 hours) is consistent with the timescale of granulation on the surface of a Sun-like star, but this process does not create sharp periodicities like we detected (see Figure \ref{fig:residual_periodogram}, which shows a clearly defined sharp peak in the periodogram of the sector 4 \textit{TESS} residuals). Stellar pulsations can sometimes create such sharp periodicities, but main sequence stars of this type should not exhibit any pulsations on similar amplitudes or timescales.

We finally searched for evidence that the signal originated from another star on the \textit{TESS} detectors and contaminated the lightcurve of HD\,15906 through a process other than direct overlap of the PSFs. This was a frequent occurrence during the \textit{Kepler} and \textit{K2} missions \citep{Coughlin2014} but is much less common during the \textit{TESS} mission due to differences in the design of the telescopes, electronics and optics. The bright contact binary DY Cet (TIC\,441128066; \textit{TESS} magnitude = 9.23) was observed on the same CCD as HD\,15906 during \textit{TESS} sectors 4 and 31. This EB has a period of 0.4408 days and the \textit{TESS} lightcurves show a sinusoidal-like variability with a period of 0.2204 days \citep{2022DYCet}. This is consistent with the period of the signal we detected in the \textit{TESS} sector 4 lightcurve of HD\,15906. During sector 4, DY Cet was in the same CCD columns as HD\,15906, but during sector 31 it was not. We therefore conclude that the flux from DY Cet contaminated that of HD\,15906 in \textit{TESS} sector 4 during CCD readout, although the exact mechanism of contamination is currently unknown. We note that DY Cet cannot be the source of the transits of HD\,15906\,b and c, which have been independently observed by \textit{CHEOPS} and LCOGT, and we reiterate that this periodic signal does not affect our fitted planet parameters (see Section \ref{sec:tess-detrend}). 

\subsection{Transit Timing Variation Predictions}\label{sec:ttv_predictions}
In Section \ref{sec:ttv-results}, we reported marginal evidence for TTVs in the HD\,15906 system. Here, we compute the expected TTV signals for each planet and compare them with our observations. We reiterate that modelling the TTVs is beyond the scope of this work due to the small amount of data.

\begin{figure}
\begin{center}
\includegraphics[width=\columnwidth]{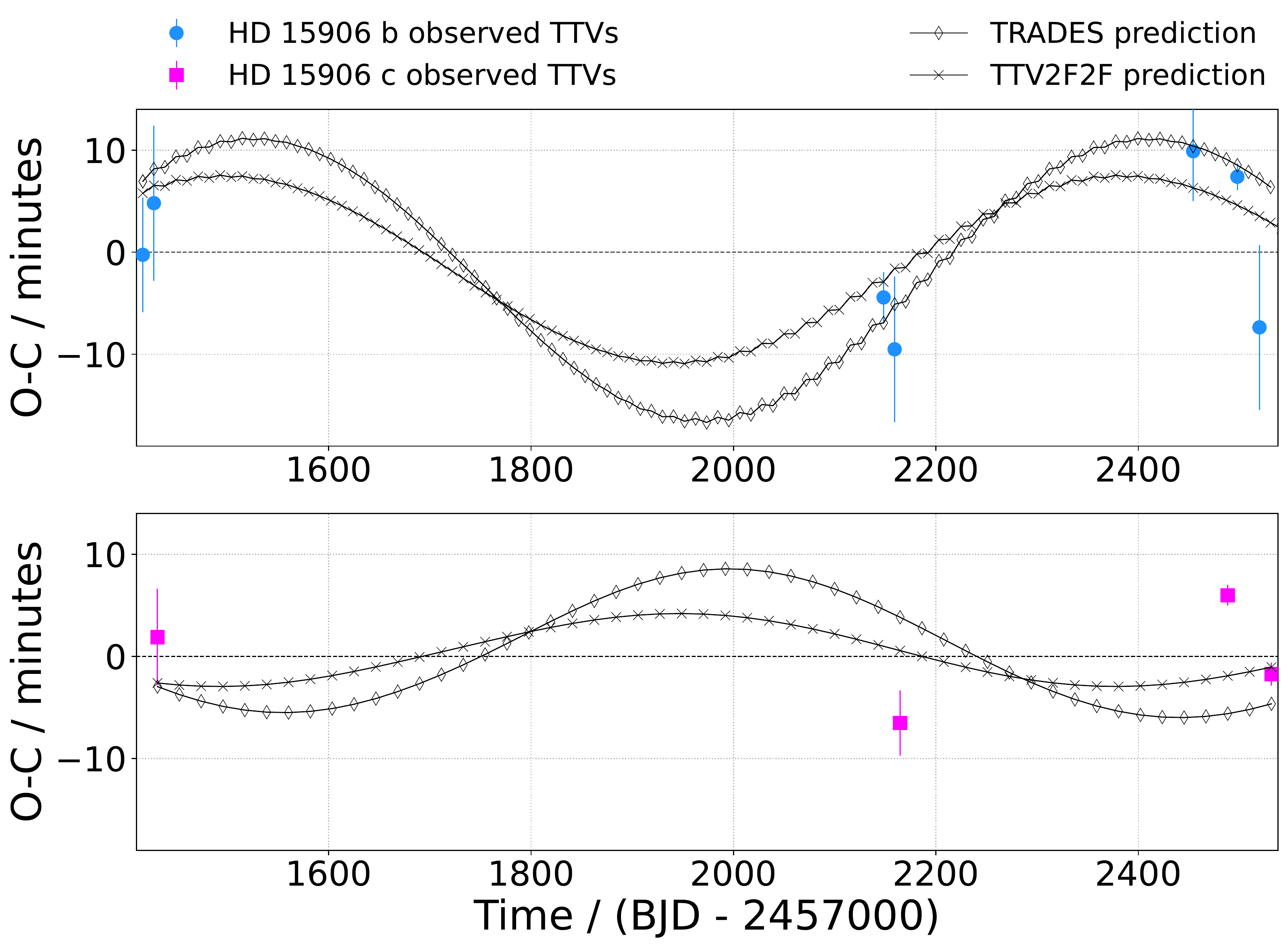}
\caption{The predicted TTV signal, as computed with \texttt{TTV2Fast2Furious} (\texttt{TTV2F2F}, crosses and solid line) and \texttt{TRADES} (open diamonds and solid line), for HD\,15906\,b (upper panel) and HD\,15906\,c (lower panel). We have also included the observed TTVs for each planet (blue circles and pink squares).}
\label{fig:ttv_predictions}
\end{center}
\end{figure}

We compared two methods for simulating the TTV signals of HD\,15906\,b and c. The first uses an approximated estimation of the TTV signal by modelling it as a linear combination of basis functions as described in \citet{Hadden2019} and implemented in the \texttt{TTV2Fast2Furious}\footnote{\url{https://github.com/shadden/TTV2Fast2Furious}} package. The second approach is a direct N-body simulation with the \texttt{TRADES}\footnote{\url{https://github.com/lucaborsato/trades}} code \citep{Borsato2014, Borsato2019, Nascimbeni2023}. For each planet, we used the orbital period and inclination presented in Table \ref{tab:photo-planet-posteriors}, the predicted mass (see Section \ref{sec:discussion-rv-future}) and we assumed a circular orbit for simplicity. We simulated the TTV signal for planets b and c for a time range that covers the full range of transit observations and we present the results in Figure \ref{fig:ttv_predictions}.

The predictions for HD\,15906\,b are generally in good agreement with the observations. We note that there is a slight difference between the amplitudes of the two simulated TTV signals, but both are consistent with most of the observations given their uncertainties. However, the predictions for HD\,15906\,c seem to be in anti-phase with the observations. As expected for a two planet configuration close to a first-order period commensurability, the predicted TTVs of planets b and c are anti-correlated \citep{2018TTVbook}. Contrary to this expectation, the observed TTVs of HD\,15906\,b and c appear to be correlated. This could suggest that there is an additional, undetected planet in the system perturbing the orbits of the two observed planets. Alternatively, the TTVs might be spurious or affected by excess systematic noise from, for example, stellar activity \citep[e.g.,][]{2013Oshagh, 2016Ioannidis}. Given the sparse sampling of the TTV signals, future observations are required to assess the true nature of the TTVs.

\subsection{Comparison with Confirmed Exoplanets}\label{sec:discussion-compare}
HD\,15906\,b and c have radii of \photoradoneshort\ R$_\oplus$ and \photoradtwoshort\ R$_\oplus$, respectively, meaning they cannot have a purely rocky composition \citep{rogers2015,Lozovsky2018}. They both fall on the upper side of the radius gap and we therefore classify them as sub-Neptunes. Furthermore, with insolation fluxes of 33 S$_\oplus$ and 13 S$_\oplus$, and equilibrium temperatures of \photoeqmToneshort\ K and \photoeqmTtwoshort\ K, both planets are in the warm regime (T$_{\text{eq}} \lesssim$ 700 K). 

Of more than 5300 confirmed exoplanets\footnote{NASA Exoplanet Archive, accessed 29/03/2023: \url{https://exoplanetarchive.ipac.caltech.edu/index.html}}, there are 66 sub-Neptune sized planets (1.75 $<$ R$_{\text{P}}$ / R$_\oplus$ $<$ 3.5) transiting bright stars (G $\leq$ 10 mag). Only 18 of these have an insolation flux less than HD\,15906\,b and only 5 have an insolation flux lower than HD\,15906\,c -- GJ\,143\,b \citep{gj143b}, $\nu^2$\,Lupi\,d \citep{nu2lupid}, HD\,23472\,c \citep{HD23472}, HD\,73583\,c \citep{HD73583} and \textit{Kepler}-37\,d \citep{Kepler37}. HD\,15906\,c is therefore one of the most lowly irradiated sub-Neptune planets transiting such a bright star. Furthermore, there are only 5 other multiplanet systems with two warm (T$_{\text{eq}} \leq$ 700 K) sub-Neptune sized planets (1.75 $<$ R$_{\text{P}}$ / R$_\oplus$ $<$ 3.5) transiting a bright (G $\leq$ 10 mag) star -- HD\,108236 \citep{bonfanti21}, $\nu^2$\,Lupi \citep{nu2lupid}, HD\,191939 \citep{2023Orell}, HD\,23472 \citep{HD23472} and TOI\,2076 \citep{TOI-2076}. The HD\,15906 system is therefore an interesting target for future follow-up studies, discussed further in Section \ref{sec:discussion-future}.

\begin{figure}
\begin{center}
\includegraphics[width=\columnwidth]{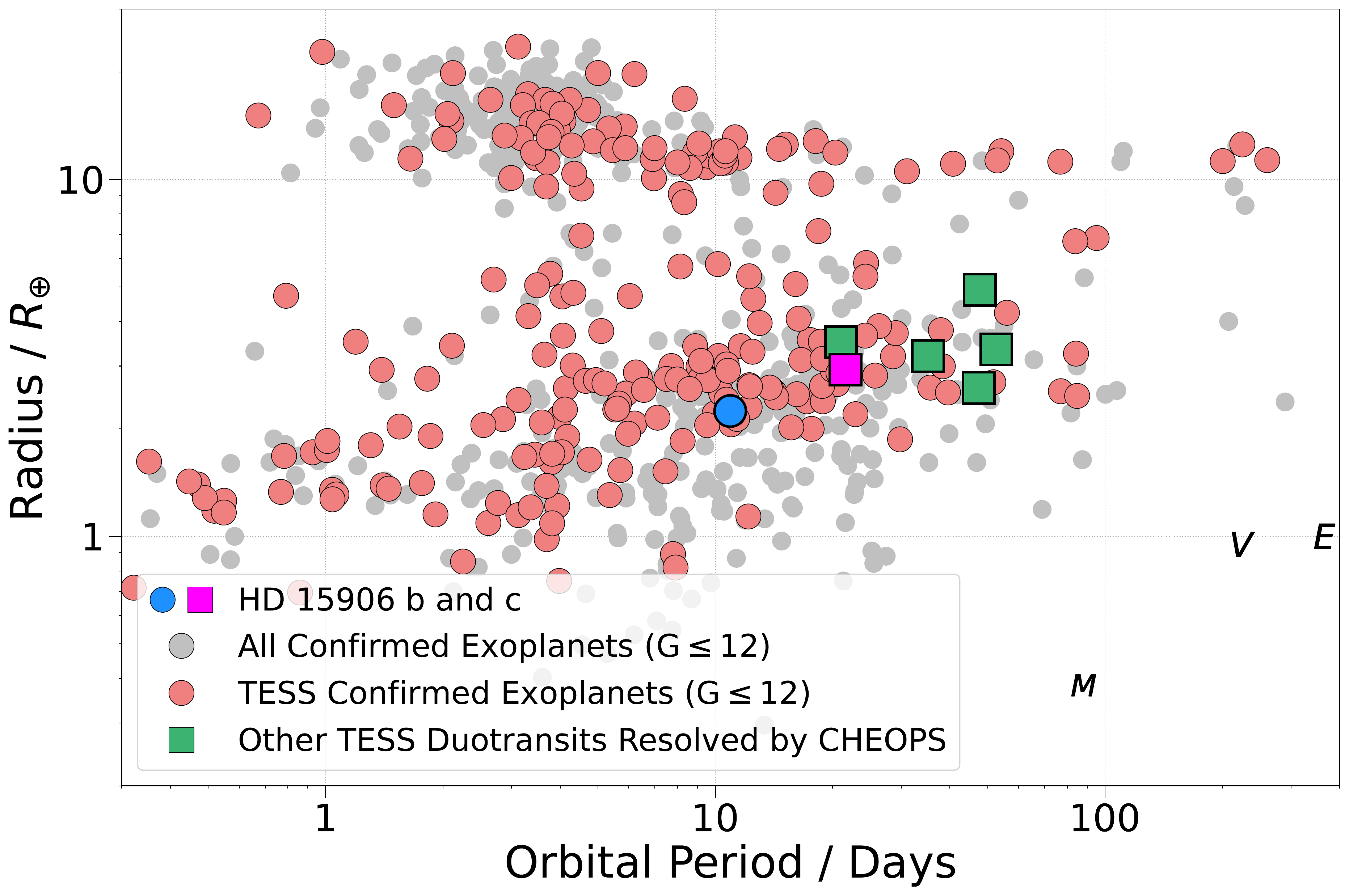}
\caption{Period-radius diagram of confirmed exoplanets with a \textit{Gaia} magnitude brighter than 12, where discoveries made by \textit{TESS} are highlighted in red. HD\,15906\,b (blue circle) and c (pink square) are included, alongside the five additional \textit{TESS} duotransits resolved by \textit{CHEOPS} (green squares).}
\label{fig:period-radius}
\end{center}
\end{figure}

Due to the nature of its observing strategy, \textit{TESS} is biased towards the discovery of short-period planets; less than 14\% of planets confirmed by \textit{TESS} have periods longer than 20 days, of which only half have radii smaller than 4 R$_\oplus$. This work has demonstrated how \textit{CHEOPS} can be used to follow-up \textit{TESS} duotransits to expand the sample of long-period planets. Figure \ref{fig:period-radius} presents a period-radius diagram comparing the two planets in the HD\,15906 system to the confirmed exoplanet population. The other \textit{TESS} duotransits resolved by \textit{CHEOPS} have also been included -- TOI\,2076\,c and d \citep{TOI-2076}, HIP\,9618\,c  \citep{Osborn2023}, TOI\,5678\,b \citep{UlmerMoll2023} and HD\,22946\,d \citep{Garai2023}. Through our \textit{CHEOPS} duotransit programme, we have contributed to the discovery of 6 planets with periods longer than 20 days, radii smaller than 5 R$_\oplus$ and host stars brighter than G = 12 mag. There are only 18 other planets confirmed by \textit{TESS} in this parameter space, illustrating the power of the \textit{TESS} and \textit{CHEOPS} synergy for the discovery of small, long-period planets transiting bright stars. 

\subsection{Potential for Future Follow-Up}\label{sec:discussion-future}
With two warm sub-Neptunes transiting a bright (G $\sim$ 9.5 mag) K-dwarf, the HD\,15906 system is an excellent target for future observations to measure the masses of the planets and perform atmospheric characterisation. Warm sub-Neptunes are less affected by radiation from their host star than their hot counterparts, meaning their atmospheres will not have been sculpted so heavily by photoevaporation and they will more closely resemble their primordial state. Observations of these planets are therefore crucial in testing models of the formation and evolution of sub-Neptune planets. In addition, as a multiplanet system, HD\,15906 will allow for comparative studies of internal structure and composition as a function of stellar irradiation.

\subsubsection{Radial Velocity}\label{sec:discussion-rv-future}
We were unable to detect HD\,15906\,b and c in the current HARPS and FIES RV data due to the small number of sparsely sampled observations (see Figure \ref{fig:rv_periodogram} and Section \ref{sec:rv-results}). Here, we use the results of our global photometric analysis (Table \ref{tab:photo-planet-posteriors}) to predict the expected mass and semi-amplitude of the two planets. \citet{OBH-MR} present a mass-radius relation that is dependent upon the density of the planet ($\rho_{\text{P}}$):
\begin{equation}
\text{M}_\text{P} = 
\begin{cases}
(0.90 \pm 0.06) {\text{R}_\text{P}}^{3.45 \pm 0.12}, & \text{if } \rho_\text{P} > 3.3 gcm^{-3}\\
(1.74 \pm 0.38) {\text{R}_\text{P}}^{1.58 \pm 0.10}, & \text{if } \rho_\text{P} < 3.3 gcm^{-3}
\end{cases}
\label{eqn:OBH-MR}
\end{equation}
The high density case is applicable when the planet has a rocky composition and the low density case is for when the planet has a volatile-rich composition. Assuming a volatile-rich composition, the inner and outer planet would have masses of \photomassOBHvolatileone M$_\oplus$ and \photomassOBHvolatiletwo M$_\oplus$, respectively, leading to semi-amplitudes of \photoKOBHvolatileone ms$^{-1}$ and \photoKOBHvolatiletwo ms$^{-1}$. 

The predicted semi-amplitudes of HD\,15906\,b and c are greater than the average HARPS RV uncertainty ($\sim$ 1.5 ms$^{-1}$). This means that the planetary signals should be detectable with sufficient observations from a high-resolution spectrograph. Since HD\,15906 is visible from both hemispheres, there are many instruments that would be capable of doing this. We note that there is a relatively large scatter in the current RV data ($\sim$ 10 ms$^{-1}$) which could make a precise mass measurement challenging. It will require a high sampling rate and a large number of RV observations to adequately model the planetary and stellar signals.

\subsubsection{Atmospheric Characterisation}
\begin{figure}
\begin{center}
\includegraphics[width=\columnwidth]{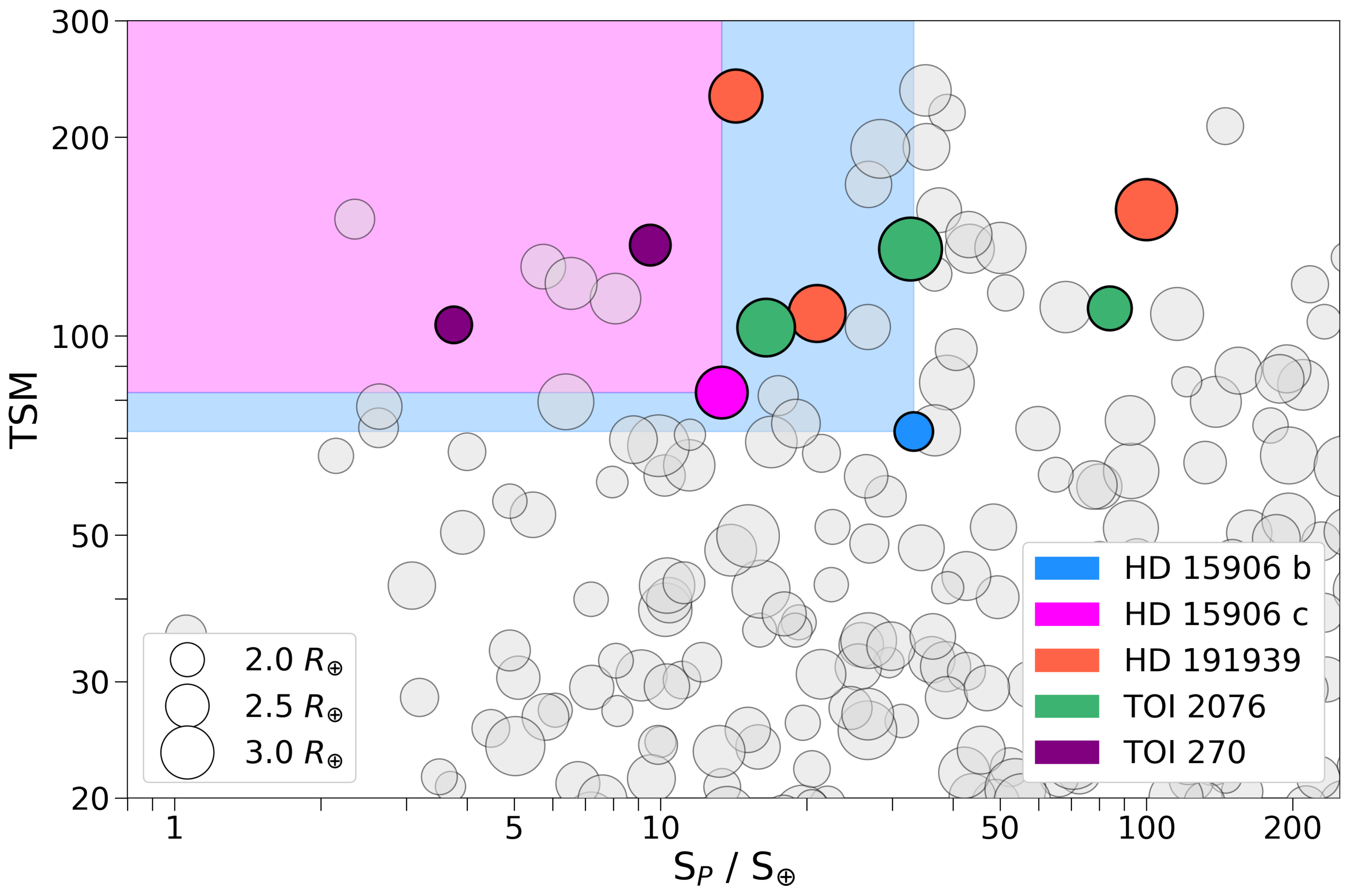}
\caption{Transmission spectroscopy metric (TSM) as a function of stellar irradiation for all confirmed sub-Neptune sized planets (1.75 $<$ R$_{\text{P}}$ / R$_\oplus$ $<$ 3.5). The marker size scales with planet radius and HD\,15906\,b (blue) and c (pink) are plotted alongside a shaded region to show planets with a lower irradiation and higher TSM. HD\,15906 is one of 4 multiplanet systems with 2 sub-Neptune sized planets with irradiation lower than 35 S$_\oplus$ and TSM higher than 70. The other 3 systems are highlighted in red (HD\,191939), green (TOI\,2076) and purple (TOI\,270).}
\label{fig:tsm-plot}
\end{center}
\end{figure}

Theory predicts a wide variety of possible chemical compositions for sub-Neptunes \citep[e.g.,][]{Moses2013,2022MNRAS.513.4015G}. Atmospheric characterisation can constrain their composition and, thanks to their bright host star, HD\,15906\,b and c are amenable to such observations. The Transmission Spectroscopy Metric \citep[TSM;][]{KemptonTSM} can be used to rank transiting planets based on their suitability for transmission spectroscopy. It quantifies the expected SNR of the spectral features for a 10 hour observation with \textit{JWST}/NIRISS, assuming a cloud-free atmosphere. Using the stellar parameters from Table \ref{tab:stellar-params} and planet parameters from Table \ref{tab:photo-planet-posteriors}, we find that HD\,15906\,b and c have TSM values of 71.7 and 82.1, respectively. This puts them in the top 3\% of all confirmed transiting planets smaller than 4 R$_\oplus$, where in the absence of a measured mass we computed the expected mass according to the empirical mass-radius relation used by \citet{KemptonTSM}. Furthermore, HD\,15906\,b and c have amongst the highest TSMs for such small and lowly irradiated planets, illustrated in Figure \ref{fig:tsm-plot}. There are only 6 sub-Neptune sized planets (1.75 $<$ R$_{\text{P}}$ / R$_\oplus$ $<$ 3.5) with a higher TSM and lower irradiation than HD\,15906\,c, of which HD\,15906 is the second brightest host star. In addition, there are only 3 other multiplanet systems which host 2 sub-Neptune sized planets with irradiation lower than 35 S$_\oplus$ and TSM higher than 70 \citep[HD\,191939, TOI\,2076 and TOI\,270;][]{2023Orell,TOI-2076,TOI-270}. The HD\,15906 system is therefore an interesting target for comparative studies of warm sub-Neptune composition.

\section{Conclusions}\label{sec:conclusion}
In this paper, we have reported the discovery and validation of two warm sub-Neptune planets transiting the bright (G = 9.5 mag) K-dwarf HD\,15906 (TOI\,461, TIC\,4646810). During \textit{TESS} sectors 4 and 31, four transits of the inner planet, HD\,15906\,b, were observed, but there were only two transits of the outer planet, HD\,15906\,c, separated by $\sim$ 734 days. The period of the outer planet was ambiguous, with 36 possible values, and we used \textit{CHEOPS} follow-up to determine the true period. Using \textit{TESS}, \textit{CHEOPS} and LCOGT photometry, we precisely characterised the two planets -- HD\,15906\,b and c have periods of \photoperiodoneshort\ days and \photoperiodtwoshort\ days and radii of \photoradoneshort\ R$_\oplus$ and \photoradtwoshort\ R$_\oplus$, respectively. We found marginal evidence for TTVs in the system and, comparing the observations to simulations, we showed that more observations are required to understand the nature of the TTV signals. Both planets are in the warm regime, with insolation fluxes of \photoinsoloneshort\ S$_\oplus$ and \photoinsoltwoshort\ S$_\oplus$ and equilibrium temperatures of \photoeqmToneshort\ K and \photoeqmTtwoshort\ K. We find that HD\,15906\,c is one of the most lowly irradiated sub-Neptune sized planets transiting such a bright star. 

Both HD\,15906\,b and c are prime targets for future detailed characterisation studies. They are amenable to precise mass measurement and they are amongst the top warm sub-Neptune candidates for atmospheric characterisation with \textit{JWST}. These studies will allow us to constrain the compositions of HD\,15906\,b and c, test planet formation and evolution models and improve our limited understanding of sub-Neptune planets as a whole. 

\section*{Acknowledgements}
We would like to thank the anonymous referee for their helpful comments and suggestions.

This paper includes data collected by the \textit{TESS} mission that are publicly available from the Mikulski Archive for Space Telescopes (MAST). Funding for the \textit{TESS} mission is provided by NASA's Science Mission Directorate. We acknowledge the use of public \textit{TESS} data from pipelines at the \textit{TESS} Science Office and at the \textit{TESS} Science Processing Operations Center. Resources supporting this work were provided by the NASA High-End Computing (HEC) Program through the NASA Advanced Supercomputing (NAS) Division at Ames Research Center for the production of the SPOC data products. This research has also made use of the Exoplanet Follow-up Observation Program website, which is operated by the California Institute of Technology, under contract with the National Aeronautics and Space Administration under the Exoplanet Exploration Program.

\textit{CHEOPS} is an ESA mission in partnership with Switzerland with important contributions to the payload and the ground segment from Austria, Belgium, France, Germany, Hungary, Italy, Portugal, Spain, Sweden, and the United Kingdom. The \textit{CHEOPS} Consortium would like to gratefully acknowledge the support received by all the agencies, offices, universities, and industries involved. Their flexibility and willingness to explore new approaches were essential to the success of this mission. \textit{CHEOPS} data analysed in this article will be made available in the \textit{CHEOPS} mission archive (\url{https://cheops.unige.ch/archive_browser/}).

This work makes use of observations from the LCOGT network. Part of the LCOGT telescope time was granted by NOIRLab through the Mid-Scale Innovations Program (MSIP). MSIP is funded by NSF. This research has made use of the Exoplanet Follow-up Observation Program (ExoFOP; DOI: 10.26134/ExoFOP5) website, which is operated by the California Institute of Technology, under contract with the National Aeronautics and Space Administration under the Exoplanet Exploration Program.

This paper includes observations made with the ESO 3.6\,m telescope at La Silla Observatory under programmes 072.C-0488, 183.C-0972, 089.C-0732, 090.C-0421, 096.C-0460, and 0100.C-0097.

This paper uses observations made with the Nordic Optical Telescope, owned in collaboration by the University of Turku and Aarhus University, and operated jointly by Aarhus University, the University of Turku and the University of Oslo, representing Denmark, Finland and Norway, the University of Iceland and Stockholm University at the Observatorio del Roque de los Muchachos, La Palma, Spain, of the Instituto de Astrofisica de Canarias.

This work has made use of data from the European Space Agency (ESA) mission \textit{Gaia} (\url{https://www.cosmos.esa.int/gaia}), processed by the \textit{Gaia} Data Processing and Analysis Consortium (DPAC, \url{https://www.cosmos.esa.int/web/gaia/dpac/consortium}). Funding for the DPAC has been provided by national institutions, in particular the institutions participating in the \textit{Gaia} Multilateral Agreement.

This paper utilizes data collected at Lick Observatory and analyzed using funds provided by the NASA Exoplanets Research Program (XRP) through grant 80NSSC20K0250 (PI: Courtney D. Dressing).

Some of the observations in the paper made use of the High-Resolution Imaging instrument ‘Alopeke obtained under Gemini LLP Proposal Number: GN/S-2021A-LP-105. ‘Alopeke was funded by the NASA Exoplanet Exploration Program and built at the NASA Ames Research Center by Steve B. Howell, Nic Scott, Elliott P. Horch, and Emmett Quigley. Alopeke was mounted on the Gemini North (and/or South) telescope of the international Gemini Observatory, a program of NSF's OIR Lab, which is managed by the Association of Universities for Research in Astronomy (AURA) under a cooperative agreement with the National Science Foundation. on behalf of the Gemini partnership: the National Science Foundation (United States), National Research Council (Canada), Agencia Nacional de Investigación y Desarrollo (Chile), Ministerio de Ciencia, Tecnología e Innovación (Argentina), Ministério da Ciência, Tecnologia, Inovações e Comunicações (Brazil), and Korea Astronomy and Space Science Institute (Republic of Korea).

Some of the data presented herein were obtained at the W. M. Keck Observatory, which is operated as a scientific partnership among the California Institute of Technology, the University of California and the National Aeronautics and Space Administration. The Observatory was made possible by the generous financial support of the W. M. Keck Foundation. The authors wish to recognize and acknowledge the very significant cultural role and reverence that the summit of Maunakea has always had within the indigenous Hawaiian community. We are most fortunate to have the opportunity to conduct observations from this mountain.

The Digitized Sky Surveys were produced at the Space Telescope Science Institute under U.S. Government grant NAG W-2166. The images of these surveys are based on photographic data obtained using the Oschin Schmidt Telescope on Palomar Mountain and the UK Schmidt Telescope. The plates were processed into the present compressed digital form with the permission of these institutions. The National Geographic Society - Palomar Observatory Sky Atlas (POSS-I) was made by the California Institute of Technology with grants from the National Geographic Society. The Oschin Schmidt Telescope is operated by the California Institute of Technology and Palomar Observatory. The UK Schmidt Telescope was operated by the Royal Observatory Edinburgh, with funding from the UK Science and Engineering Research Council (later the UK Particle Physics and Astronomy Research Council), until 1988 June, and thereafter by the Anglo-Australian Observatory. The blue plates of the southern Sky Atlas and its Equatorial Extension (together known as the SERC-J), as well as the Equatorial Red (ER), and the Second Epoch [red] Survey (SES) were all taken with the UK Schmidt.

The Pan-STARRS1 Surveys (PS1) and the PS1 public science archive have been made possible through contributions by the Institute for Astronomy, the University of Hawaii, the Pan-STARRS Project Office, the Max-Planck Society and its participating institutes, the Max Planck Institute for Astronomy, Heidelberg and the Max Planck Institute for Extraterrestrial Physics, Garching, The Johns Hopkins University, Durham University, the University of Edinburgh, the Queen's University Belfast, the Harvard-Smithsonian Center for Astrophysics, the Las Cumbres Observatory Global Telescope Network Incorporated, the National Central University of Taiwan, the Space Telescope Science Institute, the National Aeronautics and Space Administration under Grant No. NNX08AR22G issued through the Planetary Science Division of the NASA Science Mission Directorate, the National Science Foundation Grant No. AST-1238877, the University of Maryland, Eotvos Lorand University (ELTE), the Los Alamos National Laboratory, and the Gordon and Betty Moore Foundation

AT thanks the Science and Technology Facilities Council (STFC) for a PhD studentship. 
This work was also partially supported by a grant from the Simons Foundation (PI Queloz, grant number 327127). 
This work has been carried out within the framework of the NCCR PlanetS supported by the Swiss National Science Foundation under grants 51NF40\_182901 and 51NF40\_205606. 
ACC and TW acknowledge support from STFC consolidated grant numbers ST/R000824/1 and ST/V000861/1, and UKSA grant number ST/R003203/1. 
ML acknowledges support of the Swiss National Science Foundation under grant number PCEFP2\_194576. 
ABr was supported by the SNSA. 
DRC acknowledges partial support from the National Aeronautics and Space Administration through the Exoplanet Research Program grant 18-2XRP18\_2-0007. 
KAC and DWL acknowledge support from the \textit{TESS} mission via subaward s3449 from MIT. 
DG and LMS gratefully acknowledge financial support from the CRT foundation under Grant No. 2018.2323 ``Gaseous or rocky? Unveiling the nature of small worlds''. 
ZG acknowledges the support of the Hungarian National Research, Development and Innovation Office (NKFIH) grant K-125015, the PRODEX Experiment Agreement No. 4000137122 between the ELTE Eötvös Loránd University and the European Space Agency (ESA-D/SCI-LE-2021-0025), the VEGA grant of the Slovak Academy of Sciences No. 2/0031/22, the Slovak Research and Development Agency contract No. APVV-20-0148, and the support of the city of Szombathely. 
SGS acknowledges support from FCT through FCT contract nr. CEECIND/00826/2018 and POPH/FSE (EC). 
YAl acknowledges the support of the Swiss National Fund under grant 200020\_172746. 
We acknowledge support from the Spanish Ministry of Science and Innovation and the European Regional Development Fund through grants ESP2016-80435-C2-1-R, ESP2016-80435-C2-2-R, PGC2018-098153-B-C33, PGC2018-098153-B-C31, ESP2017-87676-C5-1-R, MDM-2017-0737 Unidad de Excelencia Maria de Maeztu-Centro de Astrobiología (INTA-CSIC), as well as the support of the Generalitat de Catalunya/CERCA programme. The MOC activities have been supported by the ESA contract No. 4000124370. 
SCCB acknowledges support from FCT through FCT contracts nr. IF/01312/2014/CP1215/CT0004. 
XB, SCh, DG, MF and JL acknowledge their role as ESA-appointed \textit{CHEOPS} science team members. 
LBo, VNa, IPa, GPi, RRa and GSc acknowledge support from \textit{CHEOPS} ASI-INAF agreement n. 2019-29-HH.0. 
This project was supported by the CNES. 
The Belgian participation to \textit{CHEOPS} has been supported by the Belgian Federal Science Policy Office (BELSPO) in the framework of the PRODEX Program, and by the University of Liège through an ARC grant for Concerted Research Actions financed by the Wallonia-Brussels Federation. 
LD is an F.R.S.-FNRS Postdoctoral Researcher. 
This work was supported by FCT - Fundação para a Ciência e a Tecnologia through national funds and by FEDER through COMPETE2020 - Programa Operacional Competitividade e Internacionalizacão by these grants: UID/FIS/04434/2019, UIDB/04434/2020, UIDP/04434/2020, PTDC/FIS-AST/32113/2017 \& POCI-01-0145-FEDER- 032113, PTDC/FIS-AST/28953/2017 \& POCI-01-0145-FEDER-028953, PTDC/FIS-AST/28987/2017 \& POCI-01-0145-FEDER-028987, ODSD is supported in the form of work contract (DL 57/2016/CP1364/CT0004) funded by national funds through FCT. 
B-OD acknowledges support from the Swiss State Secretariat for Education, Research and Innovation (SERI) under contract number MB22.00046. 
DD acknowledges support from the \textit{TESS} Guest Investigator Program grants 80NSSC21K0108 and 80NSSC22K0185. 
This project has received funding from the European Research Council (ERC) under the European Union’s Horizon 2020 research and innovation programme (project {\sc Four Aces}. 
grant agreement No 724427). It has also been carried out in the frame of the National Centre for Competence in Research PlanetS supported by the Swiss National Science Foundation (SNSF). DE acknowledges financial support from the Swiss National Science Foundation for project 200021\_200726. 
MF and CMP gratefully acknowledge the support of the Swedish National Space Agency (DNR 65/19, 174/18). 
MGi is an F.R.S.-FNRS Senior Research Associate. 
SeHo gratefully acknowledges CNES funding through the grant 837319. 
KGI and MNG are the ESA \textit{CHEOPS} Project Scientists and are responsible for the ESA \textit{CHEOPS} Guest Observers Programme. They do not participate in, or contribute to, the definition of the Guaranteed Time Programme of the \textit{CHEOPS} mission through which observations described in this paper have been taken, nor to any aspect of target selection for the programme. 
This work was granted access to the HPC resources of MesoPSL financed by the Region Ile de France and the project Equip@Meso (reference ANR-10-EQPX-29-01) of the programme Investissements d'Avenir supervised by the Agence Nationale pour la Recherche. 
RL acknowledges funding from University of La Laguna through the Margarita Salas Fellowship from the Spanish Ministry of Universities ref. UNI/551/2021-May 26, and under the EU Next Generation funds. 
PFLM acknowledges support from STFC research grant number ST/M001040/1. 
IR acknowledges support from the Spanish Ministry of Science and Innovation and the European Regional Development Fund through grant PGC2018-098153-B- C33, as well as the support of the Generalitat de Catalunya/CERCA programme. 
NCS acknowledges funding by the European Union (ERC, FIERCE, 101052347). Views and opinions expressed are however those of the author(s) only and do not necessarily reflect those of the European Union or the European Research Council. Neither the European Union nor the granting authority can be held responsible for them. 
GyMSz acknowledges the support of the Hungarian National Research, Development and Innovation Office (NKFIH) grant K-125015, a a PRODEX Experiment Agreement No. 4000137122, the Lend\"ulet LP2018-7/2021 grant of the Hungarian Academy of Science and the support of the city of Szombathely. 
VVG is an F.R.S-FNRS Research Associate. 
NAW acknowledges UKSA grant ST/R004838/1. 

\section*{Data Availability}
The \textit{TESS} data used in this paper is publicly available on the MAST (\url{https://mast.stsci.edu/portal/Mashup/Clients/Mast/Portal.html}). The \textit{CHEOPS} DRP data products are publicly available on the \textit{CHEOPS} mission archive (\url{https://cheops.unige.ch/archive_browser/}). The raw and detrended \textit{CHEOPS} PIPE lightcurves and the WASP photometry have been made available online at the CDS (\url{https://cdsarc.cds.unistra.fr/viz-bin/cat/J/MNRAS/523/3090}). The LCOGT data and high-resolution imaging are publicly available on ExoFOP (\url{https://exofop.ipac.caltech.edu/tess/}). The HARPS spectra are publicly available on the ESO Science Archive (\url{http://archive.eso.org/cms.html}) and the HARPS RVs were taken from \citet{Trifonov-HARPS-RVs}. The HARPS and FIES RVs used in our analysis are presented in Table \ref{tab:rv-observations}. The archival imaging is publicly available at the Digitized Sky Survey (\url{https://archive.stsci.edu/cgi-bin/dss_form}) and the MAST. 




\bibliographystyle{mnras}
\bibliography{bibliography} 



\appendix

\section*{Affiliations}\label{sec:affiliations}
$^{1}$Astrophysics Group, Cavendish Laboratory, University of Cambridge, JJ Thomson Avenue, Cambridge CB3 0HE, UK\\
$^{2}$ETH Zurich, Department of Physics, Wolfgang-Pauli-Strasse 2, CH-8093 Zurich, Switzerland\\
$^{3}$Center for Space and Habitability, University of Bern, Gesellschaftsstrasse 6, 3012 Bern, Switzerland\\
$^{4}$Department of Physics and Kavli Institute for Astrophysics and Space Research, Massachusetts Institute of Technology, Cambridge, MA 02139, USA\\
$^{5}$Centre for Exoplanet Science, SUPA School of Physics and Astronomy, University of St Andrews, North Haugh, St Andrews KY16 9SS, UK\\
$^{6}$Physikalisches Institut, University of Bern, Sidlerstrasse 5, 3012 Bern, Switzerland\\
$^{7}$Observatoire Astronomique de l'Université de Genève, Chemin Pegasi 51, CH-1290 Versoix, Switzerland\\
$^{8}$Department of Astronomy, Stockholm University, AlbaNova University Center, 10691 Stockholm, Sweden\\
$^{9}$Space Research Institute, Austrian Academy of Sciences, Schmiedlstrasse 6, A-8042 Graz, Austria\\
$^{10}$DTU Space, National Space Institute, Technical University of Denmark, Elektrovej 328, DK-2800 Kgs. Lyngby, Denmark\\
$^{11}$NASA Exoplanet Science Institute, Caltech/IPAC, Pasadena, CA 91125, USA\\
$^{12}$Center for Astrophysics \textbar \ Harvard \& Smithsonian, 60 Garden Street, Cambridge, MA 02138, USA\\
$^{13}$Dipartimento di Fisica, Universita degli Studi di Torino, via Pietro Giuria 1, I-10125, Torino, Italy\\
$^{14}$Astronomical Institute, Slovak Academy of Science, Stellar Department, Tatranská Lomnica, 05960 Vysoké Tatry, Slovakia\\
$^{15}$ELTE E\"otv\"os Lor\'and University, Gothard Astrophysical Observatory, 9700 Szombathely, Szent Imre h. u. 112, Hungary\\
$^{16}$MTA-ELTE Exoplanet Research Group, 9700 Szombathely, Szent Imre h. u. 112, Hungary\\
$^{17}$Department of Astronomy, University of California Berkeley, Berkeley, CA 94720, USA\\
$^{18}$Instituto de Astrofisica e Ciencias do Espaco, Universidade do Porto, CAUP, Rua das Estrelas, 4150-762 Porto, Portugal\\
$^{19}$NASA Ames Research Center, Moffett Field, CA 94035, USA\\
$^{20}$Department of Space, Earth and Environment, Chalmers University of Technology, Onsala Space Observatory, 439 92 Onsala, Sweden\\
$^{21}$Department of Physics, Engineering and Astronomy, Stephen F. Austin State University, 1936 North St, Nacogdoches, TX 75962, USA\\
$^{22}$Instituto de Astrofisica de Canarias, 38200 La Laguna, Tenerife, Spain\\
$^{23}$Departamento de Astrofisica, Universidad de La Laguna, 38206 La Laguna, Tenerife, Spain\\
$^{24}$Institut de Ciencies de l'Espai (ICE, CSIC), Campus UAB, Can Magrans s/n, 08193 Bellaterra, Spain\\
$^{25}$Institut d'Estudis Espacials de Catalunya (IEEC), 08034 Barcelona, Spain\\
$^{26}$Admatis, 5. Kandó Kálmán Street, 3534 Miskolc, Hungary\\
$^{27}$Depto. de Astrofisica, Centro de Astrobiologia (CSIC-INTA), ESAC campus, 28692 Villanueva de la Cañada (Madrid), Spain\\
$^{28}$Departamento de Fisica e Astronomia, Faculdade de Ciencias, Universidade do Porto, Rua do Campo Alegre, 4169-007 Porto, Portugal\\
$^{29}$Université Grenoble Alpes, CNRS, IPAG, 38000 Grenoble, France\\
$^{30}$INAF, Osservatorio Astronomico di Padova, Vicolo dell'Osservatorio 5, 35122 Padova, Italy\\
$^{31}$Institute of Planetary Research, German Aerospace Center (DLR), Rutherfordstrasse 2, 12489 Berlin, Germany\\
$^{32}$Université de Paris, Institut de physique du globe de Paris, CNRS, F-75005 Paris, France\\
$^{33}$American Association of Variable Star Observers, 49 Bay State Road, Cambridge, MA 02138, USA\\
$^{34}$INAF, Osservatorio Astrofisico di Torino, Via Osservatorio, 20, I-10025 Pino Torinese To, Italy\\
$^{35}$Centre for Mathematical Sciences, Lund University, Box 118, 221 00 Lund, Sweden\\
$^{36}$Aix Marseille Univ, CNRS, CNES, LAM, 38 rue Frédéric Joliot-Curie, 13388 Marseille, France\\
$^{37}$Astrobiology Research Unit, Université de Liège, Allée du 6 Août 19C, B-4000 Liège, Belgium\\
$^{38}$Space sciences, Technologies and Astrophysics Research (STAR) Institute, Université de Liège, Allée du 6 Août 19C, 4000 Liège, Belgium\\
$^{39}$Department of Physics and Astronomy, University of New Mexico, 210 Yale Blvd NE, Albuquerque, NM 87106, USA\\
$^{40}$Centre Vie dans l'Univers, Facult\'e des sciences, Universit\'e de Gen\`eve, Quai Ernest-Ansermet 30, CH-1211 Gen\`eve 4, Switzerland\\
$^{41}$Department of Earth, Atmospheric and Planetary Sciences, Massachusetts Institute of Technology, Cambridge, MA 02139, USA\\
$^{42}$Leiden Observatory, University of Leiden, PO Box 9513, 2300 RA Leiden, The Netherlands\\
$^{43}$Department of Astronomy and Astrophysics, University of California, Santa Cruz, CA 95064, USA\\
$^{44}$Department of Astrophysics, University of Vienna, Tuerkenschanzstrasse 17, 1180 Vienna, Austria\\
$^{45}$European Space Agency (ESA), European Space Research and Technology Centre (ESTEC), Keplerlaan 1, 2201 AZ Noordwijk, The Netherlands\\
$^{46}$Science and Operations Department - Science Division (SCI-SC), Directorate of Science, European Space Agency (ESA), European Space Research and Technology Centre (ESTEC), Keplerlaan 1, 2201 AZ Noordwijk, The Netherlands\\
$^{47}$Konkoly Observatory, Research Centre for Astronomy and Earth Sciences, 1121 Budapest, Konkoly Thege Miklós út 15-17, Hungary\\
$^{48}$ELTE E\"otv\"os Lor\'and University, Institute of Physics, P\'azm\'any P\'eter s\'et\'any 1/A, 1117 Budapest, Hungary\\
$^{49}$IMCCE, UMR8028 CNRS, Observatoire de Paris, PSL Univ., Sorbonne Univ., 77 av. Denfert-Rochereau, 75014 Paris, France\\
$^{50}$Department of Physics and Astronomy, The University of North Carolina at Chapel Hill, Chapel Hill, NC 27599-3255, USA\\
$^{51}$Institut d'astrophysique de Paris, UMR7095 CNRS, Université Pierre \& Marie Curie, 98bis blvd. Arago, 75014 Paris, France\\
$^{52}$European Southern Observatory, Av. Alonso de Cordova 3107, Casilla 19001, Santiago de Chile, Chile\\
$^{53}$Department of Astronomy \& Astrophysics, University of Chicago, Chicago, IL 60637, USA\\
$^{54}$Astrophysics Group, Keele University, Staffordshire, ST5 5BG, UK\\
$^{55}$Proto-Logic LLC, 1718 Euclid Street NW, Washington, DC 20009, USA\\
$^{56}$School of Physics \& Astronomy, University of Birmingham, Edgbaston, Birmingham B15 2TT, UK\\
$^{57}$INAF, Osservatorio Astrofisico di Catania, Via S. Sofia 78, 95123 Catania, Italy\\
$^{58}$Institute of Optical Sensor Systems, German Aerospace Center (DLR), Rutherfordstrasse 2, 12489 Berlin, Germany\\
$^{59}$Dipartimento di Fisica e Astronomia "Galileo Galilei", Universita degli Studi di Padova, Vicolo dell'Osservatorio 3, 35122 Padova, Italy\\
$^{60}$Department of Physics, University of Warwick, Gibbet Hill Road, Coventry CV4 7AL, UK\\
$^{61}$NASA Goddard Space Flight Center, 8800 Greenbelt Road, Greenbelt, MD 20771, USA\\
$^{62}$Zentrum für Astronomie und Astrophysik, Technische Universität Berlin, Hardenbergstr. 36, D-10623 Berlin, Germany\\
$^{63}$Institut für Geologische Wissenschaften, Freie Universität Berlin, 12249 Berlin, Germany\\
$^{64}$Department of Astronomy, University of Maryland, College Park, College Park, MD 20742 USA\\
$^{65}$Department of Aeronautics and Astronautics, Massachusetts Institute of Technology, Cambridge, MA 02139, USA\\
$^{66}$Hazelwood Observatory, Australia\\
$^{67}$Institute of Astronomy, University of Cambridge, Madingley Road, Cambridge, CB3 0HA, UK\\
$^{68}$Department of Astrophysical Sciences, Princeton University, 4 Ivy Lane, Princeton, NJ 08544, USA\\
$^{69}$SETI Institute, Mountain View, CA 94043, USA

\section{Priors and Posteriors of the Global Photometric Analysis}\label{sec:photo-appendix}
\begin{figure*}
\begin{center}
\includegraphics[width=\textwidth]{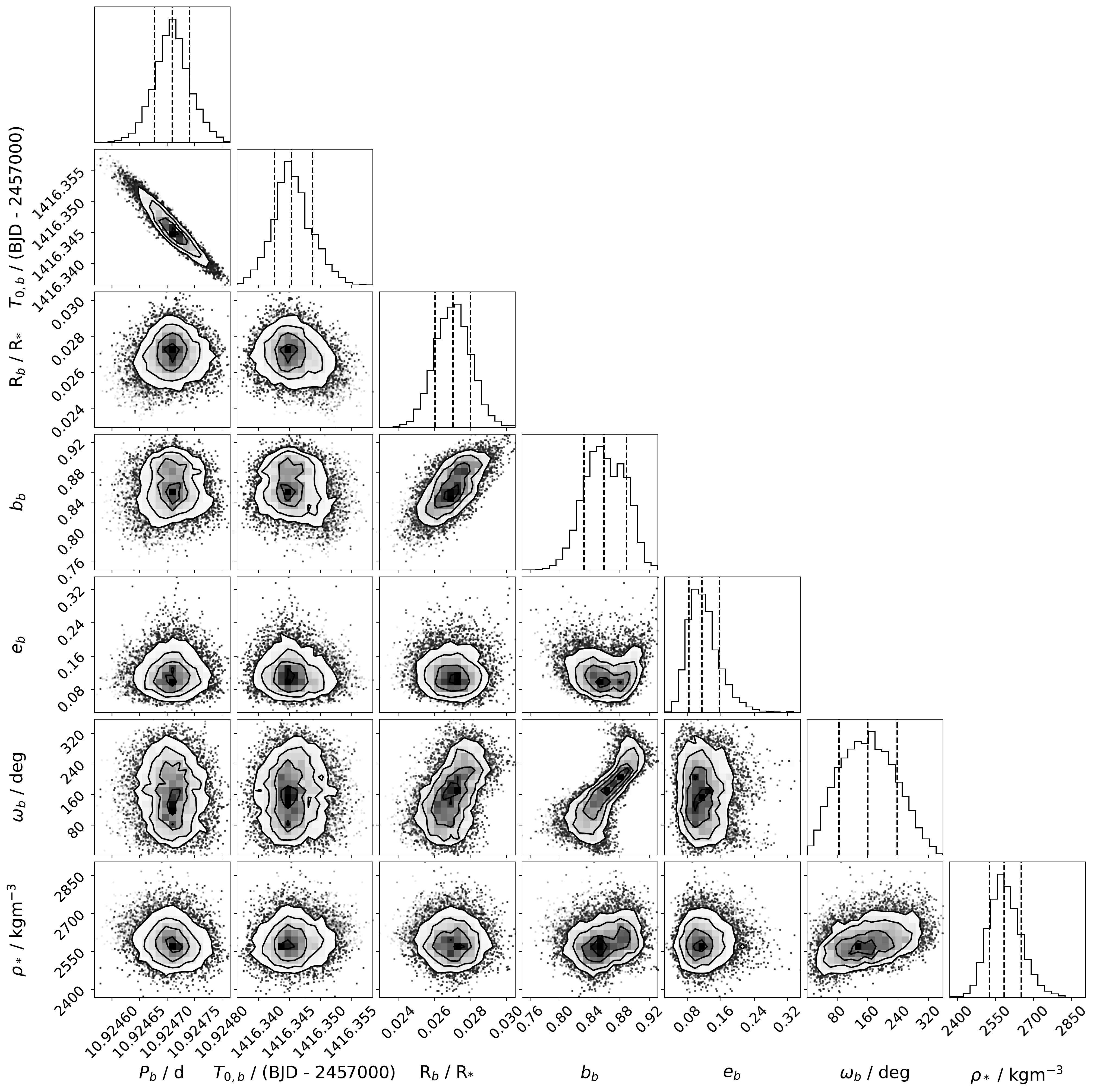}
\caption{Results of the global photometric fit. This corner plot shows the posterior distributions of the fitted planet parameters for HD\,15906\,b.}
\label{fig:photo-corner-plot-inner}
\end{center}
\end{figure*}

\begin{figure*}
\begin{center}
\includegraphics[width=\textwidth]{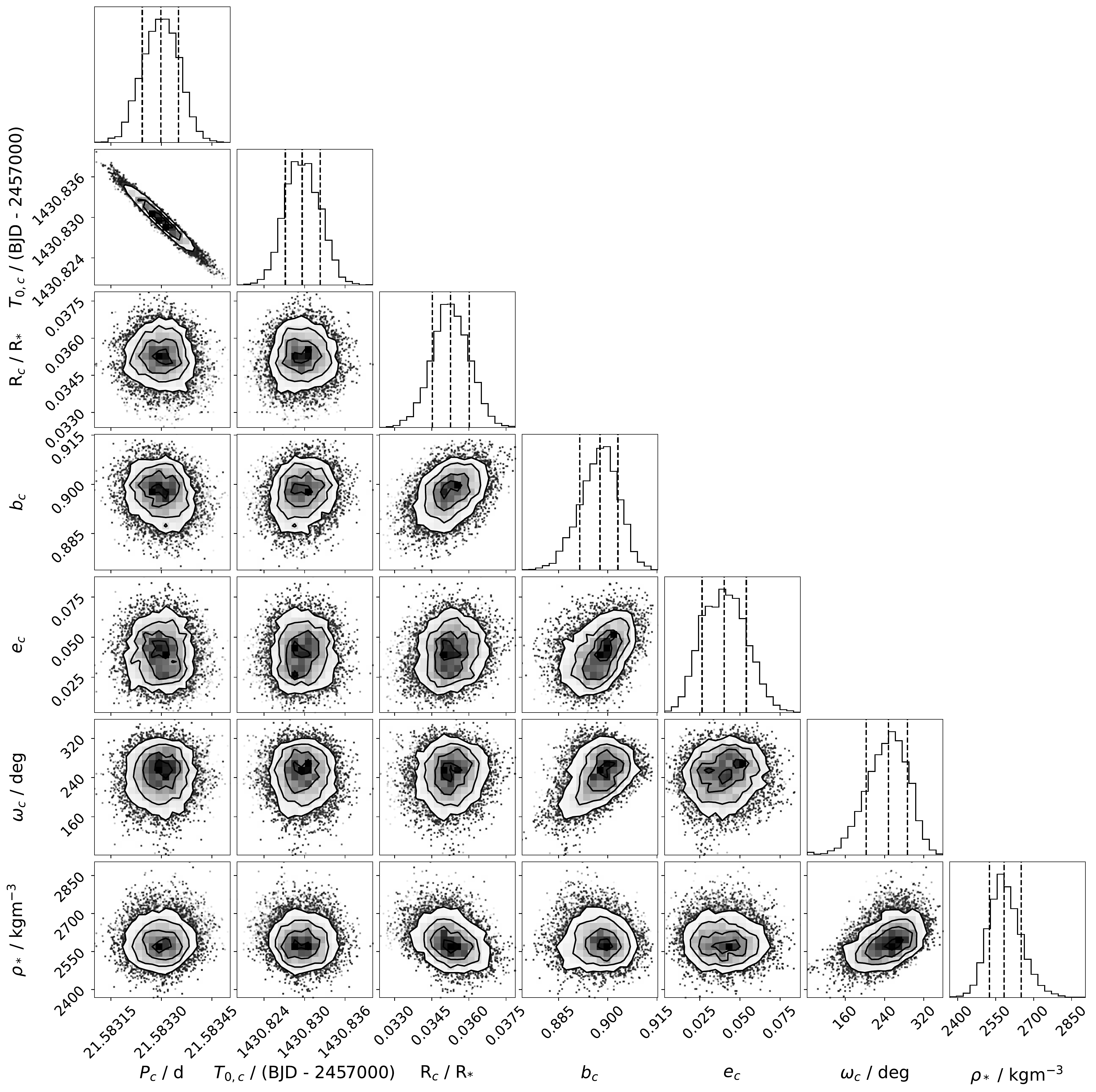}
\caption{Results of the global photometric fit. This corner plot shows the posterior distributions of the fitted planet parameters for HD\,15906\,c.}
\label{fig:photo-corner-plot-outer}
\end{center}
\end{figure*}

In Table \ref{tab:photo-priors+posteriors}, we present the priors and posterior values from the global photometric analysis as described in Section \ref{sec:photo-analysis}. In Figures \ref{fig:photo-corner-plot-inner} and \ref{fig:photo-corner-plot-outer}, we present the corner plots of the fitted planet parameters for each planet, made using \texttt{corner} \citep{cornerplot}. 

\begin{table*}
\centering
  \caption{This table presents a full list of the fitted parameters from our global photometric model, described in Section \ref{sec:photo-analysis}. We include both the prior and the posterior value of each parameter. Uniform priors are represented by $\mathcal{U}$(a,b) and log-uniform priors are written as ln$\mathcal{U}$(a,b), where a and b are the lower and upper bounds, respectively. The notation used for normal priors is $\mathcal{N}(\mu,\sigma)$, where $\mu$ and $\sigma$ are the mean and standard deviation of the distribution. Truncated normal priors are defined as $\mathcal{N}_{\mathcal{U}}$($\mu$,$\sigma$,a,b), where $\mu$ and $\sigma$ are the mean and standard deviation of the distribution and a and b are the lower and upper bounds, respectively. The limb darkening parameters were defined per instrument (\textit{TESS}, \textit{CHEOPS} and LCOGT), however the detrending was done independently for each observation: \textit{TESS}-1 is the sector 4 data, \textit{TESS}-2 is the sector 31 data, \textit{CHEOPS}-1 to -6 are \textit{CHEOPS} visits 1 to 6 and LCOGT-1 and -2 are LCOGT visits 1 and 2. The posterior values are defined by the median, 16th and 84th percentiles of the posterior distribution.}
  {\begin{tabular}{>{\centering\arraybackslash}m{6cm}>{\centering\arraybackslash}m{4cm}>{\centering\arraybackslash}m{4cm}}
    \hline
    Parameter & Prior & Posterior Value \\
    \hline
    \multicolumn{3}{c}{HD\,15906\,b} \\
    \hline
    Period ($P_{\text{b}}$) / days & $\mathcal{U}$(10.92,10.93) & \photoperiodone \vspace{1mm}\\
    Mid-transit time ($T_{0\text{,b}}$) / (BJD - 2457000) & $\mathcal{U}$(1416.3,1416.4) & \photoepochone \vspace{1mm}\\
    Radius ratio (R$_{\text{b}}$ / R$_{\star}$) & $\mathcal{U}$(0.0,0.1) & \photopone \vspace{1mm}\\
    Impact parameter ($b_{\text{b}}$) & $\mathcal{U}$(0.0,1.2) & \photobone \vspace{1mm}\\
    Eccentricity ($e_{\text{b}}$) & $\mathcal{N}_{\mathcal{U}}$(0.0,0.083,0.0,1.0) & \photoeccone \vspace{1mm}\\
    Argument of periastron ($\omega_{\text{b}}$) / deg & $\mathcal{U}$(0.0,360.0) & \photoomegaone \vspace{1mm}\\
    \hline
    \multicolumn{3}{c}{HD\,15906\,c} \\
    \hline
    Period ($P_{\text{c}}$) / days & $\mathcal{U}$(21.58,21.59) & \photoperiodtwo \vspace{1mm}\\
    Mid-transit time ($T_{0\text{,c}}$) / (BJD - 2457000) & $\mathcal{U}$(1430.8,1430.9) & \photoepochtwo \vspace{1mm}\\
    Radius ratio (R$_{\text{c}}$ / R$_{\star}$) & $\mathcal{U}$(0.0,0.1) & \photoptwo \\
    Impact parameter ($b_{\text{c}}$) & $\mathcal{U}$(0.0,1.2) & \photobtwo \vspace{1mm}\\
    Eccentricity ($e_{\text{c}}$) & $\mathcal{N}_{\mathcal{U}}$(0.0,0.083,0.0,1.0) & \photoecctwo \vspace{1mm}\\
    Argument of periastron ($\omega_{\text{c}}$) / deg & $\mathcal{U}$(0.0, 360.0) & \photoomegatwo \vspace{1mm}\\
    \hline
    \multicolumn{3}{c}{Stellar} \\
    \hline
    Stellar density ($\rho_{\star}$) / kgm$^{-3}$ & $\mathcal{N}$(2517.61,101.94) & \photorho \vspace{1mm}\\
    Quadratic LD \textit{TESS} ($q_{\text{1,\textit{TESS}}}$) & $\mathcal{N}_{\mathcal{U}}$(0.4207,0.1,0.0,1.0) & \photoqoneTESS \vspace{1mm}\\
    Quadratic LD \textit{TESS} ($q_{\text{2,\textit{TESS}}}$) & $\mathcal{N}_{\mathcal{U}}$(0.3659,0.1,0.0,1.0) & \photoqtwoTESS \vspace{1mm}\\
    Quadratic LD \textit{CHEOPS} ($q_{\text{1,\textit{CHEOPS}}}$) & $\mathcal{N}_{\mathcal{U}}$(0.5375,0.1,0.0,1.0) & \photoqoneCHEOPS \vspace{1mm}\\
    Quadratic LD \textit{CHEOPS} ($q_{\text{2,\textit{CHEOPS}}}$) & $\mathcal{N}_{\mathcal{U}}$(0.4351,0.1,0.0,1.0) & \photoqtwoCHEOPS \vspace{1mm}\\
    Quadratic LD LCOGT ($q_{\text{1,LCO}}$) & $\mathcal{N}_{\mathcal{U}}$(0.3442,0.1,0.1,0.0,1.0) & \photoqoneLCO \vspace{1mm}\\
    Quadratic LD LCOGT ($q_{\text{2,LCO}}$) & $\mathcal{N}_{\mathcal{U}}$(0.1684,0.1,0.0,1.0) & \photoqtwoLCO \vspace{1mm}\\
    \hline 
    \multicolumn{3}{c}{Instrumental} \\
    \hline 
    Flux offset \textit{TESS}-1 ($\mu_{\text{\textit{TESS}1}}$) / rel. flux & $\mathcal{N}$(0.0,0.1) & \photomfluxTESSone \vspace{1mm}\\
    Flux offset \textit{TESS}-2 ($\mu_{\text{\textit{TESS}2}}$) / rel. flux & $\mathcal{N}$(0.0,0.1) & \photomfluxTESStwo \vspace{1mm}\\
    Jitter \textit{TESS}-1 ($\sigma_{\text{WN,\textit{TESS}1}}$) / ppm & ln$\mathcal{U}$(0.1,1e3) & \photosigmawTESSone \vspace{1mm}\\
    Jitter \textit{TESS}-2 ($\sigma_{\text{WN,\textit{TESS}2}}$) / ppm & ln$\mathcal{U}$(0.1,1e3) & \photosigmawTESStwo \vspace{1mm}\\
    M32 GP amplitude \textit{TESS}-1 ($\sigma_{\text{M32,\textit{TESS}1}}$) / rel. flux & ln$\mathcal{U}$(1e-6,1e6) & \photoGPsigmaTESSone \vspace{1mm}\\
    M32 GP timescale \textit{TESS}-1 ($\rho_{\text{M32,\textit{TESS}1}}$) / days & ln$\mathcal{U}$(1e-3,1e3) & \photoGPrhoTESSone \vspace{1mm}\\
    SHO GP power \textit{TESS}-1 ($S_{\text{0,\textit{TESS}1}}$) / (rel. flux)$^{2}$days & ln$\mathcal{U}$(1e-20,1.0) & \photoGPSzeroTESSone \vspace{1mm}\\
    SHO GP frequency \textit{TESS}-1 ($\omega_{\text{0,\textit{TESS}1}}$) / days$^{-1}$ & $\mathcal{N}$(28.545,0.1) & \photoGPomegazeroTESSone \vspace{1mm}\\
    SHO GP quality factor \textit{TESS}-1 ($Q_{\text{\textit{TESS}1}}$) & ln$\mathcal{U}$(0.01,1e4) & \photoGPQTESSone \vspace{1mm}\\
    M32 GP amplitude \textit{TESS}-2 ($\sigma_{\text{M32,\textit{TESS}2}}$) / rel. flux & ln$\mathcal{U}$(1e-6,1e6) & \photoGPsigmaTESStwo \vspace{1mm}\\
    M32 GP timescale \textit{TESS}-2 ($\rho_{\text{M32,\textit{TESS}2}}$) / days & ln$\mathcal{U}$(1e-3,1e3) & \photoGPrhoTESStwo \vspace{1mm}\\
    Flux offset \textit{CHEOPS}-1 ($\mu_{\text{\textit{CHEOPS}1}}$) / rel. flux & $\mathcal{N}$(0.0,0.1) & \photomfluxCHEOPSone \vspace{1mm}\\
    Flux offset \textit{CHEOPS}-2 ($\mu_{\text{\textit{CHEOPS}2}}$) / rel. flux & $\mathcal{N}$(0.0,0.1) & \photomfluxCHEOPStwo \vspace{1mm}\\
    Flux offset \textit{CHEOPS}-3 ($\mu_{\text{\textit{CHEOPS}3}}$) / rel. flux & $\mathcal{N}$(0.0,0.1) & \photomfluxCHEOPSthree \vspace{1mm}\\
    Flux offset \textit{CHEOPS}-4 ($\mu_{\text{\textit{CHEOPS}4}}$) / rel. flux & $\mathcal{N}$(0.0,0.1) & \photomfluxCHEOPSfour \vspace{1mm}\\
    Flux offset \textit{CHEOPS}-5 ($\mu_{\text{\textit{CHEOPS}5}}$) / rel. flux & $\mathcal{N}$(0.0,0.1) & \photomfluxCHEOPSfive \vspace{1mm}\\
    Flux offset \textit{CHEOPS}-6 ($\mu_{\text{\textit{CHEOPS}6}}$) / rel. flux & $\mathcal{N}$(0.0,0.1) & \photomfluxCHEOPSsix \vspace{1mm}\\
    Jitter \textit{CHEOPS}-1 ($\sigma_{\text{WN,\textit{CHEOPS}1}}$) / ppm & ln$\mathcal{U}$(0.1,1e3) & \photosigmawCHEOPSone \vspace{1mm}\\
    Jitter \textit{CHEOPS}-2 ($\sigma_{\text{WN,\textit{CHEOPS}2}}$) / ppm & ln$\mathcal{U}$(0.1,1e3) & \photosigmawCHEOPStwo \vspace{1mm}\\
    Jitter \textit{CHEOPS}-3 ($\sigma_{\text{WN,\textit{CHEOPS}3}}$) / ppm & ln$\mathcal{U}$(0.1,1e3) & \photosigmawCHEOPSthree \vspace{1mm}\\
    Jitter \textit{CHEOPS}-4 ($\sigma_{\text{WN,\textit{CHEOPS}4}}$) / ppm & ln$\mathcal{U}$(0.1,1e3) & \photosigmawCHEOPSfour \vspace{1mm}\\
    Jitter \textit{CHEOPS}-5 ($\sigma_{\text{WN,\textit{CHEOPS}5}}$) / ppm & ln$\mathcal{U}$(0.1,1e3) & \photosigmawCHEOPSfive \vspace{1mm}\\
    Jitter \textit{CHEOPS}-6 ($\sigma_{\text{WN,\textit{CHEOPS}6}}$) / ppm & ln$\mathcal{U}$(0.1,1e3) & \photosigmawCHEOPSsix \\
  \end{tabular}}
\label{tab:photo-priors+posteriors}
\end{table*}
\begin{table*}
\ContinuedFloat
\centering
  \caption{(continued)}
  {\begin{tabular}{>{\centering\arraybackslash}m{6cm}>{\centering\arraybackslash}m{4cm}>{\centering\arraybackslash}m{4cm}}
    \hline
    Parameter & Prior & Posterior Value \\
    \hline
    bg detrending coefficient \textit{CHEOPS}-1 & $\mathcal{U}$(-1,1) & \photothetazeroCHEOPSone \vspace{1mm}\\ 
    t detrending coefficient \textit{CHEOPS}-1 & $\mathcal{U}$(-1,1) & \photothetaoneCHEOPSone \vspace{1mm}\\ 
    cos($\phi$) detrending coefficient \textit{CHEOPS}-1 & $\mathcal{U}$(-1,1) & \photothetatwoCHEOPSone \vspace{1mm}\\ 
    x detrending coefficient \textit{CHEOPS}-2 & $\mathcal{U}$(-1,1) & \photothetazeroCHEOPStwo \vspace{1mm}\\ 
    y detrending coefficient \textit{CHEOPS}-2 & $\mathcal{U}$(-1,1) & \photothetaoneCHEOPStwo \vspace{1mm}\\
    bg detrending coefficient \textit{CHEOPS}-3 & $\mathcal{U}$(-1,1) & \photothetazeroCHEOPSthree \vspace{1mm}\\ 
    x detrending coefficient \textit{CHEOPS}-3 & $\mathcal{U}$(-1,1) & \photothetaoneCHEOPSthree \vspace{1mm}\\ 
    y detrending coefficient \textit{CHEOPS}-3 & $\mathcal{U}$(-1,1) & \photothetatwoCHEOPSthree \vspace{1mm}\\  
    t detrending coefficient \textit{CHEOPS}-3 & $\mathcal{U}$(-1,1) & \photothetathreeCHEOPSthree \vspace{1mm}\\ 
    cos(3$\phi$) detrending coefficient \textit{CHEOPS}-3 & $\mathcal{U}$(-1,1) & \photothetafourCHEOPSthree \vspace{1mm}\\
    bg detrending coefficient \textit{CHEOPS}-4 & $\mathcal{U}$(-1,1) & \photothetaoneCHEOPSfour \vspace{1mm}\\
    y detrending coefficient \textit{CHEOPS}-4 & $\mathcal{U}$(-1,1) & \photothetazeroCHEOPSfour \vspace{1mm}\\  
    t detrending coefficient \textit{CHEOPS}-4 & $\mathcal{U}$(-1,1) & \photothetathreeCHEOPSfour \vspace{1mm}\\ 
    cos(3$\phi$) detrending coefficient \textit{CHEOPS}-4 & $\mathcal{U}$(-1,1) & \photothetatwoCHEOPSfour \vspace{1mm}\\
    bg detrending coefficient \textit{CHEOPS}-5 & $\mathcal{U}$(-1,1) & \photothetazeroCHEOPSfive \vspace{1mm}\\ 
    x detrending coefficient \textit{CHEOPS}-5 & $\mathcal{U}$(-1,1) & \photothetatwoCHEOPSfive \vspace{1mm}\\ 
    y detrending coefficient \textit{CHEOPS}-5 & $\mathcal{U}$(-1,1) & \photothetaoneCHEOPSfive \vspace{1mm}\\  
    t detrending coefficient \textit{CHEOPS}-5 & $\mathcal{U}$(-1,1) & \photothetathreeCHEOPSfive \vspace{1mm}\\ 
    cos(2$\phi$) detrending coefficient \textit{CHEOPS}-5 & $\mathcal{U}$(-1,1) & \photothetafourCHEOPSfive \vspace{1mm}\\
    sin(3$\phi$) detrending coefficient \textit{CHEOPS}-5 & $\mathcal{U}$(-1,1) & \photothetafiveCHEOPSfive \vspace{1mm}\\
    bg detrending coefficient \textit{CHEOPS}-6 & $\mathcal{U}$(-1,1) & \photothetaoneCHEOPSsix \vspace{1mm}\\
    y detrending coefficient \textit{CHEOPS}-6 & $\mathcal{U}$(-1,1) & \photothetazeroCHEOPSsix \vspace{1mm}\\  
    t detrending coefficient \textit{CHEOPS}-6 & $\mathcal{U}$(-1,1) & \photothetatwoCHEOPSsix \vspace{1mm}\\ 
    Flux offset LCOGT-1 ($\mu_{\text{LCO1}}$) / rel. flux & $\mathcal{N}$(0.0,0.1) & \photomfluxLCOone \vspace{1mm}\\
    Flux offset LCOGT-2 ($\mu_{\text{LCO2}}$) / rel. flux & $\mathcal{N}$(0.0,0.1) & \photomfluxLCOtwo \vspace{1mm}\\
    Jitter LCOGT-1 ($\sigma_{\text{WN,LCO1}}$) / ppm & ln$\mathcal{U}$(0.1,1e4) & \photosigmawLCOone \vspace{1mm}\\
    Jitter LCOGT-2 ($\sigma_{\text{WN,LCO2}}$) / ppm & ln$\mathcal{U}$(0.1,1e4) & \photosigmawLCOone \vspace{1mm}\\
    airmass detrending coefficient LCOGT-1 & $\mathcal{U}$(-1,1) & \photothetazeroLCOone \vspace{1mm}\\
    FWHM detrending coefficient LCOGT-1 & $\mathcal{U}$(-1,1) & \photothetaoneLCOone \vspace{1mm}\\
    airmass detrending coefficient LCOGT-2 & $\mathcal{U}$(-1,1) & \photothetazeroLCOtwo \vspace{1mm}\\
    FWHM detrending coefficient LCOGT-2 & $\mathcal{U}$(-1,1) & \photothetaoneLCOtwo \vspace{1mm}\\
    \hline
  \end{tabular}}
\end{table*}

\bsp	
\label{lastpage}

\end{document}